%% file: manuscript.tex
\documentclass[journal=jctcce,manuscript=article
,layout=twocolumn
]{achemso}
\usepackage[fontsize=9pt]{fontsize}
\usepackage{chemformula} 
\usepackage[T1]{fontenc} 
\usepackage{amsmath}
\usepackage{appendix}
\usepackage{booktabs}
\usepackage[hidelinks]{hyperref}

\author{Richard Einsele}
\author{Roland Mitri\'c}
\email{roland.mitric@uni-wuerzburg.de}
\affiliation{Institut für Physikalische und Theoretische Chemie, Emil-Fischer-Strasse 42, Julius-Maximilians-Universität, Würzburg, Germany}

\title
  {FMO-xTB: Fragment molecular orbital method with GFN1-xTB for large-scale quantum-mechanical simulations}


\let\oldmaketitle\maketitle
\let\maketitle\relax

\begin{document}

\begin{tocentry}
\includegraphics[width=\linewidth]{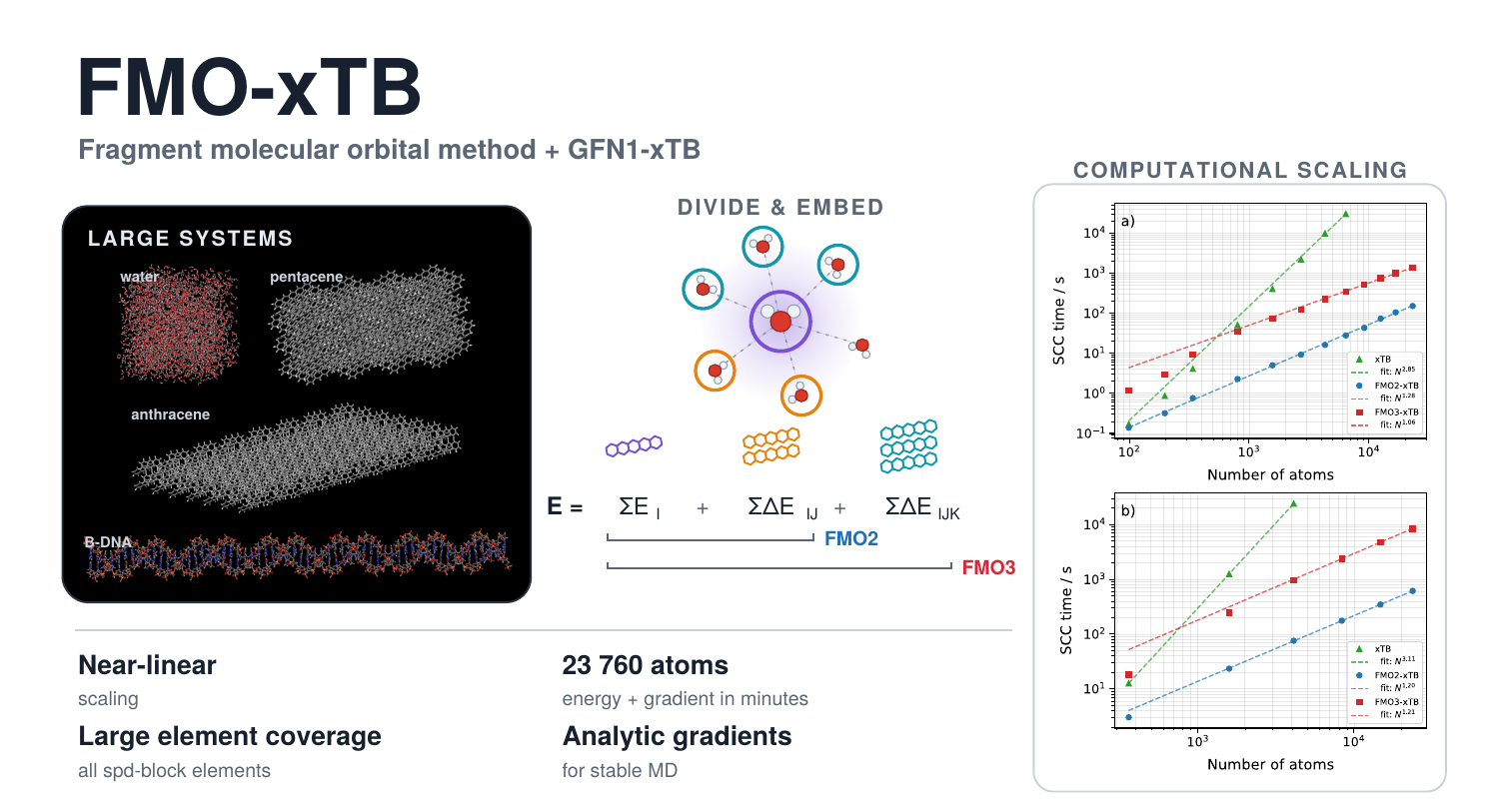}
\end{tocentry}

\twocolumn[
\begin{@twocolumnfalse}
\oldmaketitle
\begin{abstract}
We present the fragment molecular orbital method (FMO) combined with the GFN1-xTB extended tight-binding approach (FMO-xTB) for efficient quantum-mechanical calculations of large molecular systems. Both the two-body (FMO2) and three-body (FMO3) expansions are formulated, and fully analytic energy gradients including the response contribution from the self-consistent embedding potential are derived and implemented. The FMO-xTB method inherits the broad element coverage of GFN1-xTB, which employs element-specific rather than atom-pair-specific parameters and is parameterized for all spd-block elements up to radon ($Z = 86$), representing a significant practical advantage over FMO-DFTB approaches. The accuracy of FMO-xTB is systematically benchmarked against non-fragmented xTB calculations for water clusters, anthracene aggregates, and pentacene supercells. FMO3-xTB reproduces the reference energies with deviations on the order of $10^{-4}$~Hartree for organic semiconductor systems. The covalent bond fragmentation capability using the hybrid orbital projection (HOP) boundary treatment is also implemented with fully analytic gradients and validated for polyalanine $\alpha$-helices and B-DNA double helices, yielding FMO3-xTB energy deviations on the order of $10^{-6}$~Hartree for polyalanine and in the millihartree range for B-DNA. Near-linear scaling is achieved with effective scaling exponents between $b = 1.06$ and $b = 1.28$, compared to cubic scaling for non-fragmented xTB. Parallelization over multiple CPU cores yields significant speed ups, and a complete energy and gradient evaluation of a pentacene supercell containing 23\,760 atoms is feasible within minutes on a single computing node, enabling routine molecular dynamics simulations of systems with tens of thousands of atoms. The method is implemented in the DIALECT software package.
\end{abstract}
\end{@twocolumnfalse}
]

\input{introduction}
\input{theory}
\input{results}
\input{conclusion}

\begin{acknowledgement}
We gratefully acknowledge financial support by the Deutsche Forschungsgemeinschaft (DFG) via the SFB 1762, project number 551403841.
\end{acknowledgement}

\textbf{Data Availability}
The data that supports this work is available at \href{https://github.com/mitric-lab/DATA_FMO-xTB-paper}{\url{https://github.com/mitric-lab/DATA_FMO-xTB-paper}}.

\newpage
\bibliography{references}

\end{document}

%% file: introduction.tex
\section{\label{sec:level1} Introduction }
 
The quantum-mechanical description of large molecular systems remains one of the central challenges in computational chemistry.\cite{houk_holy_2017,grimme_computational_2018} While density functional theory (DFT) has become the method of choice for molecular simulations, providing a favorable balance between accuracy and computational cost, its applicability is limited to systems containing at most a few hundred atoms due to the steep scaling with system size.\cite{houk_holy_2017,grimme_computational_2018} This limitation poses significant obstacles for the theoretical investigation of large biomolecular systems, nanomaterials, and complex molecular assemblies.\cite{gordon_fragmentation_2012,raghavachari_accurate_2015}
 
Semiempirical quantum-mechanical methods offer a computationally efficient alternative by introducing systematic approximations while retaining an explicit quantum-mechanical description of the electronic structure.\cite{pople_approximate_1965,dewar_ground_1977,stewart_optimization_1989,thiel_semiempirical_2014,christensen_semiempirical_2016} Among the various semiempirical approaches, density-functional tight-binding (DFTB) methods have gained widespread popularity due to their derivation from DFT via an expansion around a reference density.\cite{elstner_self-consistent-charge_1998,cui_density_2014} The self-consistent-charge DFTB (SCC-DFTB or DFTB2) method\cite{elstner_self-consistent-charge_1998} and its third-order extension DFTB3\cite{gaus_dftb3_2011} have become established tools for large-scale simulations in biochemistry and materials science.\cite{thiel_semiempirical_2014,christensen_semiempirical_2016,cui_density_2014} However, the major drawback of DFTB lies in its atom-pair-wise parameterization, which requires the determination of thousands of empirical parameters and consequently limits its applicability to a relatively small subset of elements in the periodic table.\cite{bannwarth_extended_2021,grimme_robust_2017}
 
To address the parameterization limitations of DFTB, Grimme and coworkers developed the extended tight-binding (xTB) family of methods.\cite{grimme_robust_2017,bannwarth_gfn2-xtbaccurate_2019,grimme_non-self-consistent_2023,froitzheim_g-xtb_2025} The GFN-xTB methods (GFN standing for geometries, frequencies, and noncovalent interactions) employ primarily element-specific rather than atom-pair-specific parameters, enabling a consistent parameterization covering the periodic table up to element 86 (radon).\cite{grimme_robust_2017,bannwarth_gfn2-xtbaccurate_2019} The GFN1-xTB method incorporates the well-established D3 London dispersion correction\cite{grimme_consistent_2010} and uses coordination-number-dependent energy levels to describe diverse bonding environments.\cite{grimme_robust_2017} The successor GFN2-xTB further improved the electrostatic treatment through multipole expansion and is completely pair-parameter-free.\cite{bannwarth_gfn2-xtbaccurate_2019} Most recently, the g-xTB method\cite{froitzheim_g-xtb_2025} has been proposed as a next-generation general-purpose tight-binding approach incorporating range-separated approximate Fock exchange and higher-order charge fluctuation terms, targeting DFT-level accuracy while retaining tight-binding efficiency. This broad element coverage represents a decisive advantage over conventional DFTB and makes xTB methods attractive for simulating chemically diverse systems, particularly those containing transition metals or other elements for which DFTB parameters are unavailable.
 
Despite their computational efficiency compared to DFT, semiempirical methods still exhibit cubic scaling with system size due to the matrix diagonalization step required in each self-consistent field iteration.\cite{bannwarth_extended_2021,nishizawa_three_2016} This scaling behavior renders the direct application of semiempirical methods ultimately impractical for very large systems containing thousands of atoms. To overcome this limitation, several strategies have been developed. Hybrid quantum mechanics/molecular mechanics (QM/MM) approaches\cite{warshel_theoretical_1976,senn_qmmm_2009} and ONIOM methods\cite{dapprich_new_1999,plett_oniom_2023} treat only the chemically relevant region at the QM level while describing the environment with molecular mechanics force fields. Alternatively, subsystem DFT methods\cite{jacob_subsystem_2014} and various fragmentation schemes\cite{gordon_fragmentation_2012,raghavachari_accurate_2015} exploit the locality of electronic structure to achieve linear or near-linear scaling.
 
Among fragmentation-based approaches, the fragment molecular orbital (FMO) method introduced by Kitaura and coworkers\cite{kitaura_fragment_1999,nakano_fragment_2000,nakano_fragment_2002} has proven particularly successful for large-scale quantum-mechanical calculations. In the FMO method, a molecular system is divided into fragments, and quantum-mechanical calculations are performed for individual fragments and their pairs (dimers) in the presence of an electrostatic embedding potential generated by all other fragments.\cite{fedorov_extending_2007,fedorov_systematic_2010} The total energy and properties are then obtained by combining the fragment contributions using a many-body expansion.\cite{fedorov_importance_2004,fedorov_three-body_2006} The distinctive feature of FMO is the inclusion of the electrostatic field from the entire system in each fragment calculation, ensuring a proper description of polarization effects.\cite{fedorov_extending_2007} The FMO method has been developed for various levels of electronic structure theory, including Hartree-Fock, DFT, and correlated methods such as MP2 and coupled cluster theory,\cite{fedorov_importance_2004,fedorov_three-body_2006,fedorov_second_2004,fedorov_multilayer_2005,tanaka_electron-correlated_2014} and has been extensively applied to biological systems.\cite{fukuzawa_fragment_2022,sawada_binding_2010} Recent developments include the decomposition of polarization energies\cite{fedorov_polarization_2022} and noncovalent interaction analysis\cite{fedorov_decomposition_2025} within the FMO framework.
 
The divide-and-conquer (DC) approach, originally proposed by Yang and Lee,\cite{yang_direct_1991,yang_densitymatrix_1995} provides an alternative strategy for linear-scaling quantum-mechanical calculations.\cite{nakai_divide-and-conquer_2023} In the DC method, the system is spatially partitioned into subsystems, and the density matrix of the total system is reconstructed from subsystem contributions calculated using localized Hamiltonian matrices.\cite{akama_implementation_2007,kobayashi_extension_2008,kobayashi_divide-and-conquer_2010} The DC approach has been combined with DFTB by Nakai and coworkers,\cite{nishizawa_three_2016,nakai_divide-and-conquer-type_2016} resulting in the DC-DFTB method that enables quantum-mechanical molecular dynamics simulations of systems containing up to millions of atoms on modern supercomputers.\cite{nishimura_dcdftbmd_2019}
 
The combination of fragmentation approaches with semiempirical methods has opened new avenues for large-scale quantum-mechanical simulations. The FMO-DFTB method, developed by Nishimoto, Fedorov, and Irle,\cite{nishimoto_density-functional_2014,nishimoto_large-scale_2015,nishimoto_third-order_2015} combines the efficiency of DFTB with the linear-scaling capabilities of FMO. This approach has been extended to include three-body terms (FMO3-DFTB),\cite{nishimoto_three-body_2017} the polarizable continuum model for implicit solvation,\cite{nishimoto_fragment_2016} long-range corrected functionals (FMO-LC-DFTB),\cite{vuong_fragment_2019} analytic second derivatives,\cite{nakata_analytic_2016} and periodic boundary conditions.\cite{nishimoto_fragment_2021} Fully analytic gradients for FMO-DFTB enable efficient molecular dynamics simulations,\cite{nishimoto_large-scale_2015} and the method has demonstrated speed-ups of over two orders of magnitude compared to conventional DFTB for systems containing several hundred water molecules. Our group has recently extended the FMO-LC-DFTB methodology to enable the calculation of excited-state properties\cite{einsele_long-range_2023,einsele_fmo-lc-tddftb_2024} and nonadiabatic dynamics\cite{einsele_nonadiabatic_2024} of large molecular assemblies using the DIALECT software package.\cite{einsele_dialect_2025}
 
Very recently, the combination of the DC method with the xTB approach has been reported by Kobayashi and coworkers,\cite{nishida_divide-and-conquer_2025} demonstrating the potential of fragmentation-based xTB methods for large-scale simulations. The DC-xTB method inherits the broad element coverage of xTB while achieving linear scaling through the divide-and-conquer framework.
 
In this work, we present the combination of the fragment molecular orbital method with the GFN1-xTB method, referred to as FMO-xTB. This development extends the capabilities of the FMO framework to systems containing diverse chemical elements while maintaining the efficiency advantages of tight-binding methods. The FMO-xTB method supports both the two-body (FMO2) and three-body (FMO3) expansions, providing a systematic way to control the accuracy-cost trade-off. We derive and implement analytic gradients, enabling geometry optimizations and molecular dynamics simulations. The method is implemented in the DIALECT software package.\cite{einsele_dialect_2025}
 
In contrast to our earlier work on FMO-LC-TDDFTB for excited-state calculations,\cite{einsele_long-range_2023,einsele_fmo-lc-tddftb_2024,einsele_nonadiabatic_2024} the present work focuses exclusively on ground-state simulations of large molecular systems. Following this introduction, we present the theoretical framework of FMO-xTB, including the expressions for the total energy and analytic gradients. We then demonstrate the accuracy of the FMO approach through systematic comparisons with conventional xTB calculations for both non-covalent molecular clusters and covalently bonded biomolecular systems, assess the computational scaling and parallel efficiency of the implementation, verify the analytic gradient against numerical references, and report gradient timings for representative molecular systems.

%% file: theory.tex
\section{\label{sec:methodology} Methodology}

In the following, we present the theoretical framework of the fragment molecular orbital method combined with the GFN1-xTB method (FMO-xTB). We begin with a brief review of the GFN1-xTB method, followed by a description of the fragment molecular orbital ansatz, and conclude with the formulation of the FMO-xTB method at both the two-body (FMO2) and three-body (FMO3) levels, as well as the derivation of fully analytic energy gradients.

Throughout this section, we use the following notation: uppercase Latin letters $A$, $B$, $C$ denote atoms, while uppercase Latin letters $I$, $J$, $K$, $X$, $Y$ denote molecular fragments. Lowercase Greek letters $\mu$, $\nu$, $\lambda$ represent atomic orbital (AO) shells, and $l$, $l'$ label the angular momentum type (s, p, d) of a shell. For molecular orbitals (MOs), $i$, $j$ denote occupied and $a$, $b$ denote virtual orbital indices, with $C_{\mu i}$ being the MO expansion coefficients. The symbol $\xi$ denotes a Cartesian nuclear coordinate component in gradient expressions.

\subsection{\label{sec:gfn1xtb}GFN1-xTB}

The GFN1-xTB method\cite{grimme_robust_2017} belongs to the family of extended tight-binding (xTB) methods developed by Grimme and coworkers.\cite{bannwarth_gfn2-xtbaccurate_2019,bannwarth_extended_2021} In contrast to the density-functional tight-binding (DFTB) approach,\cite{elstner_self-consistent-charge_1998,gaus_dftb3_2011} which relies on atom-pair-specific parameterization, the GFN1-xTB method employs predominantly element-specific parameters.\cite{grimme_robust_2017}
The method was specifically designed to provide accurate structures, vibrational frequencies, and noncovalent interactions for large molecular systems. The GFN-xTB1 method uses a Slater-type basis set with angular momenta up to $l=2$, which is described in Ref. \citenum{grimme_robust_2017}.

The total energy in GFN1-xTB is expressed as\cite{grimme_robust_2017}
\begin{equation}
E=E_{\mathrm{el}}+E_{\mathrm{rep}}+E_{\mathrm{disp}}+E_{\mathrm{XB}}
\label{eq:gfn1-energy}
\end{equation}
where $E_{\mathrm{el}}$ is the electronic energy, $E_{\mathrm{rep}}$ is the repulsion energy, $E_{\mathrm{disp}}$ is the D3 dispersion correction,\cite{grimme_consistent_2010} and $E_{\mathrm{XB}}$ describes halogen-bond interactions. The electronic energy contains contributions from the zeroth-order Hamiltonian, second-order and third-order charge fluctuations, and the electronic entropy term for fractional occupation numbers,
\begin{equation}
\begin{aligned}
E_{\mathrm{el}}= & \sum_{\mu \nu}
P_{\mu \nu }H^0_{\mu \nu}+\frac{1}{2} \sum_{\mu \nu} \Delta q_{\mu}^A \Delta q_{\nu}^B \gamma_{\mu \nu} \\
& -\frac{1}{3} \sum_{\mathrm{A}} \Gamma_{\mathrm{A}} q_{\mathrm{A}}^3-T_{\mathrm{el}} S_{\mathrm{el}}
\end{aligned}
\label{eq:gfn1_energy_2}
\end{equation}
where $P_{\mu\nu}$ is the density matrix, $H^0_{\mu\nu}$ is the zeroth-order Hamiltonian matrix element, and $\gamma_{\mu\nu}$ is the Coulomb interaction kernel. The third term describes third-order charge fluctuations with the element-specific Hubbard derivative $\Gamma_A$, and the last term accounts for the electronic entropy at finite electronic temperature $T_{\mathrm{el}}$.\cite{grimme_robust_2017}

The shell-resolved Mulliken charge differences $\Delta q_{\mu}^A$ are computed as
\begin{equation}
\Delta q_{\mu}^{\mathrm{A}}=\sum_{\mu \in \mathrm{A}}\sum_\nu  S_{\mu \nu} P_{\mu \nu} -q_{\mu}^{0}
\end{equation}
where $S_{\mu\nu}$ is the overlap matrix and $q_\mu^0$ is the reference occupation of shell $\mu$. The Coulomb interaction kernel $\gamma_{\mu\nu}$ interpolates between the short-range limit determined by the chemical hardness and the long-range Coulomb interaction,
\begin{equation}
\gamma_{\mu \nu}=\left(\frac{1}{R_{\mathrm{AB}}^{k_{\mathrm{g}}}+\eta^{-k_{\mathrm{g}}}}\right)^{1 / k_{\mathrm{g}}} \quad\left(\mu \in A, \nu \in \mathrm{B}\right)
\end{equation}
where $R_{AB}$ is the interatomic distance and $k_g$ is a global parameter. The effective chemical hardness $\eta$ is computed from the element-specific hardnesses $\eta_A$ and $\eta_B$ as
\begin{equation}
\eta=2\left(\frac{1}{\left(1+\kappa_{\mathrm{A}}^l\right) \eta_{\mathrm{A}}}+\frac{1}{\left(1+\kappa_{\mathrm{B}}^{l^{\prime}}\right) \eta_{\mathrm{B}}}\right)^{-1}
\end{equation}
with shell-dependent scaling factors $\kappa_A^l$.

For systems with small HOMO-LUMO gaps or near-degeneracies, GFN1-xTB employs a Fermi-Dirac smearing of the occupation numbers,
\begin{equation}
n_i\left(T_{\mathrm{el}}\right)=\frac{1}{\exp \left[\left(\epsilon_i-\epsilon_{\mathrm{F}}\right) /\left(k_{\mathrm{B}} T_{\mathrm{el}}\right)\right]+1}
\end{equation}
where $\epsilon_i$ are the orbital energies, $\epsilon_F$ is the Fermi energy, and $k_B$ is the Boltzmann constant.\cite{grimme_robust_2017}

The self-consistent field (SCF) procedure requires the construction of the Fock matrix $F_{\mu\nu}$, which in GFN1-xTB takes the form
\begin{equation}
\begin{aligned}
F_{\mu \nu}&= H^0_{\mu \nu}+\frac{1}{2} S_{\mu \nu} \sum_{\lambda \in C} \left(\gamma_{\mu \lambda}+\gamma_{\nu \lambda}\right) \Delta q_{\lambda}^C \\
&-\frac{1}{2} S_{\mu \nu}\left(\Delta q_{\mathrm{A}}^2 \Gamma_A+\Delta q_{\mathrm{B}}^2 \Gamma_{\mathrm{B}}\right) \quad\left(\mu \in A, \nu \in \mathrm{B}\right)
\end{aligned}
\end{equation}
The first term is the zeroth-order Hamiltonian, the second term accounts for the second-order charge fluctuations, and the third term describes the third-order contributions.\cite{grimme_robust_2017} $\Delta q_A$ are the atom-centered Mulliken charge differences.

The zeroth-order Hamiltonian matrix elements $H^0_{\mu\nu}$ are constructed according to
\begin{equation}
\begin{aligned}
H^0_{\mu \nu}=&K_{\mathrm{AB}} \frac{1}{2}\left(k_l+k_{l^{\prime}}\right) \frac{1}{2}\left(h_{\mathrm{A}}^l+h_{\mathrm{B}}^{l^{\prime}}\right) S_{\mu \nu} \\
&\times \left(1+k_{EN} \Delta EN_{\mathrm{AB}}^2\right) \Pi\left(R_{\mathrm{AB}}\right) \quad\left(\mu \in A, \nu \in \mathrm{B}\right)
\end{aligned}
\label{eq:h0_hamiltonian}
\end{equation}
where $K_{AB}$ is a pair-specific scaling constant, $k_l$ and $k_{l'}$ are angular momentum-dependent parameters, and the electronegativity difference $\Delta EN_{AB}$ accounts for polar bonding effects. The polynomial distance correction $\Pi$ is given by
\begin{equation}
\begin{aligned}
\Pi\left(R_{\mathrm{AB}}\right)=&\left(1+k_{\mathrm{A}, l}^{\mathrm{poly}}\left(\frac{R_{\mathrm{AB}}}{R_{\mathrm{cov}, \mathrm{AB}}}\right)^{1 / 2}\right)\\
&\times\left(1+k_{\mathrm{B}, l^{\prime}}^{\mathrm{poly}}\left(\frac{R_{\mathrm{AB}}}{R_{\mathrm{cov}, \mathrm{AB}}}\right)^{1 / 2}\right)
\end{aligned}
\label{eq:xtb1_polynomial_correction}
\end{equation}
where $R_{\mathrm{cov},AB}$ is the sum of covalent radii. A distinctive feature of GFN1-xTB is the coordination number dependence of the diagonal Hamiltonian elements,\cite{grimme_robust_2017}
\begin{equation}
h_{\mathrm{A}}^l=H_{\mathrm{A}}^l\left(1+k_{\mathrm{CN}, l} CN_{\mathrm{A}}\right)
\label{eq:xtb1_diagonal_h0}
\end{equation}
where $H_A^l$ is the element- and shell-specific energy level parameter and $CN_A$ is the coordination number of atom $A$.

The repulsion energy $E_{\mathrm{rep}}$ accounts for the core-core repulsion and is given by
\begin{equation}
E_{\mathrm{rep}}=\sum_{A B} \frac{Z_{\mathrm{A}}^{\mathrm{eff}} Z_{\mathrm{B}}^{\mathrm{eff}}}{R_{\mathrm{AB}}} e^{-\left(\alpha_{\mathrm{A}} \alpha_{\mathrm{B}}\right)^{1 / 2}\left(R_{\mathrm{AB}}\right)^{k_f}}
\label{eq:gfn1_rep_energy}
\end{equation}
where $Z_A^{\mathrm{eff}}$ is the effective nuclear charge and $\alpha_A$ is an element-specific decay parameter.\cite{grimme_robust_2017}

For systems containing halogen atoms, an additional halogen-bond correction $E_{\mathrm{XB}}$ is included,
\begin{equation}
\begin{aligned}
E_{\mathrm{XB}}=&\sum_{\mathrm{XB}} f_{\mathrm{dmp}}^{\mathrm{AXB}} k_{\mathrm{X}}\left(1+\left(\frac{R_{\mathrm{cov}, \mathrm{AX}}}{R_{\mathrm{AX}}}\right)^{12}\right.\\
&\left.-k_{\mathrm{X} 2}\left(\frac{R_{\mathrm{cov}, \mathrm{AX}}}{R_{\mathrm{AX}}}\right)^6\right) /\left(\frac{R_{\mathrm{cov}, \mathrm{AX}}}{R_{\mathrm{AX}}}\right)^{12}
\end{aligned}
\end{equation}
where $f_{\mathrm{dmp}}^{\mathrm{AXB}}$ is a damping function that depends on the A-X-B angle.

\subsection{\label{sec:fmo}Fragment molecular orbital method}

The fragment molecular orbital (FMO) method divides a large molecular system into $N$ fragments (monomers) and computes the properties of the individual fragments and their pairs (dimers) in the presence of an electrostatic embedding potential generated by all other fragments.\cite{kitaura_fragment_1999,nakano_fragment_2000,fedorov_extending_2007} The combination of FMO with DFTB was originally developed by Nishimoto, Fedorov, and Irle,\cite{nishimoto_density-functional_2014,nishimoto_large-scale_2015} and subsequently extended to include three-body corrections,\cite{nishimoto_three-body_2017} long-range corrections,\cite{vuong_fragment_2019} and excited-state properties.\cite{einsele_long-range_2023}

In the two-body FMO expansion (FMO2), the total energy of the system is expressed as\cite{nishimoto_density-functional_2014,fedorov_importance_2004}
\begin{equation}
\begin{aligned}
E^{\mathrm{FMO} 2} &=\sum_I^N E_I+\sum_{I>J}^N \left(E_{I J}-E_I-E_J +\Delta E_{I J}^V\right)\\
&=\sum_I^N E_I+\sum_{I>J}^N \Delta E_{I J}
\end{aligned}
\label{eq:fmo2-energy-components}
\end{equation}
where $E_I$ is the energy of monomer $I$ computed in the electrostatic embedding potential of all other fragments, $E_{IJ}$ is the energy of dimer $IJ$, and $\Delta E_{IJ}$ is the pair interaction energy. The term $\Delta E_{IJ}^V$ corrects for the difference in the embedding potential between the dimer and monomer calculations,\cite{nishimoto_density-functional_2014}
\begin{equation}
\Delta E_{I J}^V=\sum_{\mu \in I J} \sum_{K \neq I, J}^N \sum_{\nu \in K} \gamma_{\mu \nu} \Delta \Delta q_{\mu}^{I J} \Delta q_{\nu}^K
\label{eq:dimer_embedding_correction}
\end{equation}
where the charge transfer upon dimer formation is given by
\begin{equation}
\Delta \Delta q_{\mu}^{I J}=\Delta q_{\mu}^{I J}-\Delta q_{\mu}^I \delta_{\mu \in I}-\Delta q_{\mu}^J \delta_{\mu \in J}
\label{eq:ddq_calc}
\end{equation}
with $\delta_{\mu \in I}$ being unity if shell $\mu$ belongs to fragment $I$ and zero otherwise.

For fragment pairs that are separated by a distance larger than a specified threshold, the electrostatic dimer (ES-DIM) approximation\cite{nakano_fragment_2002,nishimoto_density-functional_2014} can be employed to avoid explicit dimer calculations. In this approximation, the dimer energy is estimated from the monomer energies and the electrostatic interaction between the fragment charges,
\begin{equation}
E_{I J} \approx E_I+E_J+\sum_{\mu \in I} \sum_{\nu \in J} \gamma_{\mu \nu} \Delta q_{\mu}^I \Delta q_{\nu}^J
\label{eq;esdimer-energy}
\end{equation}
The ES-DIM approximation significantly reduces the computational cost for large systems while maintaining high accuracy for the total energy.\cite{nishimoto_density-functional_2014}

While the two-body expansion captures the dominant contributions to the total energy, many-body effects such as charge transfer and polarization can be important in certain systems.\cite{fedorov_importance_2004,nishimoto_three-body_2017} The three-body FMO expansion (FMO3) includes trimer corrections to systematically improve the accuracy,\cite{fedorov_importance_2004,fedorov_three-body_2006,nishimoto_three-body_2017}
\begin{equation}
\begin{aligned}
E^{\mathrm{FMO} 3} & =\sum_I^N E_I+\sum_{I>J}^N \Delta E_{I J} \\
& +\sum_{I>J>K}^N\left(\Delta E_{I J K}-\Delta E_{I J}-\Delta E_{I K}-\Delta E_{J K}\right)
\end{aligned}
\label{eq:fmo3_energy}
\end{equation}
The trimer interaction energy $\Delta E_{IJK}$ is defined as\cite{nishimoto_three-body_2017}
\begin{equation}
\Delta E_{I J K}=E_{I J K}-E_I-E_J-E_K+\Delta E_{I J K}^V
\label{eq:trimer_interaction_energy}
\end{equation}
where the correction term $\Delta E_{IJK}^V$ accounts for the change in the electrostatic embedding,
\begin{equation}
\Delta E_{I J K}^V=E_{I J K, I J K}^V-E_{I, I J K}^V-E_{J, I J K}^V-E_{K, I J K}^V
\label{eq:trimer_embedding}
\end{equation}
with the general embedding energy contribution defined as
\begin{equation}
E_{X, Y}^V=\sum_{\mu \in X} \sum_{K \neq Y}^N \sum_{\nu \in K} \gamma_{\mu \nu} \Delta q_{\mu}^X \Delta q_{\nu}^K
\end{equation}
The FMO3 expansion has been shown to significantly improve the accuracy compared to FMO2 for systems with strong many-body interactions, such as water clusters and solvated ions.\cite{fedorov_importance_2004,nishimoto_three-body_2017}

\subsection{\label{sec:fmoxtb}FMO-xTB}

In the present work, we combine the FMO approach with the GFN1-xTB method. The key feature of the FMO method is the self-consistent treatment of the electrostatic embedding potential.\cite{fedorov_extending_2007,fedorov_systematic_2010} In FMO-xTB, the embedding is incorporated through the modification of the Fock matrix for each fragment. The FMO Fock matrix for monomer $I$ is given by
\begin{equation}
F_{\mu \nu}^{\mathrm{FMO},I}= F_{\mu \nu}^I + V_{\mu \nu}^I
\end{equation}
where $F_{\mu\nu}^I$ is the Fock matrix of the isolated fragment and $V_{\mu\nu}^I$ is the embedding potential matrix. The embedding potential accounts for the electrostatic interaction with all other fragments and takes the form\cite{nishimoto_density-functional_2014,einsele_long-range_2023}
\begin{equation}
V_{\mu \nu}^I=\frac{1}{2} S_{\mu \nu}^I \sum_{K \neq I} \sum_{\lambda \in K}\left(\gamma_{\mu \lambda}^{I K}+\gamma_{\nu \lambda}^{I K}\right) \Delta q_{\lambda}^K
\label{eq:v_esp_matrix}
\end{equation}
where $\Delta q_\lambda^K$ are the Mulliken charges of fragment $K$ and $\gamma_{\mu\lambda}^{IK}$ is the Coulomb interaction kernel between shells $\mu$ and $\lambda$ belonging to fragments $I$ and $K$, respectively.

The fragment calculations are performed iteratively until self-consistency of all monomer charges is achieved. Subsequently, the dimer (and trimer in FMO3) calculations are performed using the converged embedding potential. The embedding potential ensures that each fragment experiences the electrostatic field of the entire system, leading to a proper description of polarization effects.\cite{fedorov_extending_2007,fedorov_systematic_2010}

The FMO-xTB method inherits the favorable computational scaling of the FMO approach. While conventional xTB scales cubically with system size due to the matrix diagonalization step, the FMO-xTB method achieves near-linear scaling for sufficiently large systems.\cite{nishimoto_large-scale_2015} In FMO2, the computational cost is dominated by the dimer calculations, whose number scales quadratically with the number of fragments. However, by employing the ES-DIM approximation for distant fragment pairs, the effective scaling can be reduced to near-linear.\cite{nishimoto_density-functional_2014} In FMO3, the trimer calculations constitute the most demanding step, with the number of trimers scaling as $\mathcal{O}(N^3)$. Nevertheless, the use of distance thresholds to select only nearby fragment trimers significantly reduces the computational effort in practice.\cite{nishimoto_three-body_2017}

\subsection{\label{sec:gradients}Analytic energy gradients}

Analytic energy gradients are essential for efficient geometry optimizations and molecular dynamics simulations. For the FMO method combined with Hartree--Fock theory, the fully analytic energy gradient including the response contribution from the external electrostatic potentials was derived by Nagata \textit{et al.}\cite{nagata_fully_2011} using the self-consistent Z-vector (SCZV) method. The extension to FMO-DFTB was reported by Nishimoto \textit{et al.}\cite{nishimoto_large-scale_2015} In the following, we derive the analytic gradient for FMO-xTB, highlighting the modifications required by the coordination-number-dependent Hamiltonian, the shell-resolved Coulomb interaction, and the polynomial distance correction characteristic of the GFN1-xTB method.

The FMO2-xTB gradient with respect to a nuclear coordinate $\xi$ (representing a Cartesian component of some atom) can be decomposed into an explicit and a response contribution,\cite{nagata_fully_2011}
\begin{equation}
\label{eq:grad_decomp}
\frac{\partial E^{\mathrm{FMO2}}}{\partial \xi} = \left(\frac{\partial E^{\mathrm{FMO2}}}{\partial \xi}\right)_{\!\!\mathrm{expl}} + \left(\frac{\partial E^{\mathrm{FMO2}}}{\partial \xi}\right)_{\!\!\mathrm{resp}}.
\end{equation}
The explicit gradient contains all terms that can be evaluated from the converged SCF quantities without requiring the derivatives of the MO coefficients with respect to nuclear coordinates. The response gradient accounts for the implicit dependence of the MO coefficients on the nuclear coordinates through the self-consistent embedding potential, and is derived using the Z-vector method.\cite{nagata_fully_2011}

\subsubsection{GFN1-xTB energy gradient}

We first present the gradient of the GFN1-xTB total energy for a single molecular system, which serves as the building block for the FMO gradient. Differentiating the total energy [Eq.~(\ref{eq:gfn1-energy})] with respect to a nuclear coordinate $\xi$
yields\cite{grimme_robust_2017}
\begin{equation}
\label{eq:xtb_grad}
\begin{aligned}
\frac{\partial E}{\partial \xi} &= \sum_{\mu\nu} P_{\mu\nu} \frac{\partial H^0_{\mu\nu}}{\partial \xi} + \sum_{\mu\nu} W_{\mu\nu} \frac{\partial S_{\mu\nu}}{\partial \xi} \\
&+ \frac{1}{2} \sum_{\substack{A \neq B \\ \mu \in A, \nu \in B}} \Delta q_\mu^A \Delta q_\nu^B \frac{\partial \gamma_{\mu\nu}}{\partial \xi} \\
&+ \frac{\partial E_\mathrm{rep}}{\partial \xi} + \frac{\partial E_\mathrm{disp}}{\partial \xi} + \frac{\partial E_{XB}}{\partial \xi},
\end{aligned}
\end{equation}
where the first three terms arise from the electronic energy [Eq.~(\ref{eq:gfn1_energy_2})], while $\partial E_\mathrm{rep}/\partial \xi$, $\partial E_\mathrm{disp}/\partial \xi$ and $\partial E_{XB}/\partial \xi$ are the repulsive energy [Eq.~(\ref{eq:gfn1_rep_energy})], D3 dispersion and halogen-bond correction gradients, respectively, which are straightforward sums of pairwise contributions. The matrix $W_{\mu\nu}$ is the effective overlap derivative matrix,
\begin{equation}
\label{eq:W_matrix}
W_{\mu\nu} = -\tilde{W}_{\mu\nu} + \frac{\sigma_\mu + \sigma_\nu}{2} P_{\mu\nu} - \frac{1}{2}\left(\Delta q_A^2 \Gamma_A + \Delta q_B^2 \Gamma_B\right) P_{\mu\nu},
\end{equation}
with $\mu \in A$, $\nu \in B$. Here, $\tilde{W}_{\mu\nu} = \sum_i n_i \epsilon_i C_{\mu i} C_{\nu i}$ is the energy-weighted density matrix, $\sigma_\mu = \sum_\nu \gamma_{\mu\nu} \Delta q_\nu$ is the Coulomb shift on shell $\mu$, and the last term arises from the third-order charge fluctuation. The Coulomb kernel gradient $\partial \gamma_{\mu\nu}/\partial \xi$ depends on the interatomic distance $R_{AB}$ and is evaluated at the shell level.

Since $H^0_{\mu\nu} = \mathcal{H}_{AB} S_{\mu\nu}$ is proportional to the overlap, where $\mathcal{H}_{AB}$ collects the distance- and element-dependent prefactors [Eq.~(\ref{eq:h0_hamiltonian})], the Hamiltonian derivative separates into
\begin{equation}
\label{eq:H0_deriv}
\frac{\partial H^0_{\mu\nu}}{\partial \xi} = \mathcal{H}_{AB} \frac{\partial S_{\mu\nu}}{\partial \xi} + S_{\mu\nu} \frac{\partial \mathcal{H}_{AB}}{\partial \xi}.
\end{equation}
The first term combines with the effective matrix $W_{\mu\nu}$, giving the total overlap derivative contribution $\sum_{\mu\nu} (\mathcal{H}_{AB} P_{\mu\nu} + W_{\mu\nu}) \partial S_{\mu\nu}/\partial \xi$. The second term generates two gradient contributions unique to GFN1-xTB: (i) the polynomial distance correction gradient from the derivative of $\Pi(R_{AB})$ [Eq.~(\ref{eq:xtb1_polynomial_correction})], which acts as a pairwise force between atoms $A$ and $B$; and (ii) the coordination number gradient arising from the dependence of the self-energies $h_A^l$ on $CN_A$ [Eq.~(\ref{eq:xtb1_diagonal_h0})].

The coordination number $CN_A$ entering Eq.~(\ref{eq:xtb1_diagonal_h0}) is defined as a smooth, differentiable function of the interatomic distances,\cite{grimme_robust_2017}
\begin{equation}
\label{eq:cn_def}
CN_A = \sum_{B \neq A} \frac{1}{1 + \exp\!\left[-k_1\left(k_2 \frac{R_{\mathrm{cov},A} + R_{\mathrm{cov},B}}{R_{AB}} - 1\right)\right]},
\end{equation}
where $R_{\mathrm{cov},A}$ is the covalent radius of atom $A$ and $k_1$, $k_2$ are global parameters. The coordination number gradient takes the form
\begin{equation}
\label{eq:cn_grad}
G^\xi_\mathrm{CN} = \sum_A \mathcal{F}_A \frac{\partial CN_A}{\partial \xi},
\end{equation}
where $\partial CN_A / \partial \xi$ is obtained by differentiating Eq.~(\ref{eq:cn_def}) with respect to nuclear coordinates, and $\mathcal{F}_A$ is the sensitivity of the energy to the coordination number of atom $A$,
\begin{equation}
\label{eq:cn_factor}
\begin{aligned}
\mathcal{F}_A = \sum_{l \in A} k_{\mathrm{CN},l} H_A^l &\biggl(\sum_{\mu \in l_A} P_{\mu\mu} + \sum_{\substack{B \neq A  \nu \in B}} \bar{\mathcal{H}}^\mathrm{CN}_{AB} \!\sum_{\mu \in l_A} S_{\mu\nu} P_{\mu\nu}\biggr).
\end{aligned}
\end{equation}
Here, $\bar{\mathcal{H}}^\mathrm{CN}_{AB}$ is the Hamiltonian prefactor with the self-energy term removed, and the sums run over the angular momentum shells $l$ on atom $A$ and the corresponding AO indices $\mu$. The first term arises from the diagonal Hamiltonian elements ($H^0_{\mu\mu} = h_A^l$) and the second from the off-diagonal elements via the self-energy average $\frac{1}{2}(h_A^l + h_B^{l'})$ in Eq.~(\ref{eq:h0_hamiltonian}). Since $CN_A$ depends on the positions of all neighboring atoms through the Fermi-type counting function in Eq.~(\ref{eq:cn_def}), the coordination number gradient couples atoms that may not share a direct Hamiltonian matrix element, representing a non-local contribution absent in conventional DFTB methods.

\subsubsection{Explicit FMO-xTB gradient}

The explicit FMO-xTB gradient is obtained by differentiating the FMO2 total energy with respect to nuclear coordinates while holding the MO coefficients fixed. The FMO2 energy expression (Eq.~(\ref{eq:fmo2-energy-components})) contains contributions from the internal energies $E_X$ of the fragments (monomer or dimer), which are calculted in the electrostatic potential of all other fragments, and from the embedding energy correction $\Delta E^V_{IJ}$ involving the charge transfer upon dimer formation.
Differentiation of each term in Eq.~(\ref{eq:fmo2-energy-components}) yields the following contributions.

\paragraph{Internal fragment and pair interaction gradients.}
The derivatives $\partial E_I/\partial \xi$ for monomers and $\partial E_{IJ}/\partial \xi$ for dimers are computed using the single-system GFN1-xTB gradient [Eq.~(\ref{eq:xtb_grad})].
The pair interaction gradient
\begin{equation}
\label{eq:pair_delta}
G^\xi_{\Delta,IJ} = \frac{\partial E_{IJ}}{\partial \xi} - \frac{\partial E_I}{\partial \xi} - \frac{\partial E_J}{\partial \xi}
\end{equation}
is derived from Eq.~(\ref{eq:fmo2-energy-components}); however, the gradient of the embedding energy is treated separately. For each dimer, the composite system of fragments $I$ and $J$ is treated as a single xTB system using the pair's density matrix, orbital energies, and Mulliken charges. For ES-DIM pairs, the gradient contribution is described below [cf.\ Eq.~(\ref{eq:esd_grad})].

\paragraph{Charge-transfer embedding correction.}
Differentiating the embedding correction $\Delta E^V_{IJ}$ [Eq.~(\ref{eq:dimer_embedding_correction})] with respect to nuclear coordinates at frozen charges yields
\begin{equation}
\label{eq:ctij}
G^\xi_\mathrm{CT} = \sum_{I>J} \sum_{\substack{\mu \in IJ \\ K \neq I,J}} \sum_{\nu \in K} \Delta \Delta q^{IJ}_\mu \Delta q^K_\nu \frac{\partial \gamma_{\mu\nu}}{\partial \xi}.
\end{equation}
This term describes how the Coulomb interaction energy between the charge-transfer density $\Delta \Delta q^{IJ}$ and the environment charges $\Delta q^K$ changes due to the geometry-dependent variation of $\gamma_{\mu\nu}$, and corresponds to the ESP derivative terms in Eq.~(\ref{eq;esdimer-energy}) of Nagata \textit{et al.}\cite{nagata_fully_2011} adapted to the tight-binding Coulomb interaction.

\paragraph{Electrostatic embedding potential gradient.}
The monomer energy $E_I$ includes the embedding interaction $\sum_{\mu\nu} P^I_{\mu\nu} V^I_{\mu\nu}$. Differentiating this contribution and collecting the inter-fragment Coulomb kernel derivatives gives rise to the electrostatic embedding gradient. This term involves the accumulated dimer charge transfer
\begin{equation}
\label{eq:ctmul}
\bar{q}_\mu = \sum_{\substack{\text{pairs }IJ \\ \mu \in IJ}} \Delta \Delta q^{IJ}_\mu,
\end{equation}
which represents the total charge redistribution upon formation of all dimers containing the atom hosting shell $\mu$. The electrostatic embedding gradient then takes the form
\begin{equation}
\label{eq:embed_grad}
G^\xi_\mathrm{emb} = \sum_I \sum_{\substack{\mu \in I \\ \nu \notin I}} \Delta q^I_\mu \bar{q}_\nu \frac{\partial \gamma_{\mu\nu}}{\partial \xi},
\end{equation}
where the sum extends over all inter-fragment shell pairs.

\paragraph{ES-DIM energy gradient.}
For fragment pairs treated within the ES-DIM approximation, the pair interaction gradient simplifies to the derivative of the Coulomb interaction between monomer charges [Eq.~(\ref{eq;esdimer-energy})]:
\begin{equation}
\label{eq:esd_grad}
G^\xi_\mathrm{ESD} = \sum_{\substack{(I,J) \in \mathrm{ESD} \\ \mu \in I, \nu \in J}} \Delta q^I_\mu \Delta q^J_\nu \frac{\partial \gamma_{\mu\nu}}{\partial \xi}.
\end{equation}

\paragraph{Embedding Lagrangian correction.}
When a fragment is computed in the presence of the embedding potential $V^I_{\mu\nu}$ [Eq.~(\ref{eq:v_esp_matrix})], the constraint that the MOs remain orthonormal ($\mathbf{C}^{I\,T} \mathbf{S}^I \mathbf{C}^I = \mathbf{1}$) introduces an additional overlap derivative contribution to the gradient. Since $V^I_{\mu\nu}$ is proportional to the overlap matrix $S^I_{\mu\nu}$, the embedding potential modifies the Lagrange multipliers (orbital energies) compared to the isolated-fragment case, generating an extra term in the energy-weighted density matrix. This correction, which we term the embedding Lagrangian correction, corresponds to the $W$ terms appearing in the differentiation of $\mathrm{Tr}(\Delta \mathbf{P}^{IJ} \mathbf{V}^{IJ})$ [cf. Ref. \citenum{nagata_fully_2011,nishimoto_large-scale_2015}]
\begin{equation}
\label{eq:addlag}
G^\xi_{\mathrm{lag},I} = \sum_{\mu\nu \in I} \left(\mathcal{V}^{\mathrm{emb},I}_{\mu\nu} P^I_{\mu\nu} - W^{\mathrm{emb},I}_{\mu\nu}\right) \frac{\partial S^I_{\mu\nu}}{\partial \xi},
\end{equation}
where $\mathcal{V}^{\mathrm{emb},I}_{\mu\nu} = \frac{1}{2}(\sigma^\mathrm{emb}_\mu + \sigma^\mathrm{emb}_\nu)$ is the symmetrized embedding shift, with $\sigma^\mathrm{emb}_\mu$ being the electrostatic potential on shell $\mu$ from the accumulated charge transfer $\bar{q}$ [Eq.~(\ref{eq:ctmul})], corrected by subtracting the self-interaction of pairs containing the host fragment. The matrix $W^{\mathrm{emb},I}_{\mu\nu}$ is the embedding contribution to the energy-weighted density matrix,
\begin{equation}
\label{eq:W_emb}
W^{\mathrm{emb},I}_{\mu\nu} = \frac{1}{2}\left(\mathbf{P}^I \mathbf{V}^\mathrm{emb} \mathbf{S}^I + \mathbf{S}^I \mathbf{V}^\mathrm{emb} \mathbf{P}^I\right)_{\mu\nu},
\end{equation}
where $V^\mathrm{emb}_{\mu\nu} = \mathcal{V}^{\mathrm{emb},I}_{\mu\nu} S^I_{\mu\nu}$ is the embedding shift matrix projected onto the overlap. This ensures that the total gradient is consistent with the variational condition for the embedded fragment.

\paragraph{Coordination number correction.}
In the FMO framework, the coordination numbers must be computed at the supersystem level to include contributions from atoms in neighboring fragments, ensuring consistency between the energy and gradient calculations. Since $CN_A$ for an atom $A$ in fragment $I$ depends on the positions of atoms in all fragments, the CN gradient [Eq.~(\ref{eq:cn_grad})] generates inter-fragment coupling terms absent in FMO-DFTB. The CN contribution to the FMO gradient is accumulated in global coordinates, with proper subtraction of pair delta contributions:
\begin{equation}
\label{eq:cn_fmo}
\begin{aligned}
G^\xi_\mathrm{CN,FMO} &= \sum_I G^{\xi,\mathrm{CN}}_{\mathrm{mon},I} \\
&+ \sum_{I>J}\!\left(G^{\xi,\mathrm{CN}}_{\mathrm{pair},IJ} - G^{\xi,\mathrm{CN}}_{\mathrm{mon},I} - G^{\xi,\mathrm{CN}}_{\mathrm{mon},J}\right)\!.
\end{aligned}
\end{equation}

The total explicit FMO-xTB gradient collects all contributions:
\begin{equation}
\label{eq:explicit_total}
\begin{aligned}
\left(\frac{\partial E^\mathrm{FMO2}}{\partial \xi}\right)_{\!\!\mathrm{expl}} &= \sum_I G^\xi_{\mathrm{mon},I} + \sum_{I>J} G^\xi_{\Delta,IJ} + G^\xi_\mathrm{CT} \\
&+ G^\xi_\mathrm{emb} + G^\xi_\mathrm{ESD} + \sum_I G^\xi_{\mathrm{lag},I} \\
&+ G^\xi_\mathrm{CN,FMO} + G^\xi_\mathrm{disp} + G^\xi_\mathrm{XB},
\end{aligned}
\end{equation}
where the D3 dispersion gradient $G^\xi_\mathrm{disp}$ and the halogen-bond correction gradient $G^\xi_\mathrm{XB}$ are computed at the supersystem level.

\subsubsection{Response gradient}

The response gradient accounts for the fact that the MO coefficients of each fragment depend implicitly on the nuclear coordinates through the self-consistent embedding potential.\cite{nagata_fully_2011} When the geometry changes, the embedding potential is modified, which alters the MO coefficients, which in turn changes the fragment charges and the embedding potential in a self-consistent cycle. This chain of implicit dependencies gives rise to a correction that must be evaluated to obtain fully analytic gradients.

\paragraph{Lagrangian and self-consistent Z-vector equations.}
Following Nagata \textit{et al.}\cite{nagata_fully_2011} [Eqs.~(35)--(41) therein], the response contribution to the FMO gradient is reformulated using the Z-vector method to avoid solving the coupled-perturbed equations for each nuclear coordinate separately. The response contribution arises from collecting the terms involving the occupied-virtual orbital response $U^{\xi,K}_{ai}$ (the derivative of the MO coefficients with respect to coordinate $\xi$) for each fragment $K$:\cite{nagata_fully_2011,nishimoto_large-scale_2015}
\begin{equation}
\label{eq:response_Nagata}
\begin{aligned}
    \Re^\xi &= 4 \sum_{I>J} \sum_{K \neq IJ} \sum_{\mu\nu \in IJ} \sum_{\substack{a \in K \\ i \in K}}^{\mathrm{vir/occ}} \Delta P^{IJ}_{\mu \nu} U^{\xi,K}_{ai}\, (\mu\nu | ai), \\
    & =\sum_K \sum_{a \in K}^{vir}\sum_{i \in K}^{occ} U^{\xi,K}_{ai} L_{ai}^K
\end{aligned}
\end{equation}
where $(\mu\nu|ri)$ represents the two-electron Coulomb interaction, which in the tight-binding approximation reduces to the shell-resolved $\gamma$ kernel. 
The Z-vector method eliminates the coordinate dependence by solving a single set of equations for the Z-vectors $Z^I_{ai}$ (virtual $a$, occupied $i$), with the response Lagrangian serving as the driving force:
\begin{equation}
\label{eq:lagrangian}
L^K_{ai} = -\frac{1}{2} \sum_{\lambda \in K} \mathrm{ESP}^K_\lambda \, Q^K_{ai,\lambda},
\end{equation}
where
\begin{equation}
Q^K_{ai,\lambda} = \sum_{\sigma} \left[C^K_{\lambda a} C^K_{\sigma i} + C^K_{\sigma a} C^K_{\lambda i}\right]S_{\lambda \sigma}
\end{equation}
is the occupied-virtual transition charge density and
\begin{equation}
\label{eq:esp_lagrangian}
\mathrm{ESP}^K_\lambda = \sum_{\substack{I > J \\ K \notin \{I,J\}}} \sum_{\mu \in IJ} \gamma_{\lambda\mu} \Delta \Delta q^{IJ}_\mu
\end{equation}
is the electrostatic potential on shell $\lambda$ in fragment $K$ from the charge transfer of all dimer pairs not containing $K$ [cf.\ Eq.~(36) of Ref.~\citenum{nagata_fully_2011}]. The Lagrangian $L^K_{ai}$ represents the driving force for the orbital rotation $i \to a$ in fragment $K$ induced by the charge transfer in all other dimer pairs.

The Z-vectors are obtained from the self-consistent Z-vector (SCZV) equations,\cite{nagata_fully_2011}
\begin{equation}
\label{eq:sczv}
\left(\mathbf{A}^{I,I}\right)^T \mathbf{Z}^I = \mathbf{X}'^I, \quad \mathbf{X}'^I = 4\mathbf{L}^I - \sum_{K \neq I} \left(\mathbf{A}^{K,I}\right)^T \mathbf{Z}^K,
\end{equation}
where $\mathbf{A}^{I,I}$ is the intra-fragment orbital Hessian and $\mathbf{A}^{K,I}$ describes the inter-fragment coupling through the Coulomb kernel [cf.\ Eqs.~(39)--(41) and Fig.~1 of  Ref.~\citenum{nagata_fully_2011}]. This system is solved iteratively: one obtains the Z-vectors for each fragment independently, updates the right-hand side using the latest Z-vectors from all other fragments through the off-diagonal coupling $\mathbf{A}^{K,I}$, and repeats until convergence. The factor of~4 in Eq.~(\ref{eq:sczv}) arises from the closed-shell formulation.\cite{nagata_fully_2011}

For the GFN1-xTB Hamiltonian, the intra-fragment orbital Hessian [cf. Ref.~\citenum{nagata_fully_2011}] takes the form
\begin{equation}
\label{eq:orb_hessian}
\begin{aligned}
A^{I,I}_{ai,bj} &= \delta_{ab}\delta_{ij}(\epsilon^I_i - \epsilon^I_a) \\
&- 4 \sum_{\mu\nu} Q^I_{ai,\mu}\, \gamma^I_{\mu\nu}\, Q^I_{bj,\nu} \\
&- 8 \sum_A \Gamma_A \Delta q_A \sum_{\mu \in A} Q^I_{ai,\mu} \sum_{\nu \in A} Q^I_{bj,\nu}.
\end{aligned}
\end{equation}
The first term is the orbital energy gap, the second is the second-order Coulomb coupling through the shell-resolved $\gamma$ kernel, and the third is the third-order contribution from the Hubbard derivative $\Gamma_A$. The off-diagonal blocks coupling different fragments are\cite{nagata_fully_2011}
\begin{equation}
\label{eq:orb_hessian_off}
A^{K,I}_{ai,bj} = -4 \sum_{\mu\nu} Q^K_{ai,\mu}\, \gamma^{KI}_{\mu\nu}\, Q^I_{bj,\nu},
\end{equation}
where $\gamma^{KI}_{\mu\nu}$ is the inter-fragment Coulomb kernel. The SCZV equations are solved using the preconditioned conjugate gradient method.\cite{nagata_fully_2011}

\paragraph{Response gradient evaluation.}
Once the Z-vectors $Z^I_{ai}$ are obtained, the response gradient is evaluated as \mbox{$\Re^\xi = \mathbf{Z}^T \mathbf{B}^\xi_0$}, where the vector $\mathbf{B}^\xi_0$ collects the integral derivatives of the Fock matrix projected into the occupied--virtual MO space\cite{nagata_fully_2011}. Explicitly,
\begin{equation}
\label{eq:B0_def}
B^{\xi,I}_{0,ai} = \sum_{\mu\nu \in I} C^I_{\mu a} \left(\frac{\partial F^I_{\mu\nu}}{\partial \xi}\bigg|_{\!\mathbf{C}} - \epsilon^I_i \frac{\partial S^I_{\mu\nu}}{\partial \xi}\right) C^I_{\nu i},
\end{equation}
where the notation $\partial/\partial\xi|_{\mathbf{C}}$ indicates that only the one- and two-electron integrals are differentiated while the MO coefficients are held fixed. The second term removes the orbital energy contribution from the overlap constraint.

For the GFN1-xTB Fock matrix $F_{\mu\nu} = H^0_{\mu\nu} + \frac{1}{2}S_{\mu\nu}(\sigma_\mu + \sigma_\nu)$ [Eq.~(\ref{eq:gfn1_energy_2})], the integral derivative separates into three parts:
\begin{equation}
\label{eq:Fock_deriv}
\begin{aligned}
\frac{\partial F^I_{\mu\nu}}{\partial \xi}\bigg|_{\!\mathbf{C}} &= \frac{\partial H^{0,I}_{\mu\nu}}{\partial \xi} + \frac{1}{2}\frac{\partial S^I_{\mu\nu}}{\partial \xi}\!\left(\sigma^I_\mu + \sigma^I_\nu\right) \\
&+ \frac{1}{2} S^I_{\mu\nu}\! \left(\frac{\partial \sigma^I_\mu}{\partial \xi}\bigg|_{\!\mathbf{C}} + \frac{\partial \sigma^I_\nu}{\partial \xi}\bigg|_{\!\mathbf{C}}\right)\!,
\end{aligned}
\end{equation}
where the total Coulomb shift on shell $\mu \in A$ of fragment $I$ is
\begin{equation}
\label{eq:shift_total}
\sigma^I_\mu = \sum_{\nu \in I} \gamma^I_{\mu\nu} \Delta q^I_\nu - \Gamma_A (\Delta q_A)^2 + \mathrm{ESP}^I_\mu,
\end{equation}
comprising the intra-fragment second-order Coulomb interaction, the third-order Hubbard derivative correction [$\Gamma_A$ from Eq.~(\ref{eq:orb_hessian})], and the embedding potential from surrounding fragments, $\mathrm{ESP}^I_\mu = \sum_{J \neq I} \sum_{\lambda \in J} \gamma^{IJ}_{\mu\lambda} \Delta q^J_\lambda$. When the MO coefficients are held fixed, the Mulliken charges $\Delta q^I_\nu$ are constant, so only the geometry dependence of the Coulomb kernel $\gamma_{\mu\nu}$ survives in the last term of Eq.~(\ref{eq:Fock_deriv}); this produces the Coulomb kernel derivative contribution discussed below.

The Z-vectors are transformed to the AO basis,
\begin{equation}
\label{eq:Z_AO}
\tilde{Z}^I_{\mu\nu} = \sum_{\substack{a \in \mathrm{vir} \\ i \in \mathrm{occ}}} C^I_{\mu a} \, Z^I_{ai} \, C^I_{\nu i},
\end{equation}
where $\tilde{Z}^I_{\mu\nu}$ is in general not symmetric, since the transformation to the AO basis involves virtual and occupied MO coefficients.\cite{nagata_fully_2011} The symmetrized form $\bar{Z}^I_{\mu\nu} = \frac{1}{2}(\tilde{Z}^I_{\mu\nu} + \tilde{Z}^I_{\nu\mu})$ is required for the response gradient. Similarly, the energy-weighted Z-vector density is
\begin{equation}
\label{eq:WZ_def}
\tilde{W}^{Z,I}_{\mu\nu} = \sum_{\substack{a \in \mathrm{vir} \\ i \in \mathrm{occ}}} C^I_{\mu a}\, Z^I_{ai}\, \epsilon^I_i\, C^I_{\nu i},
\end{equation}
which arises from the $\epsilon_i \partial S/\partial\xi$ term in Eq.~(\ref{eq:B0_def}).

Substituting the Fock derivative [Eq.~(\ref{eq:Fock_deriv})] into \mbox{$\Re^\xi = \sum_I \sum_{ai} Z^I_{ai} B^{\xi,I}_{0,ai}$} and transforming to the AO basis produces four contributions:
\begin{equation}
\label{eq:response_grad}
G^\xi_\mathrm{resp} = G^\xi_{\partial S} + G^\xi_{\partial \mathcal{H}} + G^\xi_{\partial\gamma,\mathrm{intra}} + G^\xi_{\partial\gamma,\mathrm{inter}}.
\end{equation}

\textit{Overlap derivative contribution.}
The first and second terms of Eq.~(\ref{eq:Fock_deriv}) both contain $\partial S_{\mu\nu}/\partial\xi$. Using the factorization $H^0_{\mu\nu} = \mathcal{H}_{AB}\, S_{\mu\nu}$ [Eq.~(\ref{eq:H0_deriv})], the Hamiltonian derivative generates $\mathcal{H}_{AB}\,\tilde{Z}_{\mu\nu}\,\partial S/\partial\xi$. Combining with the shift term and the orbital energy contribution from Eq.~(\ref{eq:B0_def}):
\begin{equation}
\label{eq:ZxHB}
G^\xi_{\partial S} = \sum_I \sum_{\mu\nu \in I} \!\left(\mathcal{H}_{AB}\, \tilde{Z}^I_{\mu\nu} + \mathcal{W}^{\mathrm{resp},I}_{\mu\nu}\right) \frac{\partial S^I_{\mu\nu}}{\partial \xi},
\end{equation}
where
\begin{equation}
\label{eq:Wresp}
\mathcal{W}^{\mathrm{resp},I}_{\mu\nu} = \frac{\sigma^I_\mu + \sigma^I_\nu}{2}\,\tilde{Z}^I_{\mu\nu} - \tilde{W}^{Z,I}_{\mu\nu}
\end{equation}
is the response work matrix. This has the same structure as the ground-state overlap derivative contribution [Eqs.~(\ref{eq:xtb_grad})--(\ref{eq:W_matrix})], with the density matrix $P_{\mu\nu}$ replaced by $\tilde{Z}_{\mu\nu}$ and the energy-weighted density $\tilde{W}_{\mu\nu}$ replaced by $\tilde{W}^Z_{\mu\nu}$. Note that the total shift $\sigma^I_\mu$ [Eq.~(\ref{eq:shift_total})] includes the embedding potential, so the response work matrix captures the effect of the electrostatic environment on the Z-vector overlap derivative.

\textit{Hamiltonian prefactor derivative contribution.}
The remaining part of the Hamiltonian derivative, $S_{\mu\nu}\,\partial\mathcal{H}_{AB}/\partial\xi$ [cf.\ Eq.~(\ref{eq:H0_deriv})], yields
\begin{equation}
\label{eq:ZxdH}
G^\xi_{\partial\mathcal{H}} = G^\xi_{\Pi,\mathrm{resp}} + G^\xi_{\mathrm{CN,resp}}.
\end{equation}
The polynomial distance correction contribution is
\begin{equation}
\label{eq:ZxPi}
G^\xi_{\Pi,\mathrm{resp}} = \sum_I \sum_{\substack{A \neq B \\ \mu \in A,\, \nu \in B}} \bar{\mathcal{H}}^{\Pi}_{AB}\, \bar{Z}^I_{\mu\nu}\, S^I_{\mu\nu}\, \frac{\partial \Pi(R_{AB})}{\partial \xi},
\end{equation}
where $\bar{\mathcal{H}}^{\Pi}_{AB}$ collects the Hamiltonian prefactors from Eq.~(\ref{eq:h0_hamiltonian}) excluding the polynomial term [cf.\ the analogous ground-state expression below Eq.~(\ref{eq:H0_deriv})]. Since $H^0_{\mu\nu}$ is symmetric but $\tilde{Z}_{\mu\nu}$ is not, the trace identity $\mathrm{Tr}(\tilde{Z}\, \partial H^0/\partial\xi) = \mathrm{Tr}(\bar{Z}\, \partial H^0/\partial\xi)$ requires that the symmetrized Z-vector $\bar{Z}_{\mu\nu}$ be used for the polynomial distance correction to ensure translational invariance.

The coordination number contribution has the same structure as the ground-state CN gradient [Eq.~(\ref{eq:cn_grad})]:
\begin{equation}
\label{eq:ZxCN}
G^\xi_{\mathrm{CN,resp}} = \sum_I \sum_A \mathcal{F}^{Z,I}_A \frac{\partial CN_A}{\partial \xi},
\end{equation}
where $\mathcal{F}^{Z,I}_A$ is obtained from Eq.~(\ref{eq:cn_factor}) by replacing $P_{\mu\nu}$ with $\tilde{Z}^I_{\mu\nu}$. The CN factor involves $\mathrm{Tr}(S\tilde{Z})$ over the shells of atom $A$, which is invariant under symmetrization: $\sum_\mu S_{\mu\nu}\tilde{Z}_{\mu\nu} = \sum_\mu S_{\mu\nu}\bar{Z}_{\mu\nu}$, since $S$ is symmetric. Because $CN_A$ depends on all neighboring atoms through Eq.~(\ref{eq:cn_def}), this contribution couples the Z-vector of fragment $I$ to atoms in other fragments, even though the CN gradient itself is computed at the supersystem level.

\textit{Coulomb kernel derivative contributions.}
The last term in Eq.~(\ref{eq:Fock_deriv}), involving the geometry dependence of the Coulomb kernel $\gamma_{\mu\nu}$, generates both intra- and inter-fragment gradient terms. Within each fragment, the gradient involves the Z-vector Mulliken charges $Q^{Z,I}_\mu$ and the ground-state shell charges:
\begin{equation}
\label{eq:ZxG}
\begin{aligned}
G^\xi_{\partial\gamma,\mathrm{intra}} &= \sum_I \sum_{\substack{A \neq B \\ \mu \in A, \nu \in B}} \left(\Delta q^I_\mu Q^{Z,I}_\nu + Q^{Z,I}_\mu \Delta q^I_\nu\right) \frac{\partial \gamma_{\mu\nu}}{\partial \xi}.
\end{aligned}
\end{equation}
The inter-fragment contribution accounts for the response of the embedding potential to geometry changes:
\begin{equation}
\label{eq:ZxG_inter}
G^\xi_{\partial\gamma,\mathrm{inter}} = \sum_I \sum_{J \neq I} \sum_{\substack{\mu \in I \\ \nu \in J}} Q^{Z,I}_\mu \Delta q^J_\nu \frac{\partial \gamma_{\mu\nu}}{\partial \xi}.
\end{equation}
This term arises because the Coulomb interaction kernel between shells on different fragments depends on the internuclear distance; when the geometry changes, the coupling between the Z-vector charges of fragment $I$ and the ground-state charges of fragment $J$ produces a force. Together with the intra-fragment term [Eq.~(\ref{eq:ZxG})], it ensures the correct response to geometry changes in the embedding potential.

\textit{Response embedding Lagrangian correction.}
In addition to the four terms above, the converged Z-vectors generate an overlap derivative contribution analogous to the ground-state embedding Lagrangian correction [Eq.~(\ref{eq:addlag})]. The Z-vector Mulliken charges at shell level,
\begin{equation}
\label{eq:qz}
Q^{Z,I}_\mu = \sum_{\nu \in \mu} (\bar{\mathbf{Z}}^I \mathbf{S}^I)_{\nu\nu},
\end{equation}
induce a shift potential
\begin{equation}
\label{eq:shiftz}
\sigma^{Z,I}_\mu = \sum_{\nu \in I} \gamma^I_{\mu\nu} Q^{Z,I}_\nu + \sum_{J \neq I} \sum_{\nu \in J} \gamma^{IJ}_{\mu\nu} Q^{Z,J}_\nu - 2\Gamma_A \Delta q_A Q^Z_A,
\end{equation}
where $Q^Z_A = \sum_{\mu \in A} Q^{Z,I}_\mu$ is the atom-level Z-vector charge and the last term is the third-order correction. The response embedding Lagrangian correction is
\begin{equation}
\label{eq:resp_addlag}
G^\xi_\mathrm{lag,resp} = \sum_I \sum_{\mu\nu \in I} \left(\boldsymbol{\Sigma}^{Z,I}_{\mu\nu} \mathbf{P}^I_{\mu\nu} - \frac{1}{2}(\mathbf{P}^I \boldsymbol{\Sigma}^{Z,I} \mathbf{S}^I \mathbf{P}^I)_{\mu\nu}\right) \frac{\partial S^I_{\mu\nu}}{\partial \xi},
\end{equation}
where $\Sigma^{Z,I}_{\mu\nu} = \frac{1}{2}(\sigma^{Z,I}_\mu + \sigma^{Z,I}_\nu) S^I_{\mu\nu}$ is the symmetrized Z-vector shift matrix. Unlike the ground-state embedding Lagrangian correction [Eq.~(\ref{eq:addlag})], which involves the energy-weighted density $W^\mathrm{emb}$, the response embedding Lagrangian correction retains the ground-state density $\mathbf{P}^I$ and constructs the work matrix from the triple product $\mathbf{P}^I \boldsymbol{\Sigma}^{Z,I} \mathbf{S}^I \mathbf{P}^I$. This term accounts for the perturbation of the embedding Fock matrix by the Z-vector charges and is required for exact agreement with numerical differentiation of the FMO energy. The complete response gradient is
\begin{equation}
\label{eq:response_grad_full}
G^\xi_\mathrm{resp} = G^\xi_{\partial S} + G^\xi_{\partial \mathcal{H}} + G^\xi_{\partial\gamma,\mathrm{intra}} + G^\xi_{\partial\gamma,\mathrm{inter}} + G^\xi_\mathrm{lag,resp}.
\end{equation}

The complete FMO2-xTB gradient is the sum of the explicit [Eq.~(\ref{eq:explicit_total})] and response [Eq.~(\ref{eq:response_grad_full})] contributions, providing a fully analytic expression that enables efficient geometry optimizations and molecular dynamics simulations.

\subsubsection{\label{sec:fmo3}Extension to FMO3}

The FMO3 gradient is obtained by differentiating the three-body energy [Eq.~(\ref{eq:fmo3_energy})],
\begin{equation}
\begin{aligned}
    \frac{\partial E^\mathrm{FMO3}}{\partial \xi} &= \frac{\partial E^\mathrm{FMO2}}{\partial \xi} \\&+  \sum_{I>J>K} \frac{\partial}{\partial \xi}\!\left(\Delta E_{IJK} - \Delta E_{IJ} - \Delta E_{IK} - \Delta E_{JK}\right)\!,
\end{aligned}
\label{eq:fmo3_grad_start}
\end{equation}
where $\Delta E_{IJK}$ is the trimer interaction energy [Eq.~(\ref{eq:trimer_interaction_energy})]. The three-body correction modifies both the explicit and response gradient contributions. In the following, we show that the resulting expressions can be cast into a compact form that generalizes each FMO2 gradient term.

\paragraph{Trimer charge transfer.}
Analogous to the dimer charge transfer $\Delta \Delta q^{IJ}_\mu$ [Eq.~(\ref{eq:ddq_calc})], the charge redistribution upon trimer formation is
\begin{equation}
\label{eq:ctijk}
\Delta \Delta q^{IJK}_\mu = \Delta q^{IJK}_\mu - \Delta q^I_\mu \delta_{\mu \in I} - \Delta q^J_\mu \delta_{\mu \in J} - \Delta q^K_\mu \delta_{\mu \in K},
\end{equation}
where $\Delta q^{IJK}_\mu$ are the Mulliken charges of the trimer computed in the embedding potential of all fragments $L \notin \{I,J,K\}$.

\paragraph{Internal trimer delta gradient.}
The three-body correction to the internal gradient follows the many-body expansion directly:
\begin{equation}
\label{eq:trimer_delta}
\begin{aligned}
G^\xi_{\Delta,IJK}{}' &= G^\xi_{\mathrm{tri},IJK} - G^\xi_{\mathrm{mon},I} - G^\xi_{\mathrm{mon},J} - G^\xi_{\mathrm{mon},K} \\
&\quad - G^\xi_{\Delta,IJ} - G^\xi_{\Delta,IK} - G^\xi_{\Delta,JK},
\end{aligned}
\end{equation}
where $G^\xi_{\mathrm{tri},IJK}$ is the internal gradient of the trimer computed using Eq.~(\ref{eq:xtb_grad}), and $G^\xi_{\Delta,IJ}$ is the pair delta gradient [Eq.~(\ref{eq:pair_delta})]. For sub-pairs treated within the ES-DIM approximation, the pair delta gradient reduces to the corresponding ES-DIM gradient [Eq.~(\ref{eq:esd_grad})]. The coordination number correction follows the same many-body subtraction pattern.

\paragraph{Embedding gradient terms.}
The three-body correction to the embedding energy involves the trimer embedding correction $\Delta E^V_{IJK}$ [Eq.~(\ref{eq:trimer_embedding})] and the subtraction of the pair corrections $\Delta E^V_{IJ}$, $\Delta E^V_{IK}$, and $\Delta E^V_{JK}$. When these terms are collected across all trimers and combined with the FMO2 embedding contributions, the accumulated charge transfer $\bar{q}_\mu$ [Eq.~(\ref{eq:ctmul})] generalizes to
\begin{equation}
\label{eq:ctmul_fmo3}
\bar{q}_\mu = \sum_{\substack{\text{pairs } IJ \\ \mu \in IJ}} s_{IJ}\, \Delta \Delta q^{IJ}_\mu + \sum_{\substack{\text{trimers } IJK \\ \mu \in IJK}} \Delta \Delta q^{IJK}_\mu,
\end{equation}
where the scaling factor
\begin{equation}
\label{eq:scal}
s_{IJ} = 1 - n_{IJ}^\mathrm{tri}
\end{equation}
arises naturally from the inclusion-exclusion structure of the many-body expansion, with $n_{IJ}^\mathrm{tri}$ being the number of trimers containing pair $(I,J)$. Each trimer introduces its own charge transfer $\Delta\Delta q^{IJK}$ while simultaneously removing the pair charge transfers $\Delta\Delta q^{IJ}$, $\Delta\Delta q^{IK}$, and $\Delta\Delta q^{JK}$ of its three sub-pairs. The net effect is that $s_{IJ}$ counts how many times pair $(I,J)$ is replaced by a trimer: $s_{IJ}=1$ for pairs in no trimer (FMO2 limit), $s_{IJ}=0$ for pairs in exactly one trimer, and $s_{IJ}=-1$ for pairs shared by two trimers.

The electrostatic embedding gradient [Eq.~(\ref{eq:embed_grad})], the embedding Lagrangian correction [Eq.~(\ref{eq:addlag})], and the charge-transfer Coulomb gradient [Eq.~(\ref{eq:ctij})] all depend on the accumulated charge transfer $\bar{q}_\mu$. Using the generalized expression [Eq.~(\ref{eq:ctmul_fmo3})], these terms automatically incorporate the three-body corrections without separate trimer embedding gradient expressions. Specifically, the charge-transfer Coulomb gradient becomes
\begin{equation}
\label{eq:ctij_fmo3}
\begin{aligned}
G^\xi_\mathrm{CT,FMO3} &= \sum_{I>J} s_{IJ} \!\!\sum_{\substack{\mu \in IJ \\ K \neq I,J}} \sum_{\nu \in K} \Delta \Delta q^{IJ}_\mu \Delta q^K_\nu \frac{\partial \gamma_{\mu\nu}}{\partial \xi} \\
&+ \sum_{I>J>K} \!\!\sum_{\substack{\mu \in IJK \\ L \neq I,J,K}} \sum_{\nu \in L} \Delta \Delta q^{IJK}_\mu \Delta q^L_\nu \frac{\partial \gamma_{\mu\nu}}{\partial \xi},
\end{aligned}
\end{equation}
where the first sum runs over all pairs with their respective scaling factors and the second sum adds the trimer charge-transfer contributions.

\paragraph{Response Lagrangian for FMO3.}
The three-body correction also modifies the response gradient through the Lagrangian that drives the Z-vector equations. The response contribution [Eq.~(\ref{eq:response_Nagata})] depends on the charge transfer $\Delta\Delta q^{IJ}$ of all dimer pairs. In FMO3, the three-body subtraction replaces each pair's charge transfer with the corresponding trimer quantity for pairs contained in trimers. Collecting all contributions, the electrostatic potential entering the Lagrangian [Eq.~(\ref{eq:esp_lagrangian})] generalizes to
\begin{equation}
\label{eq:esp_fmo3}
\begin{aligned}
\mathrm{ESP}^{\mathrm{FMO3},K}_\lambda &= \sum_{\substack{I > J \\ K \notin \{I,J\}}} s_{IJ} \sum_{\mu \in IJ} \gamma_{\lambda\mu}\, \Delta \Delta q^{IJ}_\mu \\
&+ \sum_{\substack{I > J > K' \\ K \notin \{I,J,K'\}}} \sum_{\mu \in IJK'} \gamma_{\lambda\mu}\, \Delta \Delta q^{IJK'}_\mu,
\end{aligned}
\end{equation}
and the FMO3 Lagrangian retains the form of Eq.~(\ref{eq:lagrangian}) with $\mathrm{ESP}^K$ replaced by $\mathrm{ESP}^{\mathrm{FMO3},K}$. The SCZV equations [Eq.~(\ref{eq:sczv})] and the response gradient evaluation [Eqs.~(\ref{eq:response_grad})--(\ref{eq:ZxG_inter})] remain unchanged, since they depend only on the Z-vectors and the converged monomer properties.

\paragraph{Total FMO3-xTB gradient.}
The complete explicit FMO3-xTB gradient is
\begin{equation}
\label{eq:explicit_fmo3}
\begin{aligned}
\left(\frac{\partial E^\mathrm{FMO3}}{\partial \xi}\right)_{\!\!\mathrm{expl}} &= \sum_I G^\xi_{\mathrm{mon},I} + \sum_{I>J} G^\xi_{\Delta,IJ} + \sum_{I>J>K} G^\xi_{\Delta,IJK}{}' \\
&+ G^\xi_\mathrm{CT,FMO3} + G^\xi_\mathrm{emb}(\bar{q}) + G^\xi_\mathrm{ESD} \\
&+ \sum_I G^\xi_{\mathrm{lag},I}(\bar{q}) + G^\xi_\mathrm{CN,FMO3} + G^\xi_\mathrm{disp} + G^\xi_\mathrm{XB},
\end{aligned}
\end{equation}
where $G^\xi_\mathrm{emb}(\bar{q})$ and $G^\xi_{\mathrm{lag},I}(\bar{q})$ use the generalized accumulated charge transfer [Eq.~(\ref{eq:ctmul_fmo3})], and $G^\xi_\mathrm{CN,FMO3}$ includes the three-body CN correction. The total FMO3-xTB gradient is
\begin{equation}
\label{eq:fmo3_total}
\frac{\partial E^\mathrm{FMO3}}{\partial \xi} = \left(\frac{\partial E^\mathrm{FMO3}}{\partial \xi}\right)_{\!\!\mathrm{expl}} + G^\xi_\mathrm{resp}(\mathbf{Z}^\mathrm{FMO3}),
\end{equation}
where the response gradient has the same functional form as in FMO2 [Eq.~(\ref{eq:response_grad})] but with Z-vectors determined by the FMO3 Lagrangian [Eq.~(\ref{eq:esp_fmo3})].

\subsection{\label{sec:hop}Covalent bond fragmentation}

The FMO formulation presented above applies naturally to molecular clusters where fragments correspond to individual molecules. To extend the method to systems with covalent bonding between fragments---such as proteins, polymers, and covalent frameworks---a methodology for cutting covalent bonds is required. In the FMO framework, the hybrid orbital projection (HOP) operator provides such a methodology.\cite{nakano_fragment_2000,fedorov_covalent_2008,nagata_importance_2010} In the following, we describe the adaptation of the HOP approach to the GFN1-xTB Hamiltonian.

\subsubsection{Hybrid orbital projection operator}

When a covalent bond between atoms $A$ and $B$ is cut to place them in separate fragments $I$ and $J$, the bond detachment atom (BDA) is defined as $A \in I$ and the bond attachment atom (BAA) as $B \in J$.\cite{nakano_fragment_2000,fedorov_covalent_2008} To preserve the local electronic structure at the boundary, the monomer calculation for fragment $J$ is augmented with a \emph{ghost atom}: a copy of BDA atom $A$ placed at its original position and assigned to fragment~$J$, equipped with a single valence electron.\cite{fedorov_covalent_2008} The BDA atom $A$ in fragment~$I$ correspondingly loses one valence electron. This electron redistribution is tracked through modified reference charges ($Z_\mathrm{ref}$): $Z_\mathrm{ref}(A) \to Z_\mathrm{ref}(A) - 1$ for the BDA, and $Z_\mathrm{ref}(\mathrm{ghost}) = 1$ for the ghost atom.\cite{nishimoto_density-functional_2014,nishimoto_large-scale_2015}

The HOP operator constrains the orbital space at the boundary to reproduce the bonding pattern of the intact molecule. Along the bond direction $\hat{\mathbf{b}} = (\mathbf{R}_B - \mathbf{R}_A)/|\mathbf{R}_B - \mathbf{R}_A|$, an sp$^3$-type hybrid orbital is constructed on both the BDA and the ghost atom,\cite{fedorov_covalent_2008}
\begin{equation}
\label{eq:hybrid}
|\mathbf{h}\rangle = c_s |s\rangle + c_p \left(\hat{b}_x |p_x\rangle + \hat{b}_y |p_y\rangle + \hat{b}_z |p_z\rangle\right),
\end{equation}
where $c_s = 1/2$ and $c_p = \sqrt{3}/2$ are the standard sp$^3$ hybridization coefficients. On the BDA atom in fragment $I$, the bond-pointing hybrid $|\mathbf{h}\rangle$ is projected out of the occupied space by applying the projector
\begin{equation}
\label{eq:dd_bda}
\mathbf{D}^\mathrm{BDA} = V_\mathrm{HOP}\, |\mathbf{h}\rangle\langle\mathbf{h}|,
\end{equation}
which shifts the bond-pointing orbital to very high energy ($V_\mathrm{HOP} \approx 10^6$~Hartree), effectively removing it from the occupied manifold. On the ghost atom in fragment $J$, the complement projection is applied:
\begin{equation}
\label{eq:dd_ghost}
\mathbf{D}^\mathrm{ghost} = V_\mathrm{HOP}\, \left(\mathbf{1}_{sp} - |\mathbf{h}\rangle\langle\mathbf{h}|\right),
\end{equation}
where $\mathbf{1}_{sp}$ is the identity in the $s$+$p$ subspace. This shifts all non-bonding orbitals on the ghost to high energy, leaving only the bond-pointing hybrid $|\mathbf{h}\rangle$ available for occupation. Together, the BDA and ghost projections ensure that exactly one electron occupies the bond-pointing hybrid on each side, mimicking the covalent bond of the intact system.\cite{nakano_fragment_2000,fedorov_covalent_2008}

The HOP projector enters the Fock matrix through the overlap-weighted form\cite{fedorov_covalent_2008}
\begin{equation}
\label{eq:vhop}
V^\mathrm{HOP}_{\mu\nu} = \sum_{\lambda\sigma \in \mathrm{BDA/ghost}} S_{\mu\lambda}\, D_{\lambda\sigma}\, S_{\sigma\nu},
\end{equation}
where $\mathbf{D}$ is either $\mathbf{D}^\mathrm{BDA}$ or $\mathbf{D}^\mathrm{ghost}$ depending on the atom, $S_{\mu\lambda}$ is the overlap matrix, and the summation extends over the basis functions of the BDA or ghost atom. This form ensures that $V^\mathrm{HOP}$ is Hermitian and compatible with the non-orthogonal basis. The modified Fock matrix for a fragment with HOP-constrained boundaries is
\begin{equation}
\label{eq:fock_hop}
F^\mathrm{HOP}_{\mu\nu} = F_{\mu\nu} + V^\mathrm{HOP}_{\mu\nu},
\end{equation}
where $F_{\mu\nu}$ is the standard GFN1-xTB Fock matrix including the electrostatic embedding.

\subsubsection{FMO-xTB HOP energy}

The HOP boundary treatment modifies the FMO energy expression in several ways. Each monomer that hosts a detached bond is computed with an extended basis set comprising the real atoms and one or more ghost atoms. The monomer SCC is performed using the augmented Fock matrix [Eq.~(\ref{eq:fock_hop})], and the resulting electronic energy $E^\mathrm{el}_I$ is evaluated \emph{without} the HOP projector contribution:\cite{nishimoto_density-functional_2014}
\begin{equation}
\label{eq:hop_energy}
E_I = E^\mathrm{el}_I + E^\mathrm{rep}_I,
\end{equation}
where $E^\mathrm{rep}_I$ is the repulsion energy computed with the reference charge scaling $Z_\mathrm{ref}$ to account for the modified electron counts at BDA and ghost atoms.\cite{nishimoto_density-functional_2014} The HOP projector serves as a constraint potential to enforce the correct orbital occupations but does not contribute to the physical energy.\cite{fedorov_covalent_2008}

For dimer calculations, the treatment of the detached bond depends on whether the bond is internal or external to the pair. If both BDA and BAA atoms belong to the dimer, the bond is \emph{healed}: the ghost atom is removed from the pair calculation and the bond is treated normally. If only the BAA belongs to the dimer and the BDA is external, the bond remains detached and the ghost atom is retained in the pair, subject to the same HOP constraint as in the monomer.\cite{fedorov_covalent_2008}

The embedding energy $\Delta E^V_{IJ}$ [Eq.~(\ref{eq:dimer_embedding_correction})] requires the charge transfer $\Delta\Delta q^{IJ}_\mu$, which for fragments with ghost atoms includes contributions from the ghost shells. For healed bonds, the ghost disappears from the pair but exists in the monomer; the charge transfer at the ghost shells is
\begin{equation}
\label{eq:ctij_ghost}
\Delta\Delta q_{\mu}^{IJ,\mathrm{ghost}} = -\Delta q_\mu^{I,\mathrm{ghost}},
\end{equation}
where $\Delta q_\mu^{I,\mathrm{ghost}}$ is the monomer Mulliken charge at the ghost shell, reflecting the charge ``lost'' when the ghost is removed in the pair.\cite{nishimoto_density-functional_2014} The accumulated charge transfer $\bar{q}_\mu$ [Eq.~(\ref{eq:ctmul})] incorporates these ghost contributions along with the standard real-shell charge transfers.

\subsubsection{FMO-xTB HOP gradient}

The analytic gradient for FMO-xTB with HOP boundary treatment follows the same decomposition into explicit and response contributions [Eq.~(\ref{eq:grad_decomp})], with modifications arising from the extended basis, the HOP projector, and the ghost atom forces.

\paragraph{Ground-state gradient modifications.}
The monomer and pair gradients [Eq.~(\ref{eq:xtb_grad})] are evaluated over the extended basis set (real + ghost atoms). The pair delta gradient [Eq.~(\ref{eq:pair_delta})] accounts for the healed/partial bond structure: for healed bonds, the ghost gradient from the monomer is subtracted at the BDA position; for partial bonds, the pair ghost gradient is included at the BDA position. All ghost atom forces are accumulated at the corresponding BDA atom's coordinates, since the ghost atom is constrained to occupy the BDA position.\cite{nishimoto_large-scale_2015}

The embedding gradient [Eq.~(\ref{eq:embed_grad})] and the embedding Lagrangian correction [Eq.~(\ref{eq:addlag})] use the accumulated charge transfer $\bar{q}_\mu$ that includes ghost shell contributions [Eq.~(\ref{eq:ctij_ghost})].

\paragraph{HOP projector gradient.}
The HOP projector introduces an additional gradient contribution from the bond-direction dependence of the hybrid orbital and the overlap dependence of the projector. Differentiating $\mathrm{Tr}(\mathbf{P}\, \mathbf{V}^\mathrm{HOP})$ [Eq.~(\ref{eq:vhop})] with respect to nuclear coordinates yields two terms:\cite{nishimoto_large-scale_2015}
\begin{equation}
\label{eq:hop_grad}
G^\xi_\mathrm{HOP} = G^\xi_\mathrm{HOP,S} + G^\xi_\mathrm{HOP,h}.
\end{equation}
The overlap derivative contribution is
\begin{equation}
\label{eq:hopsder}
G^\xi_\mathrm{HOP,S} = 2 \sum_{\mu,\lambda} \frac{\partial S_{\mu\lambda}}{\partial \xi}\, (\mathbf{D}\, \mathbf{S}^T \mathbf{P})_{\lambda\mu},
\end{equation}
where $\mathbf{S}^T\mathbf{P}$ denotes the product of the transpose of the overlap block with the density matrix. This term arises from the overlap-weighted form of the projector [Eq.~(\ref{eq:vhop})] and accounts for the change in the projector as the overlap integrals vary with geometry. The coefficient derivative contribution captures the dependence of the hybrid orbital coefficients on the bond direction,
\begin{equation}
\label{eq:hopcoder}
G^\xi_\mathrm{HOP,h} = V_\mathrm{HOP}\!\sum_{\lambda\sigma} \left(\frac{\partial h_\lambda}{\partial \xi} h_\sigma + h_\lambda \frac{\partial h_\sigma}{\partial \xi}\right) (\mathbf{S}^T \mathbf{P}\, \mathbf{S})_{\lambda\sigma},
\end{equation}
where $h_\lambda$ are the hybrid orbital coefficients from Eq.~(\ref{eq:hybrid}), and the derivative $\partial h_\lambda/\partial\xi$ depends on the bond direction derivative $\partial\hat{\mathbf{b}}/\partial\xi$. For the ghost atom, the coefficient sign is reversed ($-1$) relative to the BDA ($+1$), reflecting the complement relationship between Eqs.~(\ref{eq:dd_bda}) and~(\ref{eq:dd_ghost}).

In the FMO gradient, the HOP projector contribution is included in the monomer and pair gradients and enters the pair delta [Eq.~(\ref{eq:pair_delta})] through the standard many-body subtraction. The HOP projector gradient generates forces on both the BDA and BAA atoms, since the bond direction $\hat{\mathbf{b}}$ depends on both positions. The coefficient derivative forces $G^\xi_\mathrm{HOP,h}$ on the BAA atom are accumulated directly, while the overlap derivative forces $G^\xi_\mathrm{HOP,S}$ on the ghost atom are scattered to the BDA position.

\paragraph{Response gradient with HOP.}
The response gradient [Eq.~(\ref{eq:response_grad_full})] retains the same functional form when HOP is employed. The Lagrangian [Eq.~(\ref{eq:lagrangian})] and SCZV equations [Eq.~(\ref{eq:sczv})] use the extended-basis quantities (including ghost shells), and the charge transfer entering the electrostatic potential $\mathrm{ESP}^K_\lambda$ [Eq.~(\ref{eq:esp_lagrangian})] includes the ghost contributions from Eq.~(\ref{eq:ctij_ghost}). The orbital Hessian [Eq.~(\ref{eq:orb_hessian})] employs the orbital energies $\epsilon_i$, $\epsilon_a$ from the HOP-augmented Fock matrix [Eq.~(\ref{eq:fock_hop})], which ensures that the HOP-shifted virtual orbitals contribute negligibly to the Z-vectors due to the large orbital energy gap ($\epsilon_a - \epsilon_i \approx V_\mathrm{HOP}$). The inter-fragment coupling [Eq.~(\ref{eq:orb_hessian_off})] and the response gradient sub-terms [Eqs.~(\ref{eq:ZxHB})--(\ref{eq:ZxG_inter})] extend over the full extended basis of each fragment, including the ghost shells and their $\gamma$-kernel interactions with atoms in other fragments.

In the response gradient, the HOP projector derivative additionally contributes through Eq.~(\ref{eq:hop_grad}) with the density matrix $\mathbf{P}$ replaced by the symmetrized Z-vector density $\bar{\mathbf{Z}}$ [Eq.~(\ref{eq:Z_AO})]:
\begin{equation}
\label{eq:hop_resp}
G^\xi_\mathrm{HOP,resp} = G^\xi_\mathrm{HOP,S}(\bar{\mathbf{Z}}) + G^\xi_\mathrm{HOP,h}(\bar{\mathbf{Z}}).
\end{equation}
The response embedding Lagrangian correction [Eq.~(\ref{eq:resp_addlag})] extends to the ghost shells: the Z-vector-induced shift $\sigma^{Z,I}_\mu$ [Eq.~(\ref{eq:shiftz})] includes inter-fragment contributions from ghost shells through $\gamma^{IJ}_{\mu\nu}$, ensuring consistency with the SCC embedding.

\subsubsection{FMO3-xTB HOP gradient}

The combination of the three-body FMO3 expansion (Sec.~\ref{sec:fmo3}) with the HOP covalent fragmentation (Sec.~\ref{sec:hop}) requires several extensions to the gradient formulation. The key complication is that trimers may contain both \emph{healed} bonds (where both BDA and BAA belong to the trimer) and \emph{partial} bonds (where only the BAA belongs to the trimer), and the ghost shell contributions from the extended basis must be treated consistently across all gradient terms.

\paragraph{Trimer gradient with HOP.}
Trimer calculations follow the same extended-basis prescription as dimers: healed bonds are restored to full covalent bonds within the trimer and the corresponding ghost atoms are removed, while partial bonds retain the ghost atom and HOP projector. The trimer SCC is performed in the embedding potential of all fragments $L \notin \{I,J,K\}$. The internal trimer delta gradient [Eq.~(\ref{eq:trimer_delta})] subtracts monomer and pair delta contributions; for sub-pairs treated within the ES-DIM approximation, the subtracted ESD gradient is computed at the shell level as $\sum_{\mu\nu}\Delta q^I_\mu\,\Delta q^J_\nu\,\partial\gamma_{\mu\nu}/\partial\xi$.

\paragraph{Ghost shell charge transfer in trimers.}
The trimer charge transfer $\Delta\Delta q^{IJK}_\mu$ [Eq.~(\ref{eq:ctijk})] extends to the ghost shells of each fragment. A detached bond connecting BDA atom~$A$ in fragment~$I$ to BAA atom~$B$ in fragment~$J$ produces a ghost atom on fragment~$J$ at the position of~$A$. In trimers, this bond may be \emph{healed} (both $I$ and $J$ belong to the trimer, removing the ghost) or \emph{partial} (only the BAA side belongs to the trimer, retaining the ghost). The ghost shell charge transfer for fragment~$J$ is
\begin{equation}
\label{eq:ctijk_ghost}
\Delta\Delta q^{IJK}_{\mu,\mathrm{ghost}} = \begin{cases}
-\Delta q^{J}_{\mu,\mathrm{ghost}}, & \text{healed (BDA} \in IJK\text{)}, \\[4pt]
\Delta q^{IJK}_{\mu,\mathrm{ghost}} - \Delta q^{J}_{\mu,\mathrm{ghost}}, & \text{partial (BDA} \notin IJK\text{)},
\end{cases}
\end{equation}
where $\Delta q^{J}_{\mu,\mathrm{ghost}}$ denotes the monomer Mulliken charge at the ghost shell and $\Delta q^{IJK}_{\mu,\mathrm{ghost}}$ the corresponding trimer charge. For healed bonds, the ghost is absent from the trimer, analogous to the pair case [Eq.~(\ref{eq:ctij_ghost})], so the full monomer ghost charge constitutes the charge transfer. For partial bonds, the ghost is retained in both monomer and trimer, and the charge transfer reduces to their difference.

The generalized accumulated charge transfer $\bar{q}_\mu$ [Eq.~(\ref{eq:ctmul_fmo3})] incorporates these ghost contributions alongside the real-shell terms. Including the ghost charge transfer in both pair and trimer contributions to $\bar{q}_\mu$ ensures that the FMO3 embedding energy vanishes identically for a complete three-body expansion in which all fragments form a single trimer, consistent with the inclusion--exclusion structure of the many-body expansion.

\paragraph{Ghost embedding gradient correction.}
The interfragment embedding gradient [Eq.~(\ref{eq:embed_grad})] includes contributions from the ghost shells of each monomer, since the monomer basis extends over ghost atoms. To ensure proper cancellation with the trimer self-interaction subtraction, an additional correction is included in the trimer charge-transfer Coulomb gradient [Eq.~(\ref{eq:ctij_fmo3})]. For each bond whose BAA belongs to the trimer, the ghost monomer charge interacts with the trimer charge transfer through the $\gamma$-derivative evaluated at the BDA position:
\begin{equation}
\label{eq:ghost_fmode}
G^\xi_\mathrm{ghost,tri} = -\sum_{\substack{\mathrm{bonds} \\ \mathrm{BAA} \in IJK}} \sum_{\nu \in IJK} \Delta\Delta q^{IJK}_\nu\, \Delta q^I_{\mu_\mathrm{ghost}}\, \frac{\partial \gamma_{\nu\mu_\mathrm{ghost}}}{\partial \xi}.
\end{equation}
This correction applies to both healed bonds (BDA inside the trimer) and partial bonds (BDA outside the trimer). For complete three-body expansions, the ghost embedding gradient correction ensures exact cancellation of all ghost contributions.

\paragraph{HOP projector gradient in FMO3.}
The HOP projector gradient [Eq.~(\ref{eq:hop_grad})] enters the FMO3 gradient through the standard many-body subtraction. Importantly, the pair HOP delta is included at full weight and is \emph{not} scaled by $s_{IJ}$, because the HOP projector is part of the Fock matrix rather than an embedding correction. The three-body correction to the HOP projector is captured through the trimer SCC delta [Eq.~(\ref{eq:trimer_delta})], which employs the trimer density and energy-weighted density computed with the trimer's own HOP projector (active only for partial bonds). For each healed bond, a per-bond correction subtracts the monomer HOP contribution that is absent in the trimer:
\begin{equation}
\label{eq:hop_fmo3}
\begin{aligned}
    G^\xi_{\mathrm{HOP,FMO3}} &= G^\xi_{\mathrm{HOP,mono}} + G^\xi_{\mathrm{HOP,pair\;delta}} \\ &- \sum_\text{healed bonds} (1 - s_{IJ})\,G^\xi_{\mathrm{HOP,mono,bond}},
\end{aligned}
\end{equation}
where the sum runs over all bonds healed in at least one trimer. The factor $(1 - s_{IJ})$ equals the number of trimers containing the pair that heals the bond, ensuring that each healed bond is corrected exactly once regardless of how many trimers share it.

\paragraph{Response Lagrangian with ghost shells.}
The trimer contribution to the FMO3 response Lagrangian [Eq.~(\ref{eq:esp_fmo3})] must include the ghost shell charge transfers alongside the real-shell contributions to maintain consistency with the accumulated charge transfer $\bar{q}_\mu$ [Eq.~(\ref{eq:ctmul_fmo3})]. For trimers not containing fragment~$K$, the electrostatic potential entering the Lagrangian generalizes to
\begin{equation}
\label{eq:esp_fmo3_ghost}
\begin{aligned}
    \mathrm{ESP}^{\mathrm{FMO3},K}_{\lambda,\mathrm{tri}} &= \sum_{\substack{I > J > K' \\ K \notin \{I,J,K'\}}} \left(\sum_{\mu \in IJK'} \gamma_{\lambda\mu}\, \Delta \Delta q^{IJK'}_\mu \right.\\
    &\left. + \sum_{\mu_\mathrm{ghost}} \gamma_{\lambda\mu_\mathrm{ghost}}\, \Delta \Delta q^{IJK'}_{\mu_\mathrm{ghost}}\right)\!.
\end{aligned}
\end{equation}
The inclusion of the ghost shell terms ensures that the Lagrangian computed from the pair-by-pair accumulation is identical to the Lagrangian derived from the accumulated charge transfer $\bar{q}_\mu$ [Eq.~(\ref{eq:ctmul_fmo3})], which is required for a consistent response gradient.

%% file: results.tex
\section{Results}
In this section, we assess the accuracy and computational efficiency of the FMO2-xTB and FMO3-xTB methods developed in this work. The validation follows a strategy established in prior FMO-DFTB studies,\cite{nishimoto_density-functional_2014,nishimoto_three-body_2017} in which the fragmentation error is isolated by comparing FMO energies and gradients against unfragmented reference calculations performed at the same level of theory.

First, we evaluate the accuracy of the FMO2-xTB and FMO3-xTB total energies by comparing them with full GFN1-xTB reference calculations for water, anthracene, and pentacene clusters. Water clusters serve as a well-established benchmark for fragmentation methods because of the prevalence of cooperative many-body polarization and charge-transfer effects in hydrogen-bonded networks, which pose a significant challenge for two-body expansions.\cite{fedorov_importance_2004, fedorov_three-body_2006} Anthracene and pentacene clusters, in which $\pi$-stacking and dispersion forces dominate the intermolecular interactions, represent prototypical organic semiconductor systems. The comparison of FMO2 and FMO3 results for these structurally distinct systems allows us to assess the importance of three-body corrections across different interaction regimes.

Beyond molecular clusters, we validate the covalent bond fragmentation capability (Sec.~\ref{sec:hop}) by benchmarking FMO-xTB with the hybrid orbital projection (HOP) boundary treatment against unfragmented GFN1-xTB reference calculations for polyalanine chains and B-DNA double helices of increasing size.

In addition, we investigate the computational scaling of the FMO-xTB methods with respect to system size. To this end, wall-clock timings are recorded for water clusters and pentacene clusters of increasing size, and the effective power-law scaling exponents are determined. For FMO-DFTB, near-linear scaling behavior has been demonstrated in several studies,\cite{nishimoto_density-functional_2014,vuong_fragment_2019} and it is of interest to determine whether similar scaling is achieved for the FMO-xTB combination. We also assess the parallel efficiency of the implementation by comparing serial and parallel timings for pentacene supercells on a multi-core computing node.

Furthermore, we examine the accuracy of the analytic energy gradient by comparison with the numerical gradient, with particular focus on the role of the response terms. In the FMO framework, the response terms arise from the self-consistent coupling of fragment electron densities through the embedding electrostatic potential, and their inclusion is required to obtain a fully analytic gradient.\cite{nagata_fully_2011} Prior work on FMO-DFTB has shown
that neglecting the response contribution introduces gradient errors on the order of $10^{-4}$~hartree/bohr,\cite{nishimoto_large-scale_2015} which can lead to energy drift in molecular dynamics simulations and unreliable geometry optimizations.\cite{nakata_analytic_2020} We therefore compare FMO2-xTB and FMO3-xTB analytic gradients computed with and without the response terms to the numerical reference gradient, in order to quantify the magnitude of the response contribution and to verify the correctness of the implementation.

Finally, we report the computational cost of evaluating the analytic gradient for the pentacene supercell series, decomposed into explicit and response-gradient contributions, to assess the practical feasibility of FMO-xTB molecular dynamics simulations for large systems.

\subsection{Computational Details}
The FMO2-xTB and FMO3-xTB methods were implemented in our software program DIALECT\cite{einsele_dialect_2025}, which is available from GitHub. The separation of fragment pairs and trimers into those treated by full SCC calculations and those approximated by the ES-DIM scheme is controlled by van der Waals (vdW) scaling radii. A dimer is computed with a full SCC calculation if the shortest interatomic distance between its two constituent monomers is below the sum of the corresponding atomic vdW radii multiplied by a pair scaling constant, which is set to 2.0; otherwise, the pair interaction energy is obtained from the ES-DIM approximation. For trimers, a full SCC calculation is performed if at least two of the three constituent monomer pairs satisfy an analogous distance criterion with a trimer scaling constant of 1.5.

All calculations were performed on a computing node with dual E5-2680 Xeon CPUs (2.4 GHz, 14 cores each), 188 GB of DDR4 memory, and a SATA hard drive. The serial benchmarks employed a single CPU core, while the parallel benchmarks utilized all 28 cores of the same node.

The Mercury\cite{macrae_mercury_2020} program was used to create the various clusters of the anthracene and pentacene systems from their respective crystal structures\cite{mason_crystallography_1964,holmes_nature_1999}.

We employed packmol\cite{martinez_packmol_2009} to create the various water clusters used in this work.

\subsection{Energy accuracy}

\begin{figure}[t!]
    \centering
    \includegraphics[width=\linewidth]{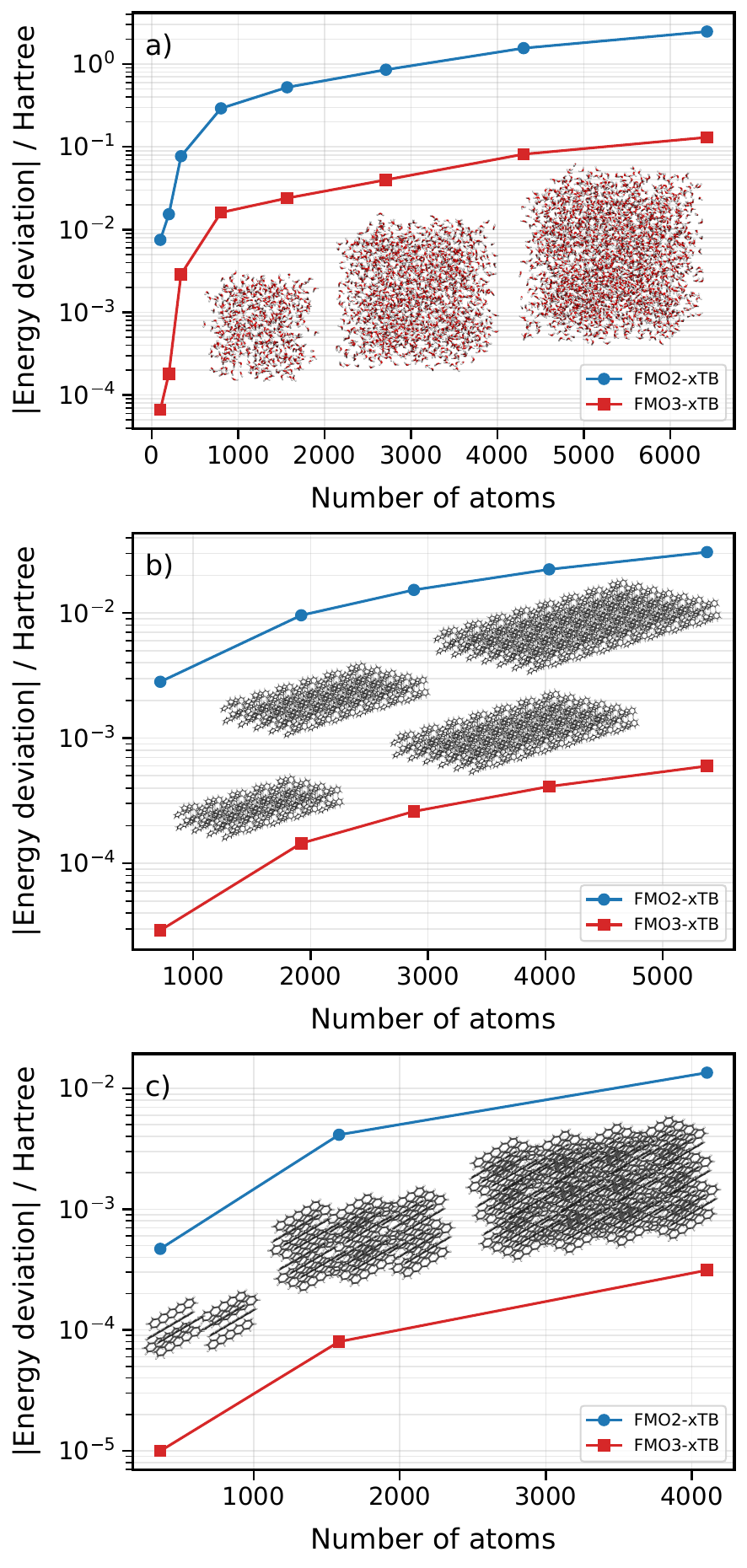}
    \caption{Absolute energy deviation of FMO2-xTB and FMO3-xTB from the non-fragmented xTB reference for (a) water clusters, (b) anthracene planes, and (c) pentacene supercells. One molecule per monomer fragment is used in all calculations.}
    \label{fig:energy_deviation}
\end{figure}

The accuracy of the FMO2-xTB and FMO3-xTB total energies is evaluated by comparison with non-fragmented GFN1-xTB reference calculations for water clusters, anthracene clusters, and pentacene supercells. In all FMO calculations, one molecule is used as the monomer fragment. The absolute energy deviations $|\Delta E| = |E_\mathrm{FMO} - E_\mathrm{xTB}|$ are shown in Fig.~\ref{fig:energy_deviation} as a function of system size for the three benchmark systems.

For the water clusters [Fig.~\ref{fig:energy_deviation}(a)], the FMO2-xTB energy deviation grows rapidly with system size, reaching $2.47$~Hartree (1\,550~kcal/mol) for the largest cluster containing 6\,423 atoms. This behavior is expected, as water clusters feature dense hydrogen-bonding networks with significant cooperative many-body polarization and charge-transfer effects that are not fully captured by the two-body expansion.\cite{fedorov_importance_2004,nishimoto_three-body_2017} The FMO3-xTB method reduces the energy deviation by roughly one to two orders of magnitude across all system sizes, yielding a deviation of $0.13$~Hartree (81.6~kcal/mol) for the 6\,423-atom cluster. This substantial improvement confirms the importance of three-body corrections for hydrogen-bonded systems, consistent with earlier findings for FMO-RHF\cite{fedorov_importance_2004} and FMO-DFTB.\cite{nishimoto_three-body_2017}

For the anthracene clusters [Fig.~\ref{fig:energy_deviation}(b)], the energy deviations are considerably smaller than for the water clusters. The anthracene systems are constructed as planar arrangements in which the cluster grows in two dimensions, representing a layered molecular packing motif. The FMO2-xTB deviation for the largest anthracene system (5\,376 atoms, 224 molecules) amounts to $3.07 \times 10^{-2}$~Hartree (19.3~kcal/mol), while FMO3-xTB reduces this to $5.99 \times 10^{-4}$~Hartree (0.38~kcal/mol), corresponding to an improvement of approximately two orders of magnitude. This reduced fragmentation error reflects the weaker many-body character of $\pi$-stacking interactions compared to hydrogen bonding. The dominant intermolecular interactions in anthracene clusters are of dispersion type, which are treated as pairwise-additive contributions in GFN1-xTB via the D3 correction\cite{grimme_robust_2017} and therefore do not contribute to the fragmentation error.

The pentacene supercells [Fig.~\ref{fig:energy_deviation}(c)] exhibit the smallest energy deviations among the three benchmark systems. In contrast to the two-dimensional anthracene planes, the pentacene systems are constructed as three-dimensional supercells from the pentacene crystal structure, in which the molecules adopt a herringbone arrangement with neighbors along all three crystallographic axes. For the $3 \times 3 \times 3$ supercell (4\,104 atoms), the FMO2-xTB and FMO3-xTB deviations are $1.35 \times 10^{-2}$~Hartree (8.5~kcal/mol) and $3.12 \times 10^{-4}$~Hartree (0.20~kcal/mol), respectively. Despite the three-dimensional growth, the energy deviations per molecule remain smaller than for the anthracene planes, which can be attributed to the herringbone packing: the tilted molecular arrangement results in weaker orbital overlap between neighboring molecules compared to the cofacial stacking in the anthracene planes, leading to reduced charge-transfer contributions and consequently smaller many-body effects.

\begin{figure}[t!]
    \centering
    \includegraphics[width=\linewidth]{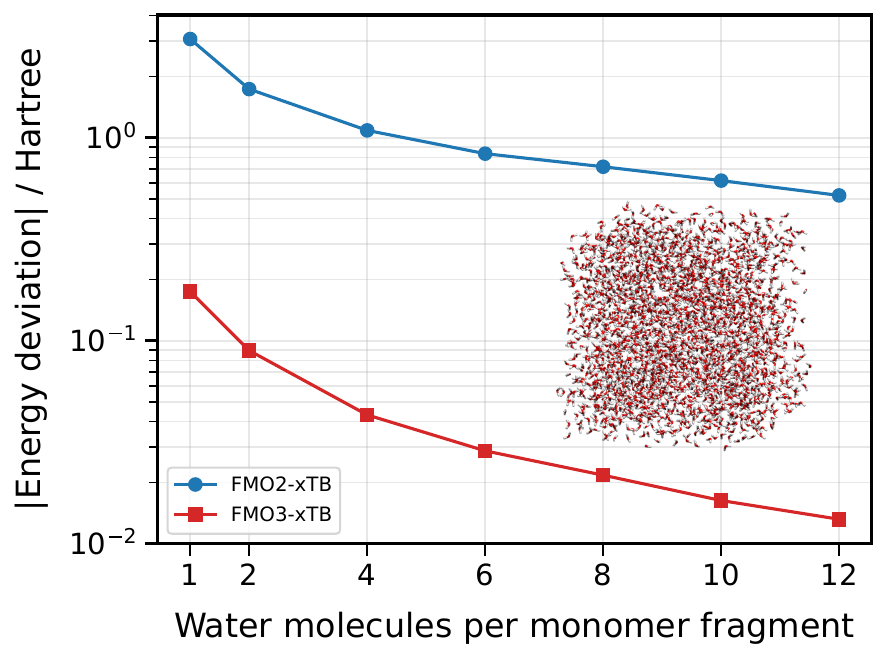}
    \caption{Absolute energy deviation of FMO2-xTB and FMO3-xTB from the non-fragmented xTB reference for a water cluster with 2\,400 molecules (7\,200 atoms) as a function of the number of water molecules per monomer fragment.}
    \label{fig:accuracy_fragment_size}
\end{figure}

In addition to the system-size dependence, we investigated the effect of the monomer fragment size on the fragmentation error. For this purpose, we computed the FMO2-xTB and FMO3-xTB energies of a water cluster containing 2\,400 molecules (7\,200 atoms) using monomer fragments consisting of 1 to 12 water molecules. The results are shown in Fig.~\ref{fig:accuracy_fragment_size}. As expected, increasing the fragment size systematically reduces the energy deviation for both FMO2 and FMO3, since larger fragments capture a greater share of the short-range many-body interactions within the monomer calculation itself.\cite{fedorov_importance_2004,fedorov_three-body_2006} At the smallest fragment size of one water molecule, the FMO2-xTB and FMO3-xTB deviations are $3.07$~Hartree (1\,926~kcal/mol) and $0.17$~Hartree (107~kcal/mol), respectively. Increasing the fragment size to 12 water molecules reduces these values to $0.52$~Hartree (326~kcal/mol) and $0.013$~Hartree (8.2~kcal/mol), respectively. The FMO3-xTB energy deviation decreases more steeply with fragment size, as the three-body expansion benefits from the reduced number of trimers with significant interactions. Notably, even with a single water molecule per fragment, the FMO3-xTB deviation remains more than one order of magnitude smaller than the FMO2-xTB result, confirming that the inclusion of three-body terms is more effective than increasing fragment size for improving accuracy in strongly interacting systems.\cite{fedorov_importance_2004}

To assess how the fragmentation error distributes across the system, Table~\ref{tab:per_frag_noncov} collects the energy deviations normalized by the number of fragments. For the water clusters, the per-fragment FMO2-xTB deviation grows from 228 to 1155~$\mu$Ha as the cluster size increases, reflecting the expanding coordination shell of each water molecule in the increasingly bulk-like interior. The FMO3-xTB per-fragment deviation follows the same trend but is one to two orders of magnitude smaller (2--61~$\mu$Ha). For the anthracene planes and pentacene supercells, the per-fragment errors are smaller overall and grow more slowly with system size, consistent with the weaker many-body character of the $\pi$-stacking interactions.

\begin{table}[t!]
  \centering
  \caption{Energy deviation per fragment ($\mu$Ha/frag) for FMO2-xTB and FMO3-xTB with respect to the unfragmented xTB reference for non-covalently bound molecular clusters. One molecule per monomer fragment is used in all calculations.}
  \label{tab:per_frag_noncov}
  \small
  \begin{tabular*}{\linewidth}{@{\extracolsep{\fill}}lrrrr}
    \toprule
    System & $N_{\text{atoms}}$ & $N_{\text{frag}}$
           & FMO2 & FMO3 \\
    \midrule
    \multicolumn{5}{l}{\textit{Water clusters}} \\
    (H$_2$O)$_{33}$    &   99 &   33 &  227.8 &   2.0 \\
    (H$_2$O)$_{66}$    &  198 &   66 &  232.4 &   2.8 \\
    (H$_2$O)$_{113}$   &  339 &  113 &  682.4 &  25.5 \\
    (H$_2$O)$_{267}$   &  801 &  267 & 1093.6 &  60.3 \\
    (H$_2$O)$_{522}$   & 1566 &  522 & 1004.8 &  46.0 \\
    (H$_2$O)$_{903}$   & 2709 &  903 &  947.5 &  44.2 \\
    (H$_2$O)$_{1434}$  & 4302 & 1434 & 1087.3 &  56.9 \\
    (H$_2$O)$_{2141}$  & 6423 & 2141 & 1155.4 &  60.8 \\
    \midrule
    \multicolumn{5}{l}{\textit{Anthracene clusters}} \\
    Cluster 1     &  720 &   30 &   94.1 &   1.0 \\
    Cluster 2     & 1920 &   80 &  120.0 &   1.8 \\
    Cluster 3     & 2880 &  120 &  127.6 &   2.2 \\
    Cluster 4     & 4032 &  168 &  133.0 &   2.5 \\
    Cluster 5     & 5376 &  224 &  137.0 &   2.7 \\
    \midrule
    \multicolumn{5}{l}{\textit{Pentacene supercells}} \\
    $1\!\times\!1\!\times\!1$ &  360 &  10 &   46.9 &   1.0 \\
    $2\!\times\!2\!\times\!2$ & 1584 &  44 &   94.0 &   1.8 \\
    $3\!\times\!3\!\times\!3$ & 4104 & 114 &  118.6 &   2.7 \\
    \bottomrule
  \end{tabular*}
\end{table}

We note that the remaining FMO3-xTB energy deviation for the water clusters could, in principle, be further reduced by extending the many-body expansion to include four-body and higher-order terms. The FMO4 expansion has been developed for \textit{ab initio} wave function methods\cite{fedorov_importance_2004} and has been shown to provide additional accuracy improvements for systems with strong many-body polarization effects. However, the rapidly increasing number of tetramers makes the FMO4 approach considerably more expensive than FMO3, and the trade-off between accuracy and computational cost would need to be carefully evaluated for the tight-binding framework. For the organic semiconductor systems studied here, the FMO3-xTB accuracy is already excellent, and the inclusion of higher-order terms would not be expected to yield significant improvements.

\subsection{Energy accuracy of covalent bond fragmentation}
\begin{figure}[b!]
    \centering
    \includegraphics[width=\linewidth]{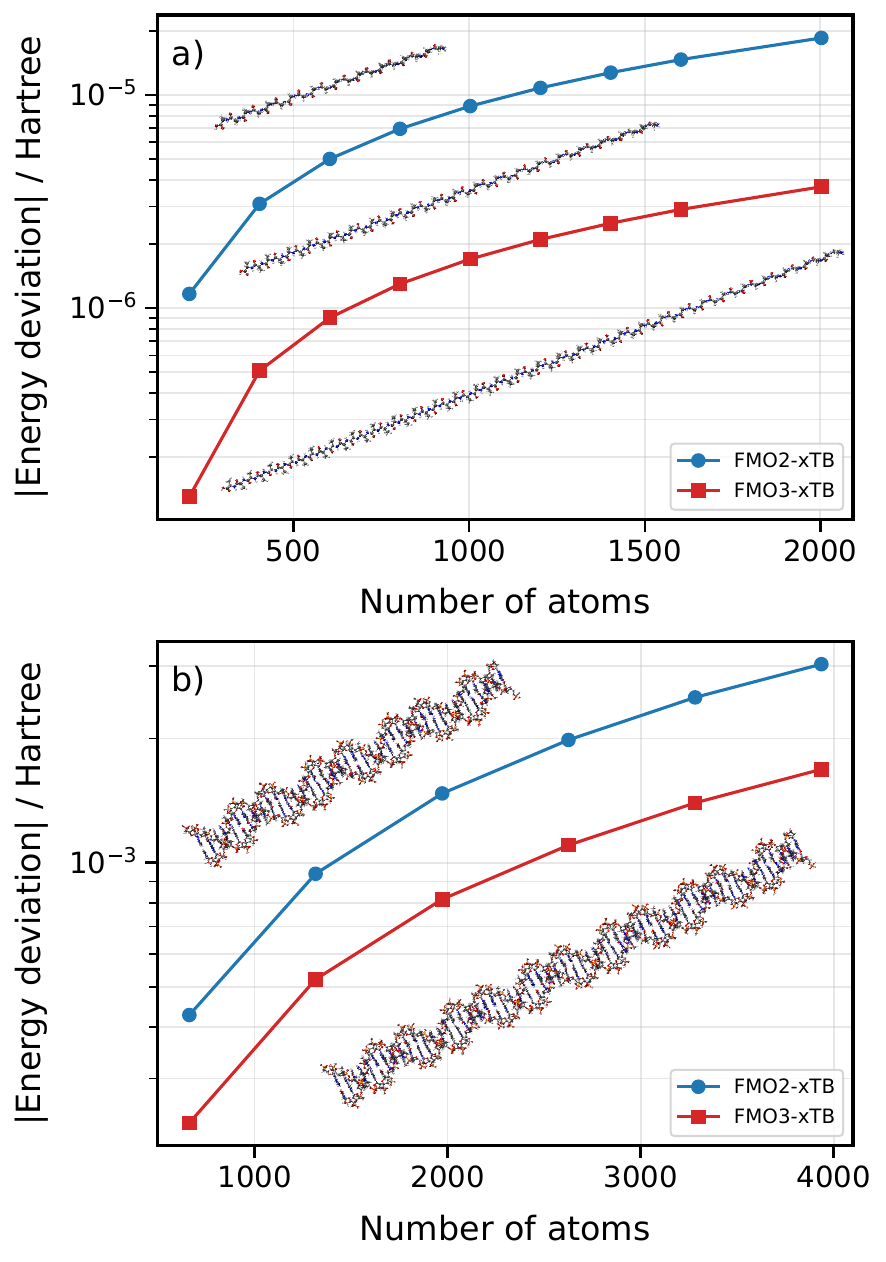}
    \caption{Absolute energy deviation of FMO2-xTB and FMO3-xTB from the non-fragmented xTB reference for (a) polyalanine chains with HOP covalent fragmentation ($\sim$40 atoms per fragment) and (b) B-DNA double helices with HOP covalent fragmentation (2 base pairs per strand fragment).}
    \label{fig:covalent_energy}
\end{figure}

The energy accuracy benchmarks presented above employ non-covalently bound molecular clusters in which each molecule defines a natural monomer fragment. To validate the HOP covalent bond fragmentation developed in Sec.~\ref{sec:hop}, we now consider two representative covalently bonded systems: polyalanine chains and B-DNA double helices. In both cases, the FMO-xTB energies are compared to unfragmented GFN1-xTB reference calculations to isolate the fragmentation error introduced by the bond detachment. The absolute energy deviations are shown in Fig.~\ref{fig:covalent_energy}.

For the polyalanine systems, covalent bonds are detached at the C$_\alpha$--C bonds using the HOP boundary treatment. Fig.~\ref{fig:covalent_energy}(a) shows the energy deviations for a fragment size of approximately 40 atoms (2 residues per fragment). The fragmentation errors are remarkably small: for the longest chain (Ala200, 2\,003 atoms), the FMO2-xTB deviation is $1.85 \times 10^{-5}$~Hartree (0.012~kcal/mol) and FMO3-xTB reduces this to $3.71 \times 10^{-6}$~Hartree (0.002~kcal/mol). These errors are several orders of magnitude smaller than those observed for the non-covalent water clusters at comparable system sizes, reflecting the predominantly local character of the backbone interactions in $\alpha$-helical polypeptides. The deviations grow linearly with chain length, indicating that the per-residue error is approximately constant.

For the B-DNA double helices [Fig.~\ref{fig:covalent_energy}(b)], the fragmentation is performed with 2 base pairs per strand fragment, with covalent bond detachment at the phosphodiester backbone. The energy deviations are larger than for the polyalanine systems, with FMO2-xTB reaching $-3.03$~millihartree ($-1.90$~kcal/mol) and FMO3-xTB reaching $+1.69$~millihartree ($+1.06$~kcal/mol) for the largest system (60 base pairs, 3\,936 atoms). The opposite signs of the FMO2 and FMO3 deviations indicate that the two-body expansion underestimates the total energy, while the three-body correction overshoots slightly. The larger fragmentation errors compared to polyalanine can be attributed to the more complex electronic structure of the DNA double helix, in which the hydrogen bonding between complementary base pairs, the $\pi$-stacking interactions between adjacent bases, and the charged phosphate backbone give rise to stronger many-body polarization effects across fragment boundaries. Nevertheless, the deviations grow approximately linearly with system size, corresponding to a per-base-pair error of approximately 0.051~millihartree (0.032~kcal/mol) for FMO2 and 0.028~millihartree (0.018~kcal/mol) for FMO3.

The per-fragment energy deviations are collected in Table~\ref{tab:per_frag_cov}. In contrast to the non-covalent systems (Table~\ref{tab:per_frag_noncov}), the per-fragment errors for both polyalanine and DNA converge rapidly with system size. For polyalanine, the FMO3-xTB per-fragment deviation stabilizes at approximately 0.07~$\mu$Ha (0.00005~kcal/mol) beyond Ala$_{60}$, while the FMO2-xTB value converges to 0.37~$\mu$Ha (0.00023~kcal/mol). This rapid convergence reflects the local character of the backbone interactions in $\alpha$-helical polypeptides, where interior fragments quickly reach a bulk-like environment. For the B-DNA systems, the per-fragment FMO2-xTB deviation converges to approximately 51~$\mu$Ha (0.032~kcal/mol) and the FMO3-xTB value to 28~$\mu$Ha (0.018~kcal/mol), with both quantities being nearly constant beyond 30 base pairs.

\begin{table}[t!]
  \centering
  \caption{Absolute energy deviation per fragment ($\mu$Ha/frag) for FMO2-xTB and FMO3-xTB with respect to the unfragmented xTB reference for covalently fragmented systems using the HOP boundary treatment.}
  \label{tab:per_frag_cov}
  \small
    \begin{tabular*}{\linewidth}{@{\extracolsep{\fill}}lrrrr}
      \toprule
      System & $N_{\text{atoms}}$ & $N_{\text{frag}}$ & FMO2 & FMO3 \\
      \midrule
      \multicolumn{5}{l}{\textit{Polyalanine} ($\sim$40 atoms/fragment)} \\
      Ala$_{20}$  &  203 &  5 & 0.23 & 0.03 \\
      Ala$_{40}$  &  403 & 10 & 0.31 & 0.05 \\
      Ala$_{60}$  &  603 & 15 & 0.33 & 0.06 \\
      Ala$_{80}$  &  803 & 20 & 0.35 & 0.06 \\
      Ala$_{100}$ & 1003 & 25 & 0.35 & 0.07 \\
      Ala$_{120}$ & 1203 & 30 & 0.36 & 0.07 \\
      Ala$_{140}$ & 1403 & 35 & 0.36 & 0.07 \\
      Ala$_{160}$ & 1603 & 40 & 0.37 & 0.07 \\
      Ala$_{200}$ & 2003 & 50 & 0.37 & 0.07 \\
      \midrule
      \multicolumn{5}{l}{\textit{B-DNA double helix} (1 bp/fragment)} \\
      10\,bp & 662  & 10 & 42.7 & 23.4 \\
      20\,bp & 1316 & 20 & 47.0 & 26.1 \\
      30\,bp & 1972 & 30 & 49.1 & 27.2 \\
      40\,bp & 2626 & 40 & 49.6 & 27.6 \\
      50\,bp & 3282 & 50 & 50.4 & 27.9 \\
      60\,bp & 3936 & 60 & 50.5 & 28.1 \\
      \bottomrule
    \end{tabular*}
\end{table}
Overall, the HOP covalent fragmentation introduces only a small systematic error that scales linearly with the number of cut bonds and saturates at a per-fragment limit. The FMO3-xTB correction reduces this error by a factor of 2--10 across all systems studied, demonstrating that the combination of HOP with the three-body expansion yields accurate energies for covalently bonded macromolecules at a fraction of the cost of the unfragmented calculation.

\subsection{Computational scaling}
\begin{figure}[b!]
    \centering
    \includegraphics[width=\linewidth]{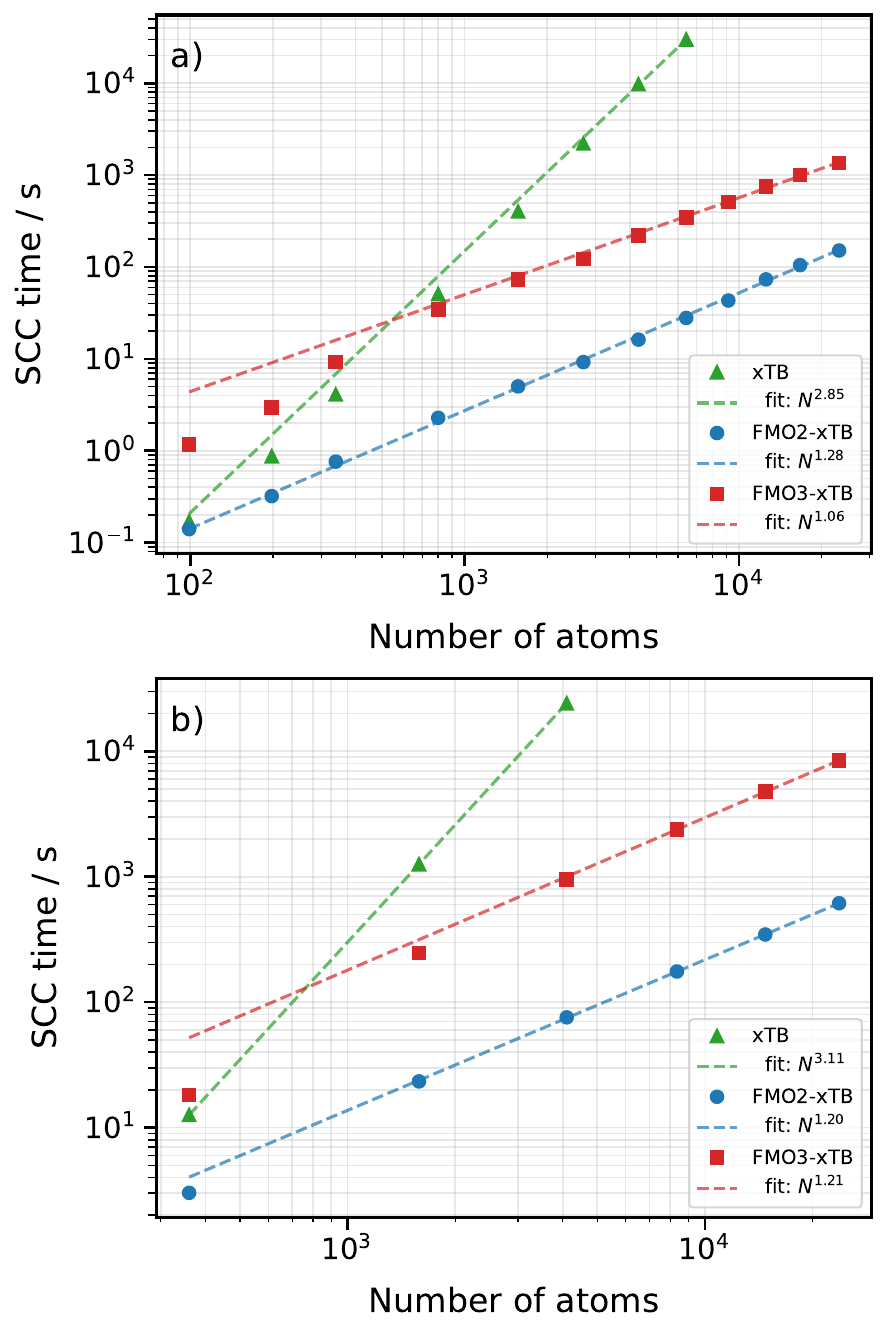}
    \caption{Computational scaling of the SCC wall-clock time for (a) water clusters and (b) pentacene supercells. The dashed lines indicate power-law fits $t \propto N_\mathrm{atoms}^b$ with the fitted scaling exponents shown in the legend. All calculations were performed on a single CPU core.}
    \label{fig:scaling}
\end{figure}

The computational efficiency of the FMO-xTB methods is assessed by measuring the wall-clock times for single-point SCC calculations of water clusters and pentacene supercells of increasing size. All timings were obtained on a single CPU core to allow for a direct determination of the scaling behavior. The results are shown in Fig.~\ref{fig:scaling} and summarized in Table~\ref{tab:scc_times}. The effective scaling exponent $b$ was determined by fitting a power law $t = a \cdot N_\mathrm{atoms}^b$ to the measured timings.

For the water clusters [Fig.~\ref{fig:scaling}(a)], the non-fragmented xTB calculation exhibits a scaling exponent of $b = 2.85$, which is close to the expected cubic scaling arising from the matrix diagonalization step. The crossover point at which FMO2-xTB becomes faster than full xTB lies at around 100 atoms. Beyond this point, the FMO approach provides substantial acceleration; the additional performance gains from parallelization are discussed below.

\begin{table}[t!]
    \centering
    \caption{SCC computation times for water clusters and pentacene supercells on a single CPU core. The scaling exponent $b$ is obtained from a power-law fit $t = a \cdot N_\mathrm{atoms}^b$.}
    \label{tab:scc_times}
      \small
    \begin{tabular*}{\linewidth}{@{\extracolsep{\fill}} r r r r}
        \hline\hline
        \multicolumn{4}{c}{\textbf{Water clusters}} \\
        \hline
        $N_\text{atoms}$ & xTB / s & FMO2-xTB / s & FMO3-xTB / s \\
        \hline
           99 &       0.17 &      0.14 &       1.17 \\
          198 &       0.87 &      0.32 &       2.95 \\
          339 &       4.10 &      0.76 &       9.22 \\
          801 &      50.63 &      2.28 &      34.63 \\
        1\,566 &     402.30 &      5.01 &      72.75 \\
        2\,709 &   2\,212.11 &      9.22 &     122.69 \\
        4\,302 &   9\,792.56 &     16.19 &     221.67 \\
        6\,423 &  29\,934.17 &     27.86 &     346.85 \\
        9\,144 &       --- &     43.15 &     512.05 \\
       12\,543 &       --- &     73.24 &     750.30 \\
       16\,698 &       --- &    104.70 &     999.17 \\
       23\,178 &       --- &    151.12 &   1\,355.95 \\
        \hline
        $b$    &      2.85 &      1.28 &       1.06 \\
        \hline\hline
        \multicolumn{4}{c}{\textbf{Pentacene supercells}} \\
        \hline
        $N_\text{atoms}$ & xTB / s & FMO2-xTB / s & FMO3-xTB / s \\
        \hline
          360 &      12.63 &      3.02 &      18.18 \\
        1\,584 &   1\,256.25 &     23.37 &     246.57 \\
        4\,104 &  24\,160.17 &     75.65 &     954.90 \\
        8\,352 &       --- &    175.64 &   2\,381.81 \\
       14\,760 &       --- &    346.33 &   4\,798.32 \\
       23\,760 &       --- &    614.76 &   8\,429.15 \\
        \hline
        $b$    &      3.11 &      1.20 &       1.22 \\
        \hline\hline
    \end{tabular*}
\end{table}

For the cluster containing 6\,423 atoms, the full xTB calculation requires approximately 29\,934~s ($\sim$8.3~h), whereas FMO2-xTB and FMO3-xTB complete in 27.9 and 346.9~s, respectively. This corresponds to speed-up factors of approximately 1\,073 and 86 for FMO2 and FMO3, respectively. For the largest water cluster studied (23\,178 atoms), where the extrapolated time of a full xTB calculation is approximately 320 hours, FMO2-xTB requires only 151~s and FMO3-xTB completes in 1\,356~s. The fitted scaling exponents are $b = 1.28$ for FMO2-xTB and $b = 1.06$ for FMO3-xTB. The near-linear scaling of the FMO approach is consistent with the behavior reported for FMO-DFTB,\cite{nishimoto_density-functional_2014,vuong_fragment_2019} and reflects the use of distance-based approximations (ES-DIM) that limit the number of explicitly computed dimer and trimer pairs to scale linearly with the number of fragments.\cite{nishimoto_density-functional_2014}

The observation that FMO3-xTB exhibits a slightly lower scaling exponent than FMO2-xTB may appear counterintuitive, since trimer calculations introduce an additional computational step. However, this behavior has been reported previously for FMO-LC-DFTB\cite{vuong_fragment_2019} and can be attributed to the fact that the number of SCC trimers and dimers both scale linearly with system size due to the distance thresholds, while the observed total scaling is a combined effect of several steps with different individual scaling behaviors. In particular, the linear-scaling trimer calculation is the predominant factor in the total FMO3 timing, thereby lowering the effective scaling exponent.

For the pentacene supercells [Fig.~\ref{fig:scaling}(b)], the non-fragmented xTB method shows a scaling exponent of $b = 3.11$. The FMO2-xTB and FMO3-xTB methods reduce the scaling to $b = 1.20$ and $b = 1.22$, respectively. For the $3 \times 3 \times 3$ supercell (4\,104 atoms), the full xTB calculation takes 24\,160~s ($\sim$6.7~h), while FMO2-xTB and FMO3-xTB require only 76 and 955~s, representing speed-up factors of approximately 319 and 25, respectively. The slightly higher scaling exponents relative to water clusters can be attributed to the larger fragment size of pentacene (36 atoms per molecule versus 3 atoms for water), which increases the computational cost of the embedding potential evaluation and dimer calculations. For the largest pentacene system ($6 \times 6 \times 6$ supercell, 23\,760 atoms), the FMO2-xTB calculation completes in approximately 615~s ($\sim$10~min), demonstrating the practical applicability of the method to systems with tens of thousands of atoms.

\begin{figure}[t!]
    \centering
    \includegraphics[width=\linewidth]{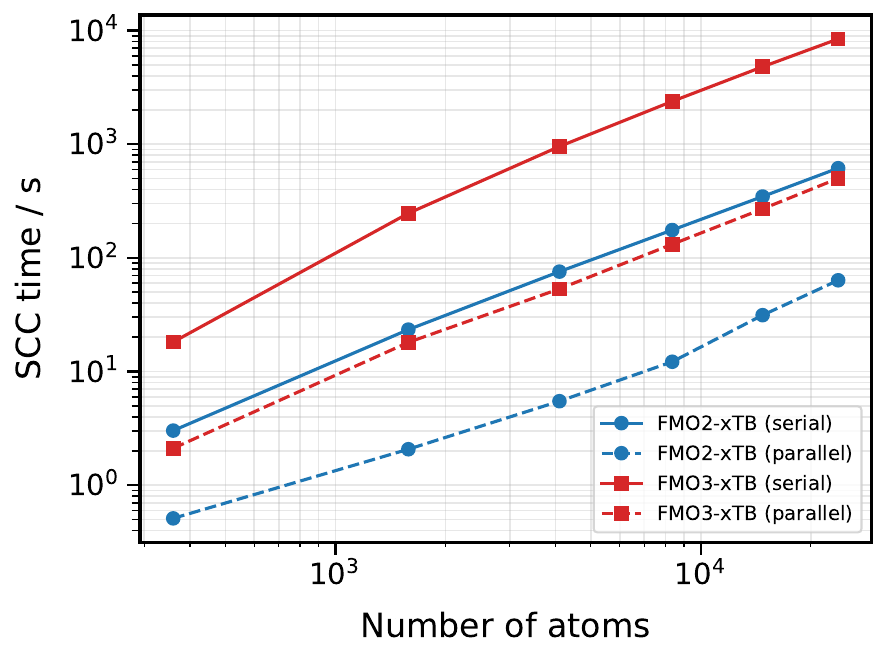}
    \caption{Comparison of serial (solid lines) and parallel (dashed lines, 28 cores) SCC wall-clock times for FMO2-xTB and FMO3-xTB calculations of pentacene supercells.}
    \label{fig:parallel}
\end{figure}

A key advantage of the FMO approach is its inherent suitability for parallelization, since the individual monomer, dimer, and trimer calculations are largely independent and can be distributed across multiple CPU cores.\cite{nishimoto_density-functional_2014} To assess the parallel efficiency of the DIALECT implementation, we repeated the pentacene supercell calculations using all 28 cores of the same computing node. The resulting serial and parallel SCC timings are compared in Fig.~\ref{fig:parallel}. Overall SCC speed-ups of $9.7$--$14.4\times$ are obtained for FMO2-xTB and $13.7$--$18.1\times$ for FMO3-xTB on 28 cores. The higher parallel efficiency of FMO3-xTB reflects the fact that the computationally dominant trimer and dimer SCC calculations parallelize very efficiently. In contrast, the monomer SCC step---which includes the assembly of the embedding electrostatic potential from the full-system $\gamma$-matrix---is not parallelized and therefore limits the overall speed-up for FMO2-xTB, where this step constitutes a larger fraction of the total time. With parallelization, the FMO3-xTB calculation of the $6 \times 6 \times 6$ pentacene supercell (23\,760 atoms) completes in approximately 499~s ($\sim$8~min), which is comparable to the serial FMO2-xTB timing of 615~s for the same system. The FMO2-xTB parallel calculation of this system requires only approximately 63~s ($\sim$1~min), demonstrating that routine quantum-mechanical calculations of systems with tens of thousands of atoms are feasible on a single computing node.

\subsection{Gradient accuracy}

The accuracy of the analytic FMO-xTB gradient is verified by comparison with numerical gradients obtained by central finite differences. As a test system, we use a water cluster consisting of 33 molecules (99 atoms) with one water molecule per fragment. To investigate the role of the response terms derived in Sec.~\ref{sec:gradients}, analytic gradients were computed both with and without the response contribution.

\begin{figure}[t!]
    \centering
    \includegraphics[width=\linewidth]{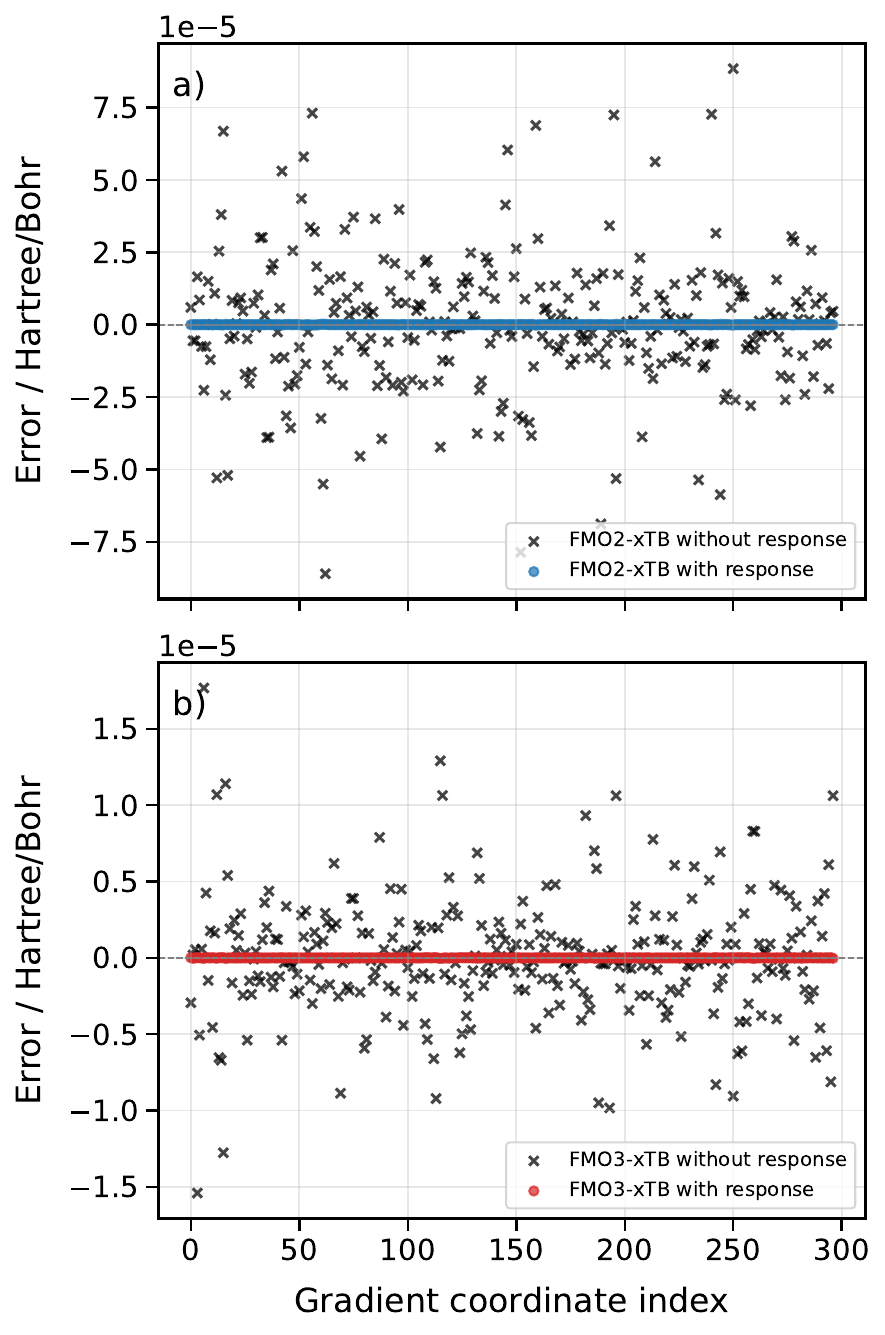}
    \caption{Signed error of the analytic gradient with respect to the numerical reference for (a) FMO2-xTB and (b) FMO3-xTB for the (H$_2$O)$_{33}$ cluster. Filled circles show the gradient with response terms included; crosses show the gradient without response terms.}
    \label{fig:gradient_error}
\end{figure}

The signed errors of the analytic gradient elements with respect to the numerical reference are shown in Fig.~\ref{fig:gradient_error} for FMO2-xTB [panel (a)] and FMO3-xTB [panel (b)]. The root-mean-square deviations (RMSD) and maximum absolute errors are summarized in Table~\ref{tab:gradient_errors}. When the response terms are included, the FMO2-xTB analytic gradient agrees with the numerical reference to an RMSD of $4.62 \times 10^{-8}$~hartree/bohr, with a maximum error of $1.88 \times 10^{-7}$~hartree/bohr. For FMO3-xTB, the agreement is even better, with an RMSD of $8.77 \times 10^{-9}$~hartree/bohr and a maximum error of $3.72 \times 10^{-8}$~hartree/bohr. These residual deviations are on the same order as the intrinsic errors of the numerical gradient itself and confirm the correctness of the analytic gradient implementation.
\begin{table}[b!]
  \centering
  \caption{Gradient errors of FMO2-xTB and FMO3-xTB analytic gradients compared to the numerical reference for the (H$_{2}$O)$_{33}$ cluster. All values are in hartree/bohr.}
  \label{tab:gradient_errors}
  \small
  \begin{tabular*}{\linewidth}{@{\extracolsep{\fill}} l c c c}
      \hline\hline
      Method & Response & RMSD & Max.\ error \\
      \hline
      FMO2-xTB & no  & $2.33 \times 10^{-5}$ & $8.84 \times 10^{-5}$ \\
      FMO2-xTB & yes & $4.62 \times 10^{-8}$ & $1.88 \times 10^{-7}$ \\
      FMO3-xTB & no  & $3.92 \times 10^{-6}$ & $1.77 \times 10^{-5}$ \\
      FMO3-xTB & yes & $8.77 \times 10^{-9}$ & $3.72 \times 10^{-8}$ \\
      \hline\hline
  \end{tabular*}
\end{table}

Without the response terms, the gradient errors increase substantially. For FMO2-xTB, the RMSD rises by a factor of approximately 500 to $2.33 \times 10^{-5}$~hartree/bohr, and the maximum error reaches $8.84 \times 10^{-5}$~hartree/bohr. As is evident from Fig.~\ref{fig:gradient_error}(a), the errors without response terms (crosses) are scattered across the full range of gradient coordinates, indicating a systematic deficiency rather than isolated outliers. For FMO3-xTB, the errors without response terms are smaller than for FMO2-xTB, with an RMSD of $3.92 \times 10^{-6}$~hartree/bohr and a maximum error of $1.77 \times 10^{-5}$~hartree/bohr. This observation is consistent with earlier findings for FMO-DFTB,\cite{nishimoto_large-scale_2015}, where the FMO3 gradient without response terms was found to be more accurate than the FMO2 gradient, since the three-body expansion indirectly captures some of the response effects through the explicit treatment of trimer interactions.\cite{nishimoto_adaptive_2018}

The magnitude of the response contribution of approximately $10^{-5}$ hartree/bohr observed for FMO2-xTB without response terms is comparable to the values reported for FMO-DFTB by Nishimoto \textit{et al.},\cite{nishimoto_large-scale_2015} where errors on the order of $10^{-4}$ hartree/bohr were found for the (H$_2$O)$_{64}$ cluster. The somewhat smaller values in the present case can be attributed to the smaller system size (33 versus 64 water molecules). These errors exceed typical convergence thresholds for geometry optimizations (usually $10^{-5}$ to $10^{-4}$~hartree/bohr) and can lead to energy oscillations during optimization and energy drift in molecular dynamics simulations.\cite{nishimoto_large-scale_2015,nakata_analytic_2020} The inclusion of response terms is therefore essential for reliable FMO-xTB geometry optimizations and molecular dynamics simulations.

\subsection{Gradient timings}
\begin{figure}[b!]
    \centering
    \includegraphics[width=\linewidth]{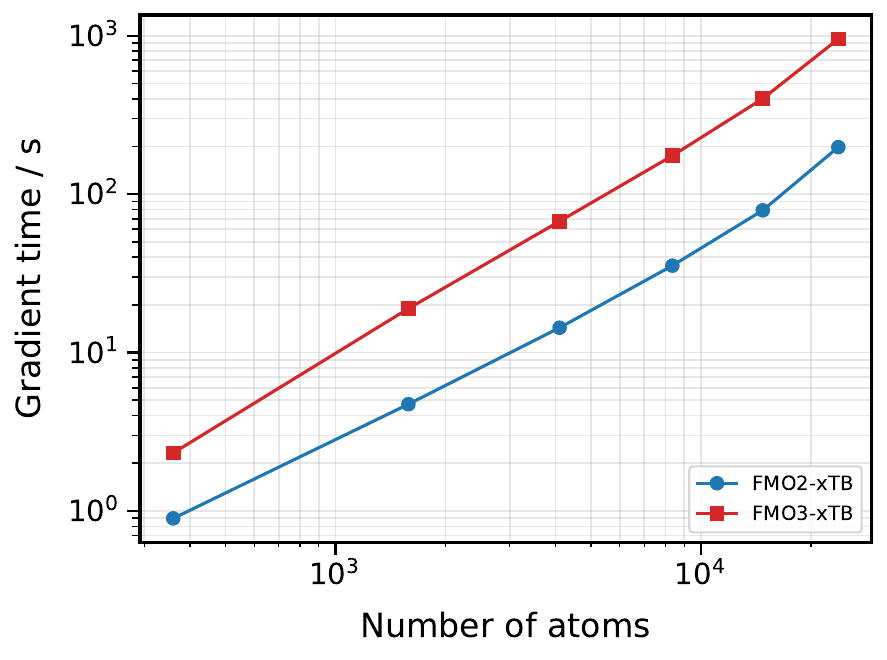}
    \caption{Wall-clock times for the analytic gradient evaluation of pentacene supercells with FMO2-xTB and FMO3-xTB using 28 CPU cores.}
    \label{fig:gradient_times}
\end{figure}

Having established the accuracy of the analytic gradient, we now assess its computational cost. The wall-clock times for the gradient evaluation of the pentacene supercell series are summarized in Table~\ref{tab:gradient_times} and illustrated in Fig.~\ref{fig:gradient_times}. The total gradient time is decomposed into the explicit and response gradient contributions.

For FMO2-xTB, the explicit gradient is dominated by the dimer gradient calculations, which account for approximately 80\% of the explicit gradient time for the largest supercell. The response gradient contributes roughly 29\% of the total FMO2-xTB gradient time, with the Lagrangian construction and the solution of the self-consistent Z-vector equations as the main steps. For FMO3-xTB, the trimer gradient calculations constitute the dominant contribution, accounting for approximately 84\% of the explicit gradient time. The response gradient represents a smaller fraction of the total time ($\sim$12\%) for FMO3-xTB, since the trimer explicit gradient is considerably more expensive than the corresponding response contribution. The FMO3-xTB response gradient is more costly than the FMO2-xTB response gradient due to additional Lagrangian terms arising from the three-body expansion.
\begin{table}[t!]
\caption{Gradient timings for pentacene supercells with FMO2-xTB and FMO3-xTB using 28 CPU cores.}
\label{tab:gradient_times}
  \small
\begin{tabular*}{\linewidth}{@{\extracolsep{\fill}}r r r r r}
\hline\hline
Supercell & $N_\text{atoms}$ & Expl.\,/\,s & Resp.\,/\,s & Total\,/\,s \\
\hline
\multicolumn{5}{c}{\textit{FMO2-xTB}} \\
\hline
$1\!\times\!1\!\times\!1$ & 360     &   0.65 &  0.24 &   0.90 \\
$2\!\times\!2\!\times\!2$ & 1\,584  &   3.22 &  1.50 &   4.72 \\
$3\!\times\!3\!\times\!3$ & 4\,104  &  10.12 &  4.23 &  14.35 \\
$4\!\times\!4\!\times\!4$ & 8\,352  &  24.74 & 10.64 &  35.38 \\
$5\!\times\!5\!\times\!5$ & 14\,760 &  54.56 & 24.47 &  79.03 \\
$6\!\times\!6\!\times\!6$ & 23\,760 & 140.85 & 57.27 & 198.12 \\
\hline
\multicolumn{5}{c}{\textit{FMO3-xTB}} \\
\hline
$1\!\times\!1\!\times\!1$ & 360     &   2.07 &   0.25 &    2.32 \\
$2\!\times\!2\!\times\!2$ & 1\,584  &  17.06 &   1.89 &   18.96 \\
$3\!\times\!3\!\times\!3$ & 4\,104  &  61.57 &   5.84 &   67.41 \\
$4\!\times\!4\!\times\!4$ & 8\,352  & 158.97 &  15.90 &  174.87 \\
$5\!\times\!5\!\times\!5$ & 14\,760 & 353.42 &  47.45 &  400.87 \\
$6\!\times\!6\!\times\!6$ & 23\,760 & 837.51 & 116.83 &  954.34 \\
\hline\hline
\end{tabular*}
\end{table}

For the largest pentacene supercell (23\,760 atoms), the total gradient evaluation requires 198~s ($\sim$3.3~min) for FMO2-xTB and 954~s ($\sim$16~min) for FMO3-xTB on 28 CPU cores. When combined with the parallel SCC times of 63~s (FMO2-xTB) and 499~s (FMO3-xTB) for this system (Fig.~\ref{fig:parallel}), a complete single-point energy and gradient calculation takes approximately 261~s ($\sim$4.4~min) for FMO2-xTB and 1\,453~s ($\sim$24~min) for FMO3-xTB. These timings demonstrate that molecular dynamics simulations with FMO-xTB are readily feasible for systems containing tens of thousands of atoms on a single computing node.

%% file: conclusion.tex
\section{Conclusion and outlook}

In this work, we have presented the fragment molecular orbital method combined with the GFN1-xTB extended tight-binding approach (FMO-xTB) for efficient quantum-mechanical calculations of large molecular systems. The method has been formulated at both the two-body (FMO2) and three-body (FMO3) levels, and fully analytic energy gradients, including the response contribution from the self-consistent embedding potential, have been derived and implemented. The FMO-xTB method has been implemented in the DIALECT software package.\cite{einsele_dialect_2025}

The accuracy of the FMO-xTB total energies has been systematically evaluated by comparison with non-fragmented GFN1-xTB reference calculations for three representative benchmark systems: water clusters with strong hydrogen-bonding networks, two-dimensional anthracene planes with $\pi$-stacking interactions, and three-dimensional pentacene supercells in the herringbone arrangement. For organic semiconductor systems, FMO3-xTB reproduces the unfragmented reference energy with deviations on the order of $10^{-4}$~Hartree, whereas for the more challenging water clusters, deviations are larger due to the strong many-body polarization and charge-transfer effects inherent to hydrogen-bonded networks. The fragment-size dependence study of a 2\,400-molecule water cluster has demonstrated that accuracy can be systematically improved by either increasing the number of molecules per fragment or extending the many-body expansion from FMO2 to FMO3, with the latter being the more effective strategy.

The covalent bond fragmentation capability using the HOP boundary treatment has been validated for polyalanine chains and B-DNA double helices. For polyalanine, the FMO3-xTB energy deviations remain on the order of $10^{-6}$~Hartree for chains up to 2\,003 atoms, demonstrating the excellent accuracy of the HOP fragmentation for polypeptide systems. For B-DNA, the deviations are larger due to the more complex electronic environment of the double helix, but remain in the millihartree range.

The computational scaling analysis confirms that the FMO-xTB method scales nearly linearly with system size. Scaling exponents between $b = 1.06$ and $b = 1.28$ have been obtained for water clusters and pentacene supercells, compared to the cubic scaling ($b \approx 2.9$--$3.1$) of the non-fragmented xTB method. For the largest systems studied, containing over 23\,000 atoms, the FMO2-xTB calculation completes in approximately 2.5 minutes on a single CPU core, rendering quantum-mechanical calculations of systems with tens of thousands of atoms routine. Speed-up factors of over three orders of magnitude compared to full xTB have been achieved for
the largest water clusters where a direct comparison was possible.

The parallel efficiency of the implementation has been demonstrated for pentacene supercells using 28 CPU cores, yielding overall SCC speed-ups of up to $14.4\times$ for FMO2-xTB and $18.1\times$ for FMO3-xTB. With parallelization, the FMO3-xTB calculation of a pentacene supercell containing 23\,760 atoms completes in approximately 8 minutes on a single computing node, while the corresponding FMO2-xTB calculation requires only about 1 minute.

The analytic gradient implementation has been verified against numerical reference gradients. The inclusion of the response terms, which account for the self-consistent coupling of fragment densities through the embedding potential, reduces the gradient RMSD by more than two orders of magnitude for both FMO2 and FMO3. Without response terms, the gradient errors are on the order of $10^{-5}$~hartree/bohr, which may lead to energy drift in molecular dynamics and unreliable geometry optimizations.\cite{nishimoto_large-scale_2015,nakata_analytic_2020} The fully analytic gradient with response terms provides gradient accuracy well within the thresholds required for stable simulations.

The gradient timings for pentacene supercells show that a complete parallel evaluation of energy and gradients for a system containing 23\,760 atoms requires approximately 4.4 minutes for FMO2-xTB and 24 minutes for FMO3-xTB on 28 CPU cores, making FMO-xTB molecular dynamics simulations of large organic semiconductor systems readily feasible on a single computing node.

A key advantage of the FMO-xTB approach compared to the previously developed FMO-DFTB method\cite{nishimoto_density-functional_2014,nishimoto_three-body_2017} lies in the broad element coverage inherited from GFN1-xTB. While DFTB methods require atom-pair-specific Slater--Koster parameters that are available only for a limited subset of element combinations,\cite{bannwarth_extended_2021} the GFN1-xTB method employs predominantly element-specific parameters that cover all spd-block elements up to radon ($Z = 86$).\cite{grimme_robust_2017} This makes FMO-xTB directly applicable to systems containing transition metals, heavy main-group elements, and other chemically diverse compositions without the need for additional parameterization.

Several avenues for future development can be envisioned. The extension of the FMO framework to GFN2-xTB,\cite{bannwarth_gfn2-xtbaccurate_2019} which includes an improved electrostatic treatment through multipole expansion and a self-consistent D4 dispersion model, would be a natural next step. The multipole-extended Coulomb interaction in GFN2-xTB could improve the description of the electrostatic embedding potential, thereby reducing fragmentation error, particularly for polar systems. However, adapting the FMO embedding formalism to multipole-based electrostatics requires careful theoretical development.

As described in Section~\ref{sec:hop}, the FMO-xTB method supports the fragmentation across covalent bonds through the hybrid orbital projection (HOP) operator,\cite{nakano_fragment_2000,fedorov_covalent_2008,nagata_importance_2010} and its accuracy has been demonstrated for polyalanine chains and B-DNA double helices. An alternative boundary treatment, the adaptive frozen orbital (AFO) approach,\cite{fedorov_covalent_2008,nishimoto_adaptive_2018} which has been applied to inorganic materials such as zeolites and nanowires,\cite{nishimoto_adaptive_2018} could be similarly adapted to the GFN1-xTB Hamiltonian in future work.

In summary, the FMO-xTB method provides an efficient and broadly applicable approach for quantum-mechanical simulations of large molecular systems. Its near-linear scaling, efficient parallelization, broad element coverage in GFN1-xTB, and availability of analytic gradients make it a versatile tool for investigating the structure and dynamics of complex molecular assemblies in materials science and biochemistry.

%% file: manuscript.bbl
\providecommand{\latin}[1]{#1}
\makeatletter
\providecommand{\doi}
  {\begingroup\let\do\@makeother\dospecials
  \catcode`\{=1 \catcode`\}=2 \doi@aux}
\providecommand{\doi@aux}[1]{\endgroup\texttt{#1}}
\makeatother
\providecommand*\mcitethebibliography{\thebibliography}
\csname @ifundefined\endcsname{endmcitethebibliography}  {\let\endmcitethebibliography\endthebibliography}{}
\begin{mcitethebibliography}{69}
\providecommand*\natexlab[1]{#1}
\providecommand*\mciteSetBstSublistMode[1]{}
\providecommand*\mciteSetBstMaxWidthForm[2]{}
\providecommand*\mciteBstWouldAddEndPuncttrue
  {\def\EndOfBibitem{\unskip.}}
\providecommand*\mciteBstWouldAddEndPunctfalse
  {\let\EndOfBibitem\relax}
\providecommand*\mciteSetBstMidEndSepPunct[3]{}
\providecommand*\mciteSetBstSublistLabelBeginEnd[3]{}
\providecommand*\EndOfBibitem{}
\mciteSetBstSublistMode{f}
\mciteSetBstMaxWidthForm{subitem}{(\alph{mcitesubitemcount})}
\mciteSetBstSublistLabelBeginEnd
  {\mcitemaxwidthsubitemform\space}
  {\relax}
  {\relax}

\bibitem[Houk and Liu(2017)Houk, and Liu]{houk_holy_2017}
Houk,~K.~N.; Liu,~F. Holy {Grails} for {Computational} {Organic} {Chemistry} and {Biochemistry}. \emph{Acc. Chem. Res.} \textbf{2017}, \emph{50}, 539--543\relax
\mciteBstWouldAddEndPuncttrue
\mciteSetBstMidEndSepPunct{\mcitedefaultmidpunct}
{\mcitedefaultendpunct}{\mcitedefaultseppunct}\relax
\EndOfBibitem
\bibitem[Grimme and Schreiner(2018)Grimme, and Schreiner]{grimme_computational_2018}
Grimme,~S.; Schreiner,~P.~R. Computational {Chemistry}: {The} {Fate} of {Current} {Methods} and {Future} {Challenges}. \emph{Angew. Chem. Int. Ed.} \textbf{2018}, \emph{57}, 4170--4176\relax
\mciteBstWouldAddEndPuncttrue
\mciteSetBstMidEndSepPunct{\mcitedefaultmidpunct}
{\mcitedefaultendpunct}{\mcitedefaultseppunct}\relax
\EndOfBibitem
\bibitem[Gordon \latin{et~al.}(2012)Gordon, Fedorov, Pruitt, and Slipchenko]{gordon_fragmentation_2012}
Gordon,~M.~S.; Fedorov,~D.~G.; Pruitt,~S.~R.; Slipchenko,~L.~V. Fragmentation {Methods}: {A} {Route} to {Accurate} {Calculations} on {Large} {Systems}. \emph{Chem. Rev.} \textbf{2012}, \emph{112}, 632--672\relax
\mciteBstWouldAddEndPuncttrue
\mciteSetBstMidEndSepPunct{\mcitedefaultmidpunct}
{\mcitedefaultendpunct}{\mcitedefaultseppunct}\relax
\EndOfBibitem
\bibitem[Raghavachari and Saha(2015)Raghavachari, and Saha]{raghavachari_accurate_2015}
Raghavachari,~K.; Saha,~A. Accurate {Composite} and {Fragment}-{Based} {Quantum} {Chemical} {Models} for {Large} {Molecules}. \emph{Chem. Rev.} \textbf{2015}, \emph{115}, 5643--5677\relax
\mciteBstWouldAddEndPuncttrue
\mciteSetBstMidEndSepPunct{\mcitedefaultmidpunct}
{\mcitedefaultendpunct}{\mcitedefaultseppunct}\relax
\EndOfBibitem
\bibitem[Pople \latin{et~al.}(1965)Pople, Santry, and Segal]{pople_approximate_1965}
Pople,~J.~A.; Santry,~D.~P.; Segal,~G.~A. Approximate {Self}‐{Consistent} {Molecular} {Orbital} {Theory}. {I}. {Invariant} {Procedures}. \emph{J. Chem. Phys.} \textbf{1965}, \emph{43}, S129--S135\relax
\mciteBstWouldAddEndPuncttrue
\mciteSetBstMidEndSepPunct{\mcitedefaultmidpunct}
{\mcitedefaultendpunct}{\mcitedefaultseppunct}\relax
\EndOfBibitem
\bibitem[Dewar and Thiel(1977)Dewar, and Thiel]{dewar_ground_1977}
Dewar,~M. J.~S.; Thiel,~W. Ground states of molecules. 38. {The} {MNDO} method. {Approximations} and parameters. \emph{J. Am. Chem. Soc.} \textbf{1977}, \emph{99}, 4899--4907\relax
\mciteBstWouldAddEndPuncttrue
\mciteSetBstMidEndSepPunct{\mcitedefaultmidpunct}
{\mcitedefaultendpunct}{\mcitedefaultseppunct}\relax
\EndOfBibitem
\bibitem[Stewart(1989)]{stewart_optimization_1989}
Stewart,~J. J.~P. Optimization of parameters for semiempirical methods {I}. {Method}. \emph{J. Comput. Chem.} \textbf{1989}, \emph{10}, 209--220\relax
\mciteBstWouldAddEndPuncttrue
\mciteSetBstMidEndSepPunct{\mcitedefaultmidpunct}
{\mcitedefaultendpunct}{\mcitedefaultseppunct}\relax
\EndOfBibitem
\bibitem[Thiel(2014)]{thiel_semiempirical_2014}
Thiel,~W. Semiempirical quantum–chemical methods. \emph{Wiley Interdiscip. Rev. Comput. Mol. Sci.} \textbf{2014}, \emph{4}, 145--157\relax
\mciteBstWouldAddEndPuncttrue
\mciteSetBstMidEndSepPunct{\mcitedefaultmidpunct}
{\mcitedefaultendpunct}{\mcitedefaultseppunct}\relax
\EndOfBibitem
\bibitem[Christensen \latin{et~al.}(2016)Christensen, Kubař, Cui, and Elstner]{christensen_semiempirical_2016}
Christensen,~A.~S.; Kubař,~T.; Cui,~Q.; Elstner,~M. Semiempirical {Quantum} {Mechanical} {Methods} for {Noncovalent} {Interactions} for {Chemical} and {Biochemical} {Applications}. \emph{Chem. Rev.} \textbf{2016}, \emph{116}, 5301--5337\relax
\mciteBstWouldAddEndPuncttrue
\mciteSetBstMidEndSepPunct{\mcitedefaultmidpunct}
{\mcitedefaultendpunct}{\mcitedefaultseppunct}\relax
\EndOfBibitem
\bibitem[Elstner \latin{et~al.}(1998)Elstner, Porezag, Jungnickel, Elsner, Haugk, Frauenheim, Suhai, and Seifert]{elstner_self-consistent-charge_1998}
Elstner,~M.; Porezag,~D.; Jungnickel,~G.; Elsner,~J.; Haugk,~M.; Frauenheim,~T.; Suhai,~S.; Seifert,~G. Self-consistent-charge density-functional tight-binding method for simulations of complex materials properties. \emph{Phys. Rev. B} \textbf{1998}, \emph{58}, 7260--7268\relax
\mciteBstWouldAddEndPuncttrue
\mciteSetBstMidEndSepPunct{\mcitedefaultmidpunct}
{\mcitedefaultendpunct}{\mcitedefaultseppunct}\relax
\EndOfBibitem
\bibitem[Cui and Elstner(2014)Cui, and Elstner]{cui_density_2014}
Cui,~Q.; Elstner,~M. Density functional tight binding: values of semi-empirical methods in an ab initio era. \emph{Phys. Chem. Chem. Phys.} \textbf{2014}, \emph{16}, 14368--14377\relax
\mciteBstWouldAddEndPuncttrue
\mciteSetBstMidEndSepPunct{\mcitedefaultmidpunct}
{\mcitedefaultendpunct}{\mcitedefaultseppunct}\relax
\EndOfBibitem
\bibitem[Gaus \latin{et~al.}(2011)Gaus, Cui, and Elstner]{gaus_dftb3_2011}
Gaus,~M.; Cui,~Q.; Elstner,~M. {DFTB3}: {Extension} of the {Self}-{Consistent}-{Charge} {Density}-{Functional} {Tight}-{Binding} {Method} ({SCC}-{DFTB}). \emph{J. Chem. Theory Comput.} \textbf{2011}, \emph{7}, 931--948\relax
\mciteBstWouldAddEndPuncttrue
\mciteSetBstMidEndSepPunct{\mcitedefaultmidpunct}
{\mcitedefaultendpunct}{\mcitedefaultseppunct}\relax
\EndOfBibitem
\bibitem[Bannwarth \latin{et~al.}(2021)Bannwarth, Caldeweyher, Ehlert, Hansen, Pracht, Seibert, Spicher, and Grimme]{bannwarth_extended_2021}
Bannwarth,~C.; Caldeweyher,~E.; Ehlert,~S.; Hansen,~A.; Pracht,~P.; Seibert,~J.; Spicher,~S.; Grimme,~S. Extended tight‐binding quantum chemistry methods. \emph{Wiley Interdiscip. Rev. Comput. Mol. Sci.} \textbf{2021}, \emph{11}\relax
\mciteBstWouldAddEndPuncttrue
\mciteSetBstMidEndSepPunct{\mcitedefaultmidpunct}
{\mcitedefaultendpunct}{\mcitedefaultseppunct}\relax
\EndOfBibitem
\bibitem[Grimme \latin{et~al.}(2017)Grimme, Bannwarth, and Shushkov]{grimme_robust_2017}
Grimme,~S.; Bannwarth,~C.; Shushkov,~P. A {Robust} and {Accurate} {Tight}-{Binding} {Quantum} {Chemical} {Method} for {Structures}, {Vibrational} {Frequencies}, and {Noncovalent} {Interactions} of {Large} {Molecular} {Systems} {Parametrized} for {All} spd-{Block} {Elements} ({Z} = 1–86). \emph{J. Chem. Theory Comput.} \textbf{2017}, \emph{13}, 1989--2009\relax
\mciteBstWouldAddEndPuncttrue
\mciteSetBstMidEndSepPunct{\mcitedefaultmidpunct}
{\mcitedefaultendpunct}{\mcitedefaultseppunct}\relax
\EndOfBibitem
\bibitem[Bannwarth \latin{et~al.}(2019)Bannwarth, Ehlert, and Grimme]{bannwarth_gfn2-xtbaccurate_2019}
Bannwarth,~C.; Ehlert,~S.; Grimme,~S. {GFN2}-{xTB}—{An} {Accurate} and {Broadly} {Parametrized} {Self}-{Consistent} {Tight}-{Binding} {Quantum} {Chemical} {Method} with {Multipole} {Electrostatics} and {Density}-{Dependent} {Dispersion} {Contributions}. \emph{J. Chem. Theory Comput.} \textbf{2019}, \emph{15}, 1652--1671\relax
\mciteBstWouldAddEndPuncttrue
\mciteSetBstMidEndSepPunct{\mcitedefaultmidpunct}
{\mcitedefaultendpunct}{\mcitedefaultseppunct}\relax
\EndOfBibitem
\bibitem[Grimme \latin{et~al.}(2023)Grimme, Müller, and Hansen]{grimme_non-self-consistent_2023}
Grimme,~S.; Müller,~M.; Hansen,~A. A non-self-consistent tight-binding electronic structure potential in a polarized double- \textit{\${\textbackslash}zeta\$} basis set for all \textit{spd} -block elements up to {Z} = 86. \emph{J. Chem. Phys.} \textbf{2023}, \emph{158}, 124111\relax
\mciteBstWouldAddEndPuncttrue
\mciteSetBstMidEndSepPunct{\mcitedefaultmidpunct}
{\mcitedefaultendpunct}{\mcitedefaultseppunct}\relax
\EndOfBibitem
\bibitem[Froitzheim \latin{et~al.}(2025)Froitzheim, Müller, Hansen, and Grimme]{froitzheim_g-xtb_2025}
Froitzheim,~T.; Müller,~M.; Hansen,~A.; Grimme,~S. g-{xTB}: {A} {General}-{Purpose} {Extended} {Tight}-{Binding} {Electronic} {Structure} {Method} {For} the {Elements} {H} to {Lr} ({Z}=1–103). 2025; \url{https://chemrxiv.org/engage/chemrxiv/article-details/685434533ba0887c335fc974}\relax
\mciteBstWouldAddEndPuncttrue
\mciteSetBstMidEndSepPunct{\mcitedefaultmidpunct}
{\mcitedefaultendpunct}{\mcitedefaultseppunct}\relax
\EndOfBibitem
\bibitem[Grimme \latin{et~al.}(2010)Grimme, Antony, Ehrlich, and Krieg]{grimme_consistent_2010}
Grimme,~S.; Antony,~J.; Ehrlich,~S.; Krieg,~H. A consistent and accurate \textit{ab initio} parametrization of density functional dispersion correction ({DFT}-{D}) for the 94 elements {H}-{Pu}. \emph{J. Chem. Phys.} \textbf{2010}, \emph{132}, 154104\relax
\mciteBstWouldAddEndPuncttrue
\mciteSetBstMidEndSepPunct{\mcitedefaultmidpunct}
{\mcitedefaultendpunct}{\mcitedefaultseppunct}\relax
\EndOfBibitem
\bibitem[Nishizawa \latin{et~al.}(2016)Nishizawa, Nishimura, Kobayashi, Irle, and Nakai]{nishizawa_three_2016}
Nishizawa,~H.; Nishimura,~Y.; Kobayashi,~M.; Irle,~S.; Nakai,~H. Three pillars for achieving quantum mechanical molecular dynamics simulations of huge systems: {Divide}-and-conquer, density-functional tight-binding, and massively parallel computation. \emph{J. Comput. Chem.} \textbf{2016}, \emph{37}, 1983--1992\relax
\mciteBstWouldAddEndPuncttrue
\mciteSetBstMidEndSepPunct{\mcitedefaultmidpunct}
{\mcitedefaultendpunct}{\mcitedefaultseppunct}\relax
\EndOfBibitem
\bibitem[Warshel and Levitt(1976)Warshel, and Levitt]{warshel_theoretical_1976}
Warshel,~A.; Levitt,~M. Theoretical studies of enzymic reactions: {Dielectric}, electrostatic and steric stabilization of the carbonium ion in the reaction of lysozyme. \emph{J. Mol. Biol.} \textbf{1976}, \emph{103}, 227--249\relax
\mciteBstWouldAddEndPuncttrue
\mciteSetBstMidEndSepPunct{\mcitedefaultmidpunct}
{\mcitedefaultendpunct}{\mcitedefaultseppunct}\relax
\EndOfBibitem
\bibitem[Senn and Thiel(2009)Senn, and Thiel]{senn_qmmm_2009}
Senn,~H.~M.; Thiel,~W. {QM}/{MM} {Methods} for {Biomolecular} {Systems}. \emph{Angew. Chem. Int. Ed.} \textbf{2009}, \emph{48}, 1198--1229\relax
\mciteBstWouldAddEndPuncttrue
\mciteSetBstMidEndSepPunct{\mcitedefaultmidpunct}
{\mcitedefaultendpunct}{\mcitedefaultseppunct}\relax
\EndOfBibitem
\bibitem[Dapprich \latin{et~al.}(1999)Dapprich, Komáromi, Byun, Morokuma, and Frisch]{dapprich_new_1999}
Dapprich,~S.; Komáromi,~I.; Byun,~K.~S.; Morokuma,~K.; Frisch,~M.~J. A new {ONIOM} implementation in {Gaussian98}. {Part} {I}. {The} calculation of energies, gradients, vibrational frequencies and electric field derivatives1. \emph{Journal of Molecular Structure: THEOCHEM} \textbf{1999}, \emph{461-462}, 1--21\relax
\mciteBstWouldAddEndPuncttrue
\mciteSetBstMidEndSepPunct{\mcitedefaultmidpunct}
{\mcitedefaultendpunct}{\mcitedefaultseppunct}\relax
\EndOfBibitem
\bibitem[Plett \latin{et~al.}(2023)Plett, Katbashev, Ehlert, Grimme, and Bursch]{plett_oniom_2023}
Plett,~C.; Katbashev,~A.; Ehlert,~S.; Grimme,~S.; Bursch,~M. {ONIOM} meets xtb: efficient, accurate, and robust multi-layer simulations across the periodic table. \emph{Phys. Chem. Chem. Phys.} \textbf{2023}, \emph{25}, 17860--17868\relax
\mciteBstWouldAddEndPuncttrue
\mciteSetBstMidEndSepPunct{\mcitedefaultmidpunct}
{\mcitedefaultendpunct}{\mcitedefaultseppunct}\relax
\EndOfBibitem
\bibitem[Jacob and Neugebauer(2014)Jacob, and Neugebauer]{jacob_subsystem_2014}
Jacob,~C.~R.; Neugebauer,~J. Subsystem density-functional theory: {Subsystem} density-functional theory. \emph{Wiley Interdiscip. Rev. Comput. Mol. Sci.} \textbf{2014}, \emph{4}, 325--362\relax
\mciteBstWouldAddEndPuncttrue
\mciteSetBstMidEndSepPunct{\mcitedefaultmidpunct}
{\mcitedefaultendpunct}{\mcitedefaultseppunct}\relax
\EndOfBibitem
\bibitem[Kitaura \latin{et~al.}(1999)Kitaura, Ikeo, Asada, Nakano, and Uebayasi]{kitaura_fragment_1999}
Kitaura,~K.; Ikeo,~E.; Asada,~T.; Nakano,~T.; Uebayasi,~M. Fragment molecular orbital method: an approximate computational method for large molecules. \emph{Chem. Phys. Lett.} \textbf{1999}, \emph{313}, 701--706\relax
\mciteBstWouldAddEndPuncttrue
\mciteSetBstMidEndSepPunct{\mcitedefaultmidpunct}
{\mcitedefaultendpunct}{\mcitedefaultseppunct}\relax
\EndOfBibitem
\bibitem[Nakano \latin{et~al.}(2000)Nakano, Kaminuma, Sato, Akiyama, Uebayasi, and Kitaura]{nakano_fragment_2000}
Nakano,~T.; Kaminuma,~T.; Sato,~T.; Akiyama,~Y.; Uebayasi,~M.; Kitaura,~K. Fragment molecular orbital method: application to polypeptides. \emph{Chem. Phys. Lett.} \textbf{2000}, \emph{318}, 614--618\relax
\mciteBstWouldAddEndPuncttrue
\mciteSetBstMidEndSepPunct{\mcitedefaultmidpunct}
{\mcitedefaultendpunct}{\mcitedefaultseppunct}\relax
\EndOfBibitem
\bibitem[Nakano \latin{et~al.}(2002)Nakano, Kaminuma, Sato, Fukuzawa, Akiyama, Uebayasi, and Kitaura]{nakano_fragment_2002}
Nakano,~T.; Kaminuma,~T.; Sato,~T.; Fukuzawa,~K.; Akiyama,~Y.; Uebayasi,~M.; Kitaura,~K. Fragment molecular orbital method: use of approximate electrostatic potential. \emph{Chem. Phys. Lett.} \textbf{2002}, \emph{351}, 475--480\relax
\mciteBstWouldAddEndPuncttrue
\mciteSetBstMidEndSepPunct{\mcitedefaultmidpunct}
{\mcitedefaultendpunct}{\mcitedefaultseppunct}\relax
\EndOfBibitem
\bibitem[Fedorov and Kitaura(2007)Fedorov, and Kitaura]{fedorov_extending_2007}
Fedorov,~D.~G.; Kitaura,~K. Extending the {Power} of {Quantum} {Chemistry} to {Large} {Systems} with the {Fragment} {Molecular} {Orbital} {Method}. \emph{J. Phys. Chem. A} \textbf{2007}, \emph{111}, 6904--6914\relax
\mciteBstWouldAddEndPuncttrue
\mciteSetBstMidEndSepPunct{\mcitedefaultmidpunct}
{\mcitedefaultendpunct}{\mcitedefaultseppunct}\relax
\EndOfBibitem
\bibitem[Fedorov \latin{et~al.}(2010)Fedorov, Slipchenko, and Kitaura]{fedorov_systematic_2010}
Fedorov,~D.~G.; Slipchenko,~L.~V.; Kitaura,~K. Systematic {Study} of the {Embedding} {Potential} {Description} in the {Fragment} {Molecular} {Orbital} {Method}. \emph{J. Phys. Chem. A} \textbf{2010}, \emph{114}, 8742--8753\relax
\mciteBstWouldAddEndPuncttrue
\mciteSetBstMidEndSepPunct{\mcitedefaultmidpunct}
{\mcitedefaultendpunct}{\mcitedefaultseppunct}\relax
\EndOfBibitem
\bibitem[Fedorov and Kitaura(2004)Fedorov, and Kitaura]{fedorov_importance_2004}
Fedorov,~D.~G.; Kitaura,~K. The importance of three-body terms in the fragment molecular orbital method. \emph{J. Chem. Phys.} \textbf{2004}, \emph{120}, 6832--6840\relax
\mciteBstWouldAddEndPuncttrue
\mciteSetBstMidEndSepPunct{\mcitedefaultmidpunct}
{\mcitedefaultendpunct}{\mcitedefaultseppunct}\relax
\EndOfBibitem
\bibitem[Fedorov and Kitaura(2006)Fedorov, and Kitaura]{fedorov_three-body_2006}
Fedorov,~D.~G.; Kitaura,~K. The three-body fragment molecular orbital method for accurate calculations of large systems. \emph{Chem. Phys. Lett.} \textbf{2006}, \emph{433}, 182--187\relax
\mciteBstWouldAddEndPuncttrue
\mciteSetBstMidEndSepPunct{\mcitedefaultmidpunct}
{\mcitedefaultendpunct}{\mcitedefaultseppunct}\relax
\EndOfBibitem
\bibitem[Fedorov and Kitaura(2004)Fedorov, and Kitaura]{fedorov_second_2004}
Fedorov,~D.~G.; Kitaura,~K. Second order {Møller}-{Plesset} perturbation theory based upon the fragment molecular orbital method. \emph{J. Chem. Phys.} \textbf{2004}, \emph{121}, 2483--2490\relax
\mciteBstWouldAddEndPuncttrue
\mciteSetBstMidEndSepPunct{\mcitedefaultmidpunct}
{\mcitedefaultendpunct}{\mcitedefaultseppunct}\relax
\EndOfBibitem
\bibitem[Fedorov \latin{et~al.}(2005)Fedorov, Ishida, and Kitaura]{fedorov_multilayer_2005}
Fedorov,~D.~G.; Ishida,~T.; Kitaura,~K. Multilayer {Formulation} of the {Fragment} {Molecular} {Orbital} {Method} ({FMO}). \emph{J. Phys. Chem. A} \textbf{2005}, \emph{109}, 2638--2646\relax
\mciteBstWouldAddEndPuncttrue
\mciteSetBstMidEndSepPunct{\mcitedefaultmidpunct}
{\mcitedefaultendpunct}{\mcitedefaultseppunct}\relax
\EndOfBibitem
\bibitem[Tanaka \latin{et~al.}(2014)Tanaka, Mochizuki, Komeiji, Okiyama, and Fukuzawa]{tanaka_electron-correlated_2014}
Tanaka,~S.; Mochizuki,~Y.; Komeiji,~Y.; Okiyama,~Y.; Fukuzawa,~K. Electron-correlated fragment-molecular-orbital calculations for biomolecular and nano systems. \emph{Phys. Chem. Chem. Phys.} \textbf{2014}, \emph{16}, 10310--10344\relax
\mciteBstWouldAddEndPuncttrue
\mciteSetBstMidEndSepPunct{\mcitedefaultmidpunct}
{\mcitedefaultendpunct}{\mcitedefaultseppunct}\relax
\EndOfBibitem
\bibitem[Fukuzawa and Tanaka(2022)Fukuzawa, and Tanaka]{fukuzawa_fragment_2022}
Fukuzawa,~K.; Tanaka,~S. Fragment molecular orbital calculations for biomolecules. \emph{Curr. Opin. Struct. Biol.} \textbf{2022}, \emph{72}, 127--134\relax
\mciteBstWouldAddEndPuncttrue
\mciteSetBstMidEndSepPunct{\mcitedefaultmidpunct}
{\mcitedefaultendpunct}{\mcitedefaultseppunct}\relax
\EndOfBibitem
\bibitem[Sawada \latin{et~al.}(2010)Sawada, Fedorov, and Kitaura]{sawada_binding_2010}
Sawada,~T.; Fedorov,~D.~G.; Kitaura,~K. Binding of {Influenza} {A} {Virus} {Hemagglutinin} to the {Sialoside} {Receptor} {Is} {Not} {Controlled} by the {Homotropic} {Allosteric} {Effect}. \emph{J. Phys. Chem. B} \textbf{2010}, \emph{114}, 15700--15705\relax
\mciteBstWouldAddEndPuncttrue
\mciteSetBstMidEndSepPunct{\mcitedefaultmidpunct}
{\mcitedefaultendpunct}{\mcitedefaultseppunct}\relax
\EndOfBibitem
\bibitem[Fedorov(2022)]{fedorov_polarization_2022}
Fedorov,~D.~G. Polarization energies in the fragment molecular orbital method. \emph{J. Comput. Chem.} \textbf{2022}, \emph{43}, 1094--1103\relax
\mciteBstWouldAddEndPuncttrue
\mciteSetBstMidEndSepPunct{\mcitedefaultmidpunct}
{\mcitedefaultendpunct}{\mcitedefaultseppunct}\relax
\EndOfBibitem
\bibitem[Fedorov \latin{et~al.}(2025)Fedorov, Inostroza, Courbiere, Guegan, Contreras-García, and Mori]{fedorov_decomposition_2025}
Fedorov,~D.~G.; Inostroza,~D.; Courbiere,~B.; Guegan,~F.; Contreras-García,~J.; Mori,~S. Decomposition {Analysis} for {Visualization} of {Noncovalent} {Interactions} {Based} on the {Fragment} {Molecular} {Orbital} {Method}. \emph{J. Chem. Theory Comput.} \textbf{2025}, \relax
\mciteBstWouldAddEndPunctfalse
\mciteSetBstMidEndSepPunct{\mcitedefaultmidpunct}
{}{\mcitedefaultseppunct}\relax
\EndOfBibitem
\bibitem[Yang(1991)]{yang_direct_1991}
Yang,~W. Direct calculation of electron density in density-functional theory. \emph{Phys. Rev. Lett.} \textbf{1991}, \emph{66}, 1438--1441\relax
\mciteBstWouldAddEndPuncttrue
\mciteSetBstMidEndSepPunct{\mcitedefaultmidpunct}
{\mcitedefaultendpunct}{\mcitedefaultseppunct}\relax
\EndOfBibitem
\bibitem[Yang and Lee(1995)Yang, and Lee]{yang_densitymatrix_1995}
Yang,~W.; Lee,~T. A density‐matrix divide‐and‐conquer approach for electronic structure calculations of large molecules. \emph{J. Chem. Phys.} \textbf{1995}, \emph{103}, 5674--5678\relax
\mciteBstWouldAddEndPuncttrue
\mciteSetBstMidEndSepPunct{\mcitedefaultmidpunct}
{\mcitedefaultendpunct}{\mcitedefaultseppunct}\relax
\EndOfBibitem
\bibitem[Nakai \latin{et~al.}(2023)Nakai, Kobayashi, Yoshikawa, Seino, Ikabata, and Nishimura]{nakai_divide-and-conquer_2023}
Nakai,~H.; Kobayashi,~M.; Yoshikawa,~T.; Seino,~J.; Ikabata,~Y.; Nishimura,~Y. Divide-and-{Conquer} {Linear}-{Scaling} {Quantum} {Chemical} {Computations}. \emph{J. Phys. Chem. A} \textbf{2023}, \emph{127}, 589--618\relax
\mciteBstWouldAddEndPuncttrue
\mciteSetBstMidEndSepPunct{\mcitedefaultmidpunct}
{\mcitedefaultendpunct}{\mcitedefaultseppunct}\relax
\EndOfBibitem
\bibitem[Akama \latin{et~al.}(2007)Akama, Kobayashi, and Nakai]{akama_implementation_2007}
Akama,~T.; Kobayashi,~M.; Nakai,~H. Implementation of divide-and-conquer method including {Hartree}-{Fock} exchange interaction. \emph{J. Comput. Chem.} \textbf{2007}, \emph{28}, 2003--2012\relax
\mciteBstWouldAddEndPuncttrue
\mciteSetBstMidEndSepPunct{\mcitedefaultmidpunct}
{\mcitedefaultendpunct}{\mcitedefaultseppunct}\relax
\EndOfBibitem
\bibitem[Kobayashi and Nakai(2008)Kobayashi, and Nakai]{kobayashi_extension_2008}
Kobayashi,~M.; Nakai,~H. Extension of linear-scaling divide-and-conquer-based correlation method to coupled cluster theory with singles and doubles excitations. \emph{J. Chem. Phys.} \textbf{2008}, \emph{129}, 044103\relax
\mciteBstWouldAddEndPuncttrue
\mciteSetBstMidEndSepPunct{\mcitedefaultmidpunct}
{\mcitedefaultendpunct}{\mcitedefaultseppunct}\relax
\EndOfBibitem
\bibitem[Kobayashi \latin{et~al.}(2010)Kobayashi, Yoshikawa, and Nakai]{kobayashi_divide-and-conquer_2010}
Kobayashi,~M.; Yoshikawa,~T.; Nakai,~H. Divide-and-conquer self-consistent field calculation for open-shell systems: {Implementation} and application. \emph{Chem. Phys. Lett.} \textbf{2010}, \emph{500}, 172--177\relax
\mciteBstWouldAddEndPuncttrue
\mciteSetBstMidEndSepPunct{\mcitedefaultmidpunct}
{\mcitedefaultendpunct}{\mcitedefaultseppunct}\relax
\EndOfBibitem
\bibitem[Nakai \latin{et~al.}(2016)Nakai, Sakti, and Nishimura]{nakai_divide-and-conquer-type_2016}
Nakai,~H.; Sakti,~A.~W.; Nishimura,~Y. Divide-and-{Conquer}-{Type} {Density}-{Functional} {Tight}-{Binding} {Molecular} {Dynamics} {Simulations} of {Proton} {Diffusion} in a {Bulk} {Water} {System}. \emph{J. Phys. Chem. B} \textbf{2016}, \emph{120}, 217--221\relax
\mciteBstWouldAddEndPuncttrue
\mciteSetBstMidEndSepPunct{\mcitedefaultmidpunct}
{\mcitedefaultendpunct}{\mcitedefaultseppunct}\relax
\EndOfBibitem
\bibitem[Nishimura and Nakai(2019)Nishimura, and Nakai]{nishimura_dcdftbmd_2019}
Nishimura,~Y.; Nakai,~H. Dcdftbmd: {Divide}‐and‐{Conquer} {Density} {Functional} {Tight}‐{Binding} {Program} for {Huge}‐{System} {Quantum} {Mechanical} {Molecular} {Dynamics} {Simulations}. \emph{J. Comput. Chem.} \textbf{2019}, \emph{40}, 1538--1549\relax
\mciteBstWouldAddEndPuncttrue
\mciteSetBstMidEndSepPunct{\mcitedefaultmidpunct}
{\mcitedefaultendpunct}{\mcitedefaultseppunct}\relax
\EndOfBibitem
\bibitem[Nishimoto \latin{et~al.}(2014)Nishimoto, Fedorov, and Irle]{nishimoto_density-functional_2014}
Nishimoto,~Y.; Fedorov,~D.~G.; Irle,~S. Density-{Functional} {Tight}-{Binding} {Combined} with the {Fragment} {Molecular} {Orbital} {Method}. \emph{J. Chem. Theory Comput.} \textbf{2014}, \emph{10}, 4801--4812\relax
\mciteBstWouldAddEndPuncttrue
\mciteSetBstMidEndSepPunct{\mcitedefaultmidpunct}
{\mcitedefaultendpunct}{\mcitedefaultseppunct}\relax
\EndOfBibitem
\bibitem[Nishimoto \latin{et~al.}(2015)Nishimoto, Nakata, Fedorov, and Irle]{nishimoto_large-scale_2015}
Nishimoto,~Y.; Nakata,~H.; Fedorov,~D.~G.; Irle,~S. Large-{Scale} {Quantum}-{Mechanical} {Molecular} {Dynamics} {Simulations} {Using} {Density}-{Functional} {Tight}-{Binding} {Combined} with the {Fragment} {Molecular} {Orbital} {Method}. \emph{J. Phys. Chem. Lett.} \textbf{2015}, \emph{6}, 5034--5039\relax
\mciteBstWouldAddEndPuncttrue
\mciteSetBstMidEndSepPunct{\mcitedefaultmidpunct}
{\mcitedefaultendpunct}{\mcitedefaultseppunct}\relax
\EndOfBibitem
\bibitem[Nishimoto \latin{et~al.}(2015)Nishimoto, Fedorov, and Irle]{nishimoto_third-order_2015}
Nishimoto,~Y.; Fedorov,~D.~G.; Irle,~S. Third-order density-functional tight-binding combined with the fragment molecular orbital method. \emph{Chem. Phys. Lett.} \textbf{2015}, \emph{636}, 90--96\relax
\mciteBstWouldAddEndPuncttrue
\mciteSetBstMidEndSepPunct{\mcitedefaultmidpunct}
{\mcitedefaultendpunct}{\mcitedefaultseppunct}\relax
\EndOfBibitem
\bibitem[Nishimoto and Fedorov(2017)Nishimoto, and Fedorov]{nishimoto_three-body_2017}
Nishimoto,~Y.; Fedorov,~D.~G. Three-body expansion of the fragment molecular orbital method combined with density-functional tight-binding. \emph{J. Comput. Chem.} \textbf{2017}, \emph{38}, 406--418\relax
\mciteBstWouldAddEndPuncttrue
\mciteSetBstMidEndSepPunct{\mcitedefaultmidpunct}
{\mcitedefaultendpunct}{\mcitedefaultseppunct}\relax
\EndOfBibitem
\bibitem[Nishimoto and Fedorov(2016)Nishimoto, and Fedorov]{nishimoto_fragment_2016}
Nishimoto,~Y.; Fedorov,~D.~G. The fragment molecular orbital method combined with density-functional tight-binding and the polarizable continuum model. \emph{Phys. Chem. Chem. Phys.} \textbf{2016}, \emph{18}, 22047--22061\relax
\mciteBstWouldAddEndPuncttrue
\mciteSetBstMidEndSepPunct{\mcitedefaultmidpunct}
{\mcitedefaultendpunct}{\mcitedefaultseppunct}\relax
\EndOfBibitem
\bibitem[Vuong \latin{et~al.}(2019)Vuong, Nishimoto, Fedorov, Sumpter, Niehaus, and Irle]{vuong_fragment_2019}
Vuong,~V.~Q.; Nishimoto,~Y.; Fedorov,~D.~G.; Sumpter,~B.~G.; Niehaus,~T.~A.; Irle,~S. The {Fragment} {Molecular} {Orbital} {Method} {Based} on {Long}-{Range} {Corrected} {Density}-{Functional} {Tight}-{Binding}. \emph{J. Chem. Theory Comput.} \textbf{2019}, \emph{15}, 3008--3020\relax
\mciteBstWouldAddEndPuncttrue
\mciteSetBstMidEndSepPunct{\mcitedefaultmidpunct}
{\mcitedefaultendpunct}{\mcitedefaultseppunct}\relax
\EndOfBibitem
\bibitem[Nakata \latin{et~al.}(2016)Nakata, Nishimoto, and Fedorov]{nakata_analytic_2016}
Nakata,~H.; Nishimoto,~Y.; Fedorov,~D.~G. Analytic second derivative of the energy for density-functional tight-binding combined with the fragment molecular orbital method. \emph{J. Chem. Phys.} \textbf{2016}, \emph{145}, 044113\relax
\mciteBstWouldAddEndPuncttrue
\mciteSetBstMidEndSepPunct{\mcitedefaultmidpunct}
{\mcitedefaultendpunct}{\mcitedefaultseppunct}\relax
\EndOfBibitem
\bibitem[Nishimoto and Fedorov(2021)Nishimoto, and Fedorov]{nishimoto_fragment_2021}
Nishimoto,~Y.; Fedorov,~D.~G. The fragment molecular orbital method combined with density-functional tight-binding and periodic boundary conditions. \emph{J. Chem. Phys.} \textbf{2021}, \emph{154}, 111102\relax
\mciteBstWouldAddEndPuncttrue
\mciteSetBstMidEndSepPunct{\mcitedefaultmidpunct}
{\mcitedefaultendpunct}{\mcitedefaultseppunct}\relax
\EndOfBibitem
\bibitem[Einsele \latin{et~al.}(2023)Einsele, Hoche, and Mitrić]{einsele_long-range_2023}
Einsele,~R.; Hoche,~J.; Mitrić,~R. Long-range corrected fragment molecular orbital density functional tight-binding method for excited states in large molecular systems. \emph{J. Chem. Phys.} \textbf{2023}, \emph{158}, 044121\relax
\mciteBstWouldAddEndPuncttrue
\mciteSetBstMidEndSepPunct{\mcitedefaultmidpunct}
{\mcitedefaultendpunct}{\mcitedefaultseppunct}\relax
\EndOfBibitem
\bibitem[Einsele \latin{et~al.}(2024)Einsele, Philipp, and Mitrić]{einsele_fmo-lc-tddftb_2024}
Einsele,~R.; Philipp,~L.~N.; Mitrić,~R. {FMO}-{LC}-{TDDFTB} method for excited states of large molecular assemblies in the strong light-matter coupling regime. \emph{J. Chem. Phys.} \textbf{2024}, \emph{161}, 154106\relax
\mciteBstWouldAddEndPuncttrue
\mciteSetBstMidEndSepPunct{\mcitedefaultmidpunct}
{\mcitedefaultendpunct}{\mcitedefaultseppunct}\relax
\EndOfBibitem
\bibitem[Einsele and Mitrić(2024)Einsele, and Mitrić]{einsele_nonadiabatic_2024}
Einsele,~R.; Mitrić,~R. Nonadiabatic {Exciton} {Dynamics} and {Energy} {Gradients} in the {Framework} of {FMO}-{LC}-{TDDFTB}. \emph{J. Chem. Theory Comput.} \textbf{2024}, \relax
\mciteBstWouldAddEndPunctfalse
\mciteSetBstMidEndSepPunct{\mcitedefaultmidpunct}
{}{\mcitedefaultseppunct}\relax
\EndOfBibitem
\bibitem[Einsele \latin{et~al.}(2025)Einsele, Miao, Philipp, and Mitrić]{einsele_dialect_2025}
Einsele,~R.; Miao,~X.; Philipp,~L.~N.; Mitrić,~R. {DIALECT}, a {Software} {Package} for {Exciton} {Spectra} and {Dynamics} in {Large} {Molecular} {Assemblies} from {Weak} to {Strong} {Light}–{Matter} {Coupling} {Regimes}. \emph{J. Phys. Chem. A} \textbf{2025}, \relax
\mciteBstWouldAddEndPunctfalse
\mciteSetBstMidEndSepPunct{\mcitedefaultmidpunct}
{}{\mcitedefaultseppunct}\relax
\EndOfBibitem
\bibitem[Nishida \latin{et~al.}(2025)Nishida, Fujiwara, Taketsugu, and Kobayashi]{nishida_divide-and-conquer_2025}
Nishida,~M.; Fujiwara,~K.; Taketsugu,~T.; Kobayashi,~M. Divide-{And}-{Conquer} {Extended} {Tight}-{Binding} {Molecular} {Dynamics}: {A} {General}-{Purpose}, {Very} {Large}-{Scale} {Quantum} {Molecular} {Dynamics} {Method}. \emph{J. Comput. Chem.} \textbf{2025}, \emph{46}, e70255\relax
\mciteBstWouldAddEndPuncttrue
\mciteSetBstMidEndSepPunct{\mcitedefaultmidpunct}
{\mcitedefaultendpunct}{\mcitedefaultseppunct}\relax
\EndOfBibitem
\bibitem[Nagata \latin{et~al.}(2011)Nagata, Brorsen, Fedorov, Kitaura, and Gordon]{nagata_fully_2011}
Nagata,~T.; Brorsen,~K.; Fedorov,~D.~G.; Kitaura,~K.; Gordon,~M.~S. Fully analytic energy gradient in the fragment molecular orbital method. \emph{J. Chem. Phys.} \textbf{2011}, \emph{134}, 124115\relax
\mciteBstWouldAddEndPuncttrue
\mciteSetBstMidEndSepPunct{\mcitedefaultmidpunct}
{\mcitedefaultendpunct}{\mcitedefaultseppunct}\relax
\EndOfBibitem
\bibitem[Fedorov \latin{et~al.}(2008)Fedorov, Jensen, Deka, and Kitaura]{fedorov_covalent_2008}
Fedorov,~D.~G.; Jensen,~J.~H.; Deka,~R.~C.; Kitaura,~K. Covalent {Bond} {Fragmentation} {Suitable} {To} {Describe} {Solids} in the {Fragment} {Molecular} {Orbital} {Method}. \emph{J. Phys. Chem. A} \textbf{2008}, \emph{112}, 11808--11816\relax
\mciteBstWouldAddEndPuncttrue
\mciteSetBstMidEndSepPunct{\mcitedefaultmidpunct}
{\mcitedefaultendpunct}{\mcitedefaultseppunct}\relax
\EndOfBibitem
\bibitem[Nagata \latin{et~al.}(2010)Nagata, Fedorov, and Kitaura]{nagata_importance_2010}
Nagata,~T.; Fedorov,~D.~G.; Kitaura,~K. Importance of the hybrid orbital operator derivative term for the energy gradient in the fragment molecular orbital method. \emph{Chem. Phys. Lett.} \textbf{2010}, \emph{492}, 302--308\relax
\mciteBstWouldAddEndPuncttrue
\mciteSetBstMidEndSepPunct{\mcitedefaultmidpunct}
{\mcitedefaultendpunct}{\mcitedefaultseppunct}\relax
\EndOfBibitem
\bibitem[Nakata and Fedorov(2020)Nakata, and Fedorov]{nakata_analytic_2020}
Nakata,~H.; Fedorov,~D.~G. Analytic first and second derivatives of the energy in the fragment molecular orbital method combined with molecular mechanics. \emph{Int. J. Quantum Chem.} \textbf{2020}, \emph{120}\relax
\mciteBstWouldAddEndPuncttrue
\mciteSetBstMidEndSepPunct{\mcitedefaultmidpunct}
{\mcitedefaultendpunct}{\mcitedefaultseppunct}\relax
\EndOfBibitem
\bibitem[Macrae \latin{et~al.}(2020)Macrae, Sovago, Cottrell, Galek, McCabe, Pidcock, Platings, Shields, Stevens, Towler, and Wood]{macrae_mercury_2020}
Macrae,~C.~F.; Sovago,~I.; Cottrell,~S.~J.; Galek,~P. T.~A.; McCabe,~P.; Pidcock,~E.; Platings,~M.; Shields,~G.~P.; Stevens,~J.~S.; Towler,~M.; Wood,~P.~A. Mercury 4.0: from visualization to analysis, design and prediction. \emph{J. Appl. Crystallogr.} \textbf{2020}, \emph{53}, 226--235\relax
\mciteBstWouldAddEndPuncttrue
\mciteSetBstMidEndSepPunct{\mcitedefaultmidpunct}
{\mcitedefaultendpunct}{\mcitedefaultseppunct}\relax
\EndOfBibitem
\bibitem[Mason(1964)]{mason_crystallography_1964}
Mason,~R. The crystallography of anthracene at 95°{K} and 290°{K}. \emph{Acta Cryst} \textbf{1964}, \emph{17}, 547--555\relax
\mciteBstWouldAddEndPuncttrue
\mciteSetBstMidEndSepPunct{\mcitedefaultmidpunct}
{\mcitedefaultendpunct}{\mcitedefaultseppunct}\relax
\EndOfBibitem
\bibitem[Holmes \latin{et~al.}(1999)Holmes, Kumaraswamy, Matzger, and Vollhardt]{holmes_nature_1999}
Holmes,~D.; Kumaraswamy,~S.; Matzger,~A.~J.; Vollhardt,~K. P.~C. On the {Nature} of {Nonplanarity} in the [{N}]{Phenylenes}. \emph{Chem. Eur. J.} \textbf{1999}, \emph{5}, 3399--3412\relax
\mciteBstWouldAddEndPuncttrue
\mciteSetBstMidEndSepPunct{\mcitedefaultmidpunct}
{\mcitedefaultendpunct}{\mcitedefaultseppunct}\relax
\EndOfBibitem
\bibitem[Martínez \latin{et~al.}(2009)Martínez, Andrade, Birgin, and Martínez]{martinez_packmol_2009}
Martínez,~L.; Andrade,~R.; Birgin,~E.~G.; Martínez,~J.~M. {PACKMOL}: {A} package for building initial configurations for molecular dynamics simulations. \emph{J. Comput. Chem.} \textbf{2009}, \emph{30}, 2157--2164\relax
\mciteBstWouldAddEndPuncttrue
\mciteSetBstMidEndSepPunct{\mcitedefaultmidpunct}
{\mcitedefaultendpunct}{\mcitedefaultseppunct}\relax
\EndOfBibitem
\bibitem[Nishimoto and Fedorov(2018)Nishimoto, and Fedorov]{nishimoto_adaptive_2018}
Nishimoto,~Y.; Fedorov,~D.~G. Adaptive frozen orbital treatment for the fragment molecular orbital method combined with density-functional tight-binding. \emph{J. Chem. Phys.} \textbf{2018}, \emph{148}, 064115\relax
\mciteBstWouldAddEndPuncttrue
\mciteSetBstMidEndSepPunct{\mcitedefaultmidpunct}
{\mcitedefaultendpunct}{\mcitedefaultseppunct}\relax
\EndOfBibitem
\end{mcitethebibliography}
